# A program logic for higher-order procedural variables and non-local jumps

**Tristan Crolard** **Emmanuel Polonowski**

*December 2011*

**TR–LACL–2011–4**

**Laboratoire d'Algorithmique, Complexité et Logique (LACL)**
**Département d'Informatique**
**Université Paris-Est Créteil – Val de Marne, Faculté des Sciences et Technologie**
61, Avenue du Général de Gaulle, 94010 Créteil cedex, France
Tel.: (33)(1) 45 17 16 47, Fax: (33)(1) 45 17 66 01

Tristan Crolard, Emmanuel Polonowski.
*A program logic for higher-order procedural variables and non-local jumps*

# A program logic for higher-order procedural variables and non-local jumps

T. Crolard[a,1], E. Polonowski[a,1]

[a]*LACL, Université Paris Est, 61 avenue du Général de Gaulle, 94010 Créteil Cedex, France*

**Abstract**

Relying on the formulae-as-types paradigm for classical logic, we define a program logic for an imperative language with higher-order procedural variables and non-local jumps. Then, we show how to derive a sound program logic for this programming language. As a by-product, we obtain a non-dependent type system which is more permissive than what is usually found in statically typed imperative languages. As a generic example, we encode imperative versions of delimited continuations operators **shift** and **reset**.

## 1 Introduction

In his seminal series of papers [44, 45, 46], Landin proposed a direct translation of an idealized Algol into the $\lambda$-calculus. This translation required to extend the $\lambda$-calculus with a new operator **J** in order to handle non-local jumps in Algol. This operator, which was described in detail in [47] (see also [74] for an introduction), is the father to all control operators in functional languages (such as the famous **call/cc** of Scheme [40] or Standard ML of New Jersey [32]). The syntactic theory of control has subsequently been explored thoroughly by Felleisen [21].

A type system for control operators which extends the so-called Curry-Howard correspondence [16, 39] to classical logic first appeared in Griffin's pioneering work [31], and was immediately generalized to first-order dependent types (and Peano's arithmetic) by Murthy in his thesis [56]. The following years, this extension of the formulas-as-types paradigm to classical logic has then studied by several researchers, for instance in [7, 69, 19, 41, 62] and many others since.

It is thus tempting to revisit Landin's work in the light of the formulas-as-types interpretation of control. Indeed, it is notoriously difficult to derive a sound program logic for an imperative language with procedures and non-local jumps [60], especially in the presence of local variables and higher-order procedures [73]. On the other hand, adding first-order dependent types to such an imperative language, and translating type derivations into proof derivations appears more tractable. The difficult to obtain program logic is then mechanically derived. Moreover, this logic permits by construction to deal elegantly with *mutable* higher-order procedural variables.

As a stepping stone, we focus in this paper on Peano's arithmetic. The corresponding functional language (through the proofs-as-programs paradigm) is thus an extension of Gödel System T [30] with control operators as described in [56]. We shall use instead a variant which was proposed by Leivant [48, 49] (and rediscovered independently by Krivine and Parigot in the second-order framework [42]). The main advantage of this variant is that it requires no encoding in formulas (with Gödel numbers) to reason about functional programs. Moreover it can be extended to any other algebraic datatypes (such as lists or trees). In this paper, the control operators are given an indirect semantics through a call-by-value CPS transform (we do not consider any direct style semantics). As noticed in [56], this CPS transformation operates a variant of Kuroda's translation on dependent types [43].

The imperative counterpart of Gödel System T [30] (called $\text{Loop}^\omega$) which was defined by the authors in [15], is essentially an extension of Meyer and Ritchie's Loop language [51] with higher-order procedural variables. $\text{Loop}^\omega$ is a genuine imperative language as opposed to functional languages with imperative features. However, $\text{Loop}^\omega$ is a "pure" imperative language: side-effects and aliasing are forbidden. These restrictions

1. *Email addresses:* `crolard@u-pec.fr` (T. Crolard), `polonowski@u-pec.fr` (E. Polonowski)

enable simple location-free operational semantics [20]. Moreover, the type system relies on the distinction between mutable and read-only variables to prevent procedure bodies to refer to non-local mutable variables. This property is crucial to guarantee that fix-points cannot be encoded using procedural variables. Since there is no recursivity and no unbounded loop construct in $\text{Loop}^\omega$, one can prove that all $\text{Loop}^\omega$ programs terminate (note that the expressive power of system T is still attained thanks to mutable higher-order procedural variables).

In this paper, we extend $\text{Loop}^\omega$ with first-order dependent types. This led us in particular to relax the underlying static type system. Indeed, for instance, after the assignment $x := 0$, the type of $x$ is $\mathbf{nat}(0)$. The type of $x$ is thus changed by this assignment whenever the former value of $x$ is different from 0. Moreover, the type of $x$ before the assignment does not matter: there is no need to even require that $x$ be a natural number. Pushing this idea to the limit, we obtain a type system for $\text{Loop}^\omega$ where the type of any mutable variable can be changed by an assignment (or a procedure call). Although, this feature seems characteristic of a dynamic language, our type system is fully static. Moreover, since dealing with mutable variables is natural in imperative programming, global variables are easily simulated with usual state-passing style. Besides, the logical meaning of this simulation is perfectly clear.

This above remark suggests that usual static type systems for imperative languages are overly restrictive. Indeed, a pseudo-dynamic type system is quite expressive: typing an imperative program in state-passing style amounts (up to curryfication) to typing its functional image with a parameterised state monad [5]. To capture this expressivity would usually require an effect system on the imperative side [28]. Moreover, a pseudo-dynamic type system provides an elegant way to deal with uninitialized variables. Indeed, in a logical type system, a type is not necessarily inhabited and there are thus no default values for arbitrary types. Although it is possible to design a type system which track uninitialized variables, it would be awkward (and meaningless from a logical standpoint). On the other hand, in a pseudo-dynamic type system any mutable variable can be initialized to a default inhabited type with a chosen default value.

Let us summarize the main developments of this paper. We rephrase Landin's translation for a total imperative language featuring higher-order procedures and non-local jumps and then we rely on the Curry-Howard correspondence for classical logic to derive a program logic for this language. To be more specific, we define a framework which includes an imperative language **I**, a call-by-value functional language **F** and a retraction between **I** and **F** as follows:

- The functional language **F**, which is our formulation of Gödel System T, is equipped with two usual type systems, a simple type system **FS** and a dependent type system **FD** which is akin to Leivant's **M1LP** [48]. In particular, dependent types include arbitrary formulas of first-order arithmetic.
- The imperative language **I** (essentially $\text{Loop}^\omega$ from [15]) is an extension of Meyer and Ritchie's Loop language [51] with higher-order procedural variables. Language **I** is also equipped with two (unusual) type systems, a pseudo-dynamic simple type system **IS** and a dependent type system **ID**.
- A compositional translation $^\star$ from **I** to **F** is definable [15]. This translation actually provides a simulation: each evaluation step of an imperative program is simulated by a bounded number of reduction step of its functional image. In this paper, we show that this translation is type-preserving in both the pseudo-dynamic and dependent frameworks.
- We characterize the shape of the functional image of an imperative program by $^\star$: these functional terms are monadic normal forms [33] (also called $A$-normal forms [25]). A reverse translation $^\diamond$ from monadic normal forms of **F** to **I** is then defined, which is also compositional and type-preserving in both the pseudo-dynamic and dependent frameworks.
- We show that $\langle {}^\diamond, {}^\star \rangle$ forms a retraction. Consequently, from any dependently-typed functional program (and thus from any proof in Heyting arithmetic) we can derive an imperative program which implements the corresponding dependent type.
- $\mathbf{F}^c$ is then defined as an extension of **F** with control operators **callcc** and **throw** (taken from [32]). The semantics of $\mathbf{F}^c$ is given by a call-by-value CPS-transformation into **F**. Following [33], since the functional image of an imperative program is in monadic normal form, we factor the CPS transformation through Moggi's computational meta-language [53, 54].
- From $\mathbf{F}^c$ we derive $\mathbf{I}^c$ which extends **I** with two primitive procedures **callcc** and **throw**. Although we do not pretend that these control operators are natural in an imperative language, they can be used to define more conventional statements which have to interact with the control flow. It is of course not possible to encode arbitrary **goto** statements since our programming language is total.

- Finally, as a generic example, by combining a simulated global state with **callcc** and **throw**, we show how to encode **shift** and **reset** [18] (and thus any representable monad) using Filinski's decomposition [23]. As a consequence, we obtain an indirect formulas-as-types interpretation of delimited continuations in a dependently-typed framework.

## Related works

Although several program logics have been designed for higher-order procedural mutable variables or non-local jumps, we are not aware of any work which combines both in an imperative setting.

Of course, there has been much research on Floyd-Hoare logics [26, 35, 36] (see the surveys [2] and [13]). Such program logics for higher-order procedures have been defined for instance in [17] (for Clarke's language L4 [9]) or more recently for stored parameterless procedures in [70]. Program logics for jumps exists since [10], and although designing such a logic is error-prone [60], there have been successfully used recently for proving properties in low-level languages [22, 72].

A dependent type system for an imperative programming language is defined in [76], where the dependent types are restricted to ensure that type checking remains decidable. They also made the observation that imperative dependent types requires to allow the type of variables to change during evaluation. However they chose to restrict the type system in order to guarantee that the extracted program is typable in some usual static (non-dependent) type systems. On the contrary, we believe that a dynamically-flavoured static type system should be advocated.

Proofs-as-Imperative-Program [67, 68] adapts the proofs-as-programs paradigm for the synthesis of imperative SML programs with side-effect free return values. The type theory is however intrinsically constructive: it requires a strong existential quantifier which is not compatible with classical logic [34].

The Dependent Hoare Type Theory [58] and the Imperative Hoare Logic [38, 37] are frameworks for reasoning about effectful higher-order functions. The dynamic semantics of those systems are much more complicated (since aliasing is allowed) than our location-free semantics. Although the Dependent Hoare Type Theory contains control expressions and enjoys a formulas-as-types interpretation, it is not clear whether programs correspond to proofs in some deduction system for classical logic.

*Plan of the paper.* In Section 2, we present the untyped functional language **F**, the untyped imperative language **I** and and their dynamic semantics. We define also the retraction $\langle^\diamond, ^\star\rangle$ between programs of **I** and monadic normal forms of **F**. Section 3 is devoted to the definition of the pseudo-dynamic type system **IS**. Section 4 contains the definitions of the dependently-typed systems **ID** and **FD** together with their main properties. In Section 5, we extend language **F** with control operators and its type system is raised to classical arithmetic $\mathbf{FD}^c$. Finally, in Section 6, we extend **I** with non-local jumps and we derive a corresponding program logic $\mathbf{ID}^\mathbf{c}$.

# 2 Dynamic semantics of I and F

In this section, we present the untyped functional language **F** (which is a variant of Gödel System T) and the untyped imperative language **I** (which is an extension of Meyer and Ritchie's Loop language [51] with higher-order procedural variables studied in [15]). We define also the dynamic semantics of both languages and the retraction $\langle^\diamond, ^\star\rangle$ between programs of **I** and monadic normal forms of **F**.

## 2.1 Language F

Gödel System T may be defined as the simply typed $\lambda$-calculus extended with a type of natural numbers and with primitive recursion at all types [29]. The language **F** we consider in this paper a variant of System T with product types (n-ary tuples actually) and a constant-time predecessor operation (since any definition of this function as a term of System T is at least linear under the call-by-value evaluation strategy [11]). Moreover, we formulate this system directly as a context semantics (a set of reduction rules together with an inductive definition of evaluation contexts). As usual, we consider terms up to $\alpha$-conversion and the set $\mathcal{FV}(t)$ of free variables of a term $t$ is defined in the standard way. The rewriting system is summarized in Figure 2.1, where variables $x$, $x_1$, ..., $x_n$, $y$ range over a set of identifiers and $t[v_1/x_1, ..., v_n/x_n]$ denotes the usual capture-avoiding substitution.

*(terms)*

$$
\begin{array}{rcl}
t & ::= & x \\
& | & 0 \\
& | & S(t) \\
& | & \mathbf{pred}(t) \\
& | & t_1\ t_2 \\
& | & \lambda x.t \\
& | & (t_1, ..., t_n) \\
& | & \mathbf{let}\ (x_1, ..., x_n) = t_1\ \mathbf{in}\ t_2 \\
& | & \mathbf{rec}(t_1, t_2, t_3)
\end{array}
$$

*(values)*

$$
\begin{array}{rcl}
v & ::= & x \\
& | & 0 \\
& | & S(v) \\
& | & (v_1, ..., v_n) \\
& | & \lambda x.t
\end{array}
$$

*(contexts)*

$$
\begin{array}{rcl}
C[\ ] & ::= & [\ ] \\
& | & C[\ ]\ t \\
& | & v\ C[\ ] \\
& | & S(C[\ ]) \\
& | & \mathbf{pred}(C[\ ]) \\
& | & \mathbf{rec}(C[\ ], t_2, t_3) \\
& | & \mathbf{rec}(v_1, C[\ ], t_3) \\
& | & \mathbf{rec}(v_1, v_2, C[\ ]) \\
& | & (v_1, ... v_{i-1}, C[\ ], t_{i+1} ..., t_n) \\
& | & \mathbf{let}\ (x_1, ..., x_n) = C[\ ]\ \mathbf{in}\ t
\end{array}
$$

*(evaluation rules)*

$$
\begin{array}{rcl}
C[\mathbf{pred}(0)] & \rightsquigarrow & C[0] \\
C[\mathbf{pred}(S(v))] & \rightsquigarrow & C[v] \\
C[\mathbf{rec}(0, v_2, \lambda x.\lambda \vec{y}.t)] & \rightsquigarrow & C[v_2] \\
C[\mathbf{rec}(S(v_1), v_2, \lambda x.\lambda \vec{y}.t)] & \rightsquigarrow & C[\lambda x.\lambda \vec{y}.t\ v_1\ \mathbf{rec}(v_1, v_2, \lambda x.\lambda \vec{y}.t)] \\
C[\lambda x.t\ v] & \rightsquigarrow & C[t[v/x]] \\
C[\mathbf{let}\ (x_1, ..., x_n) = (v_1, ..., v_n)\ \mathbf{in}\ t] & \rightsquigarrow & C[t[v_1/x_1, ..., v_n/x_n]]
\end{array}
$$

**Figure 2.1.** Syntax and context semantics of Language **F**

**Remark 2.1.** In order to distinguish the successor $S$ (which is a constructor) from the successor seen as an operation (whose evaluation should imply a reduction step), we use the keyword **succ** as an abbreviation for $\lambda x.S(x)$.

**Remark 2.2.** We write $\lambda(x_1, ..., x_n).t$ (or $\lambda\vec{x}.t$) as an abbreviation for $\lambda z.\mathbf{let}\ (x_1, ..., x_n) = z\ \mathbf{in}\ t$ where $z$ is a fresh variable. Similarly, we write $\lambda().t$ as an abbreviation for $\lambda z.\mathbf{let}\ () = z\ \mathbf{in}\ t$ where $z$ is a fresh variable.

### 2.1.1 Example: the Ackermann function

The Ackermann function is an example of function known not to be primitive recursive [63] but which can be represented in System T. Here follows an example of a slightly modified version of the function defined by the following equations [49]:

$$
\begin{array}{llcl}
(1) & \mathbf{a}(0, n) & = & \mathbf{s}(n) \\
(2) & \mathbf{a}(\mathbf{s}(z), 0) & = & \mathbf{s}(\mathbf{s}(0)) \\
(3) & \mathbf{a}(\mathbf{s}(z), \mathbf{s}(u)) & = & \mathbf{a}(z, \mathbf{a}(\mathbf{s}(z), u))
\end{array}
$$

$$ack(m, n) = \mathbf{rec}(m, \lambda y.S(y), \lambda i.\lambda f.\lambda y.\mathbf{rec}(y, S(S(0)), \lambda j.\lambda k.(f\ k)))\ n$$

## 2.2 Language I

The untyped language **I** is essentially the $\text{LOOP}^{\omega}$ language presented in [15] except that $\text{LOOP}^{\omega}$ was explicitly typed. Moreover the loop syntax is now **for** $y := 0$ **until** $e$ $\{s\}$ where the bound $e$ is excluded from the range (since this new syntax corresponds more closely to reasoning by induction). The location-free transition semantics [20] is also the same as in [15] except that we consider only sequences. Although it is somewhat more verbose, both semantics are clearly equivalent.

### 2.2.1 Syntax

The raw syntax of imperative programs is given below. There is nothing particular to this syntax except that we annotate each block $\{s\}_{\vec{x}}$ with a list of variables $\vec{x}$ (which corresponds to the mutable variables which may occur in the block). In the following grammar, $x$, $y$, $z$ range over a set of identifiers, $\bar{q}$ ranges over natural numbers (*i.e.* constant literals), $\varepsilon$ denotes the empty sequence and $*$ denotes the singleton value. Free identifiers are defined in the standard way (see Appendix A).

$$
\begin{array}{rrcl}
(\textit{command}) & c & ::= & \{s\}_{\vec{x}} \\
 & & | & \textbf{for } y := 0 \textbf{ until } e\ \{s\}_{\vec{x}} \\
 & & | & \vec{y} := e \quad | \quad \textbf{inc}(y) \quad | \quad \textbf{dec}(y) \\
 & & | & e(\vec{e}\,;\vec{y}\,) \\
 & & & \\
(\textit{sequence}) & s & ::= & \varepsilon \\
 & & | & c\,;\ s \\
 & & | & \textbf{cst } y = e;\ s \\
 & & | & \textbf{var } y := e;\ s \\
 & & & \\
(\textit{expression}) & e & ::= & y \quad | \quad * \quad | \quad \bar{q} \quad | \quad (\vec{e}\,) \\
 & & | & \textbf{proc } (\textbf{in } \vec{y}\,; \textbf{out } \vec{z}\,)\ \{s\}_{\vec{z}}
\end{array}
$$

**Notation 2.3.** *(values). Imperative values are closed expressions,* i.e. *the singleton value, natural numbers, procedures and tuples of values. We shall use $w$ as a syntactica category for values whenever we will need to distinguish between expressions and values.*

**Remark 2.4.** *(no aliasing).* In order to avoid parameter-induced aliasing problems, we assume that all $y_i$ are pairwise distinct in a procedure call $p(\vec{e}\,;\vec{y}\,)$.

**Remark 2.5.** *(annotations).* In a block $\{s\}_{\vec{x}}$, the variables in $\vec{x}$ are visible mutable variables (according to standard C-like scoping rules). Moreover, the list $\vec{x}$ must also contain all the free mutable variables occurring in the sequence. Such annotations can automatically be inferred by taking, for instance, all the visible mutable variables.

**Remark 2.6.** *(no backpatching).* No free mutable variable is allowed in the body of a procedure (except its **out** parameters). This restriction is required to prevent the well-known technique called "tying the recursive knot" [44] which takes advantage of higher-order mutable variables (or function pointers) to define arbitrary recursive functions.

### 2.2.2 Example: the addition procedure

Here follows a procedure that computes the addition of two natural numbers:

$$
\begin{array}{l}
\textbf{cst } \textit{add} = \textbf{proc } (\textbf{in } X, Y; \textbf{out } Z)\ \{ \\
\quad Z := X; \\
\quad \textbf{for } I := 0 \textbf{ until } Y\ \{ \\
\quad\quad \textbf{inc}(Z); \\
\quad \}_Z; \\
\}_Z
\end{array}
$$

### 2.2.3 Operational semantics

The operational semantics is given as transition system [66] which defines inductively a binary relation between states. A state is a pair $(s,\ \mu)$ consisting of a sequence $s$ and a store $\mu$, where a store is a finite ordered mapping from (mutable) variables to closed imperative values (i.e. integer literals, procedures and $*$, and tuples of imperative values).

Note that expressions do not require any evaluation since they are either variables or values. We introduce thus the following notation which allows us to treat uniformly values and variables in the semantics:

**Notation 2.7.** *Given a store $\mu$, let $\varphi_\mu$ be the trivial extension of $\mu$ to expressions defined as follows $\varphi_\mu(x) = \mu(x)$ if $x$ is a variable, $\varphi_\mu(w) = w$ and $\varphi_\mu((e_1, ..., e_n)) = (\varphi_\mu(e_1), ..., \varphi_\mu(e_n))$. In the sequel, we write $e =_\mu w$ for $\varphi_\mu(e) = w$.*

**Notation 2.8.** *Let $s$ be a sequence. We write $s[x \leftarrow w]$ for the substitution of a read-only variable $x$ by a closed imperative value $w$ and $s[y \hookleftarrow z]$ for the renaming of a mutable variable $y$ by a mutable variable $z$. The formal definitions are similar to those given in [15].*

$$((\{\}_{\vec{z}}; s), \mu) \mapsto (s, \mu) \qquad \text{(S.BLOCK-I)}$$

$$\frac{(s_1, \mu) \mapsto (s_1', \mu')}{((\{s_1\}_{\vec{z}};\ s_2), \mu) \mapsto ((\{s_1'\}_{\vec{z}};\ s_2), \mu')} \qquad \text{(S.BLOCK-II)}$$

$$((\mathbf{var}\ y := e;\ \varepsilon), \mu) \mapsto (\varepsilon, \mu) \qquad \text{(S.VAR-I)}$$

$$\frac{e =_\mu w \qquad (s, (\mu, y \leftarrow w)) \mapsto (s', (\mu', y \leftarrow w'))}{((\mathbf{var}\ y := e;\ s), \mu) \mapsto ((\mathbf{var}\ y := w';\ s'), \mu')} \qquad \text{(S.VAR-II)}$$

$$\frac{e =_\mu \vec{w}}{((\vec{y} := e;\ s), \mu) \mapsto (s, \mu[\vec{y} \leftarrow \vec{w}])} \qquad \text{(S.ASSIGN)}$$

$$\frac{\mu(y) = \bar{q}}{((\mathbf{inc}(y);\ s), \mu) \mapsto ((y := \overline{q+1};\ s), \mu)} \qquad \text{(S.INC)}$$

$$\frac{\mu(y) = \bar{q}}{((\mathbf{dec}(y);\ s), \mu) \mapsto ((y := \overline{q \dot{-} 1};\ s), \mu)} \qquad \text{(S.DEC)}$$

$$\frac{\vec{e} =_\mu \vec{w} \qquad p =_\mu \mathbf{proc}\ (\mathbf{in}\ \vec{y}\,; \mathbf{out}\ \vec{z}\,)\{s'\}_{\vec{z}}}{((p(\vec{e}\,;\vec{r}\,);\ s), \mu) \mapsto ((\{s'[\vec{y} \leftarrow \vec{w}][\vec{z} \hookleftarrow \vec{r}\,]\}_{\vec{r}};\ s), \mu[\vec{r} \leftarrow *\,])} \qquad \text{(S.CALL)}$$

$$\frac{e =_\mu w}{((\mathbf{cst}\ y = e;\ s), \mu) \mapsto (s[y \leftarrow w], \mu)} \qquad \text{(S.CST)}$$

$$\frac{e =_\mu \bar{0}}{((\mathbf{for}\ y := 0\ \mathbf{until}\ e\ \{s\}_{\vec{z}};\ s'), \mu) \mapsto (s', \mu)} \qquad \text{(S.FOR-I)}$$

$$\frac{e =_\mu \overline{q+1}}{((\mathbf{for}\ y := 0\ \mathbf{until}\ e\ \{s\}_{\vec{z}};\ s'), \mu) \mapsto ((\{\mathbf{for}\ y := 0\ \mathbf{until}\ \bar{q}\ \{s\}_{\vec{z}};\ s[y \leftarrow \bar{q}]\}_{\vec{z}};\ s'), \mu)} \qquad \text{(S.FOR-II)}$$

**Figure 2.2.** Transition semantics

**Notation 2.9.** *Let $\mu$ be a store. We write $(\mu[y \leftarrow w])$ for the store update,* i.e. *$\mu[y \leftarrow w](x) = \mu(x)$ if $x \neq y$ and $\mu[y \leftarrow w](y) = \mu(y)$. We write $(\mu, y \leftarrow w)$ for the store extension with the new variable $y$ mapped to $w$.*

This definition of the transition system is summarized in Figure 2.2.

**Remark 2.10.** This semantics is clearly deterministic since there is always at most one rule which can be applied (depending on the content of the store and the shape of the command).

## 2.3 Translation from I to F and simulation

We recall the translation, similar in spirit to Landin's translation of Algol-like languages, described in [15]. The intuition behind this translation of imperative programs into functional programs is the following: a sequence $\{c_1; ...; c_n;\}_{\vec{x}}$ is translated into:

$$\mathbf{let}\ \vec{x}_1 = c_1^\star\ \mathbf{in} \ldots\ \mathbf{let}\ \vec{x}_n = c_n^\star\ \mathbf{in}\ \vec{x}$$

where each $\vec{x}_i \subseteq \vec{x}$ corresponds to the "output" of command $c_i$ and $\vec{x}$ is the output of the sequence.

**Definition 2.11.** *For any expression $e$, sequence $s$ and variables $\vec{x}$, the translations $e^\star$ and $(s)^\star_{\vec{x}}$ into terms of language* **F** *are defined by mutual induction as follows:*

- $\bar{n}^\star = S^n(0)$
- $y^\star = y$
- $*^\star = ()$
- $(e_1, ..., e_n)^\star = (e_1^\star, ..., e_n^\star)$

- $(\textbf{proc}\ (\textbf{in}\ \vec{y}\,;\textbf{out}\ \vec{z}\,)\{s\}_{\vec{z}})^{\star} = \lambda\vec{y}\,.(s)^{\star}_{\vec{z}}\ [\vec{()}/\vec{z}\,]$
- $(\varepsilon)^{\star}_{\vec{x}} = \vec{x}$
- $(\textbf{var}\ y := e;\ s)^{\star}_{\vec{x}} = (s)^{\star}_{\vec{x}}[e^{\star}/y]$
- $(\textbf{cst}\ y = e;\ s)^{\star}_{\vec{x}} = \textbf{let}\ y = e^{\star}\ \textbf{in}\ (s)^{\star}_{\vec{x}}$
- $(\vec{y} := e;\ s)^{\star}_{\vec{x}} = \textbf{let}\ \vec{y} = e^{\star}\ \textbf{in}\ (s)^{\star}_{\vec{x}}$
- $(\textbf{inc}(y);\ s)^{\star}_{\vec{x}} = \textbf{let}\ y = \textbf{succ}(y)\ \textbf{in}\ (s)^{\star}_{\vec{x}}$
- $(\textbf{dec}(y);\ s)^{\star}_{\vec{x}} = \textbf{let}\ y = \textbf{pred}(y)\ \textbf{in}\ (s)^{\star}_{\vec{x}}$
- $(p(\vec{e}\,;\vec{z}\,);\ s)^{\star}_{\vec{x}} = \textbf{let}\ \vec{z} = p^{\star}\ (\vec{e}^{\,\star})\ \textbf{in}\ (s)^{\star}_{\vec{x}}$
- $(\{s_1\}_{\vec{z}};\ s_2)^{\star}_{\vec{x}} = \textbf{let}\ \vec{z} = (s_1)^{\star}_{\vec{z}}\ \textbf{in}\ (s_2)^{\star}_{\vec{x}}$
- $(\textbf{for}\ y := 0\ \textbf{until}\ e\ \{s_1\}_{\vec{z}};\ s_2)^{\star}_{\vec{x}} = \textbf{let}\ \vec{z} = \textbf{rec}(e^{\star}, \vec{z}\,, \lambda y.\lambda\vec{z}\,.(s_1)^{\star}_{\vec{z}})\ \textbf{in}\ (s_2)^{\star}_{\vec{x}}$

### 2.3.1 Simulation

We recall the simulation theorem from [15] which states that for any sequence $s$, the evaluation of $s$ is simulated by the reduction of $(s)^{\star}_{\vec{z}}$.

**Proposition 2.12.** *For any state $(s, \mu)$, if $\vec{x} = dom(\mu)$ and $\vec{z} \subseteq \vec{x}$ we have:*

$$(s, \mu) \mapsto (s', \mu')\ \textit{implies}\ (s)^{\star}_{\vec{z}}[\mu(\vec{x}\,)^{\star}/\vec{x}\,] \rightsquigarrow^{*} (s')^{\star}_{\vec{z}}[\mu'(\vec{x}\,)^{\star}/\vec{x}\,]$$

## 2.4 Translation from F to I and retraction

In this section, we show how to translate a functional program of **F** into an imperative program of **I**. However, this translation is only defined for a sub-language $\mathcal{L}$ of monadic normal forms (terms where any non-trivial intermediate computation is named [33, 25]). This sub-language $\mathcal{L}$ characterize the image of imperative programs by $^{\star}$. We show in appendix C.5 how to transform any term of language **F** into a monadic normal form of $\mathcal{L}$.

**Definition 2.13.** *We define inductively $\mathcal{L}$ and $\mathcal{V}$, families of terms (resp. values) of* **F***, as follows:*

- $x \in \mathcal{V}$
- $() \in \mathcal{V}$
- $S^n(0) \in \mathcal{V}$
- $\lambda\vec{x}\,.t \in \mathcal{V}$
- $(v_1, ..., v_n) \in \mathcal{V}$ — *if* $v_1, ..., v_n \in \mathcal{V}$

- $v \in \mathcal{L}$ — *if* $v \in \mathcal{V}$
- $\textbf{let}\ \vec{x} = v\ \textbf{in}\ u \in \mathcal{L}$ — *if* $v \in \mathcal{V}$ *and* $u \in \mathcal{L}$
- $\textbf{let}\ x = \textbf{succ}(v)\ \textbf{in}\ u \in \mathcal{L}$ — *if* $v \in \mathcal{V}$ *and* $u \in \mathcal{L}$
- $\textbf{let}\ x = \textbf{pred}(v)\ \textbf{in}\ u \in \mathcal{L}$ — *if* $v \in \mathcal{V}$ *and* $u \in \mathcal{L}$
- $\textbf{let}\ \vec{x} = v\ v'\ \textbf{in}\ u \in \mathcal{L}$ — *if* $v \in \mathcal{V}$, $v' \in \mathcal{V}$ *and* $u \in \mathcal{L}$
- $\textbf{let}\ x = \textbf{rec}(v, v', \lambda y.\lambda z.t)\ \textbf{in}\ u \in \mathcal{L}$ — *if* $v \in \mathcal{V}$, $v' \in \mathcal{V}$ *and* $u \in \mathcal{L}$
- $\textbf{let}\ \vec{x} = t\ \textbf{in}\ u \in \mathcal{L}$ — *if* $t \in \mathcal{L}$ *and* $u \in \mathcal{L}$

**Proposition 2.14.** *For any sequence $s$, any expression $e$ and any variables $\vec{x} = (x_1, ..., x_n)$, $(s)^{\star}_{\vec{x}} \in \mathcal{L}$ and $e^{\star} \in \mathcal{V}$.*

**Proof.** Straightforward mutual induction on $s$ and $e$. ☐

**Notation 2.15.** *In the sequel, we shall use the following abbreviations:*

$$\begin{aligned}
\textbf{var}\ \vec{y}\,;\ s &= \textbf{var}\ y_1 := *\,;\ ...;\ \textbf{var}\ y_n := *\,;\ s\\
\textbf{var}\ \vec{y} := \vec{w};\ s &= \textbf{var}\ y_1 := w_1;\ ...;\ \textbf{var}\ y_n := w_n;\ s\\
\textbf{cst}\ \vec{y} = \vec{z}\,;\ s &= \textbf{cst}\ y_1 = z_1;\ ...;\ \textbf{cst}\ y_n = z_n;\ s\\
\vec{y} := \vec{w};\ s &= y_1 := w_1;\ ...;\ y_n := w_n;\ s
\end{aligned}$$

**Definition 2.16.** *For any value $w \in \mathcal{W}$ and any term $t \in \mathcal{L}_n$, the translation $w^\diamond$ and $t^\diamond_{\vec{r}}$ are defined by mutual induction, where $\vec{r} = (r_1, ..., r_n)$, $z$ and $\vec{z} = (z_1, ..., z_n)$ are fresh variables, as follows:*

- $()^\diamond \ = \ *$
- $S^n(0)^\diamond \ = \ \bar{n}$
- $y^\diamond \ = \ y$
- $(\lambda x.t)^\diamond \ = \ \textbf{proc}\ (\textbf{in}\ x; \textbf{out}\ z)\ \{t^\diamond_z\}_z$
- $(\vec{w})^\diamond \ = \ (\vec{w}^\diamond)$
- $(w)^\diamond_r \ = \ r := w;\ \varepsilon$
- $(\textbf{let}\ y = w\ \textbf{in}\ u)^\diamond_r \ = \ \textbf{cst}\ y = w^\diamond;\ (u)^\diamond_r$
- $(\textbf{let}\ y = \textbf{succ}(w)\ \textbf{in}\ u)^\diamond_r \ = \ \textbf{var}\ z := w^\diamond;\ \textbf{inc}(z);\ \textbf{cst}\ y = z;\ (u)^\diamond_r$
- $(\textbf{let}\ y = \textbf{pred}(w)\ \textbf{in}\ u)^\diamond_r \ = \ \textbf{var}\ z := w^\diamond;\ \textbf{dec}(z);\ \textbf{cst}\ y = z;\ (u)^\diamond_r$
- $(\textbf{let}\ \vec{x} = w\ w'\ \textbf{in}\ u)^\diamond_r \ = \ \textbf{var}\ \vec{z};\ w^\diamond(w'^\diamond; \vec{z});\ \textbf{cst}\ \vec{x} = \vec{z};\ (u)^\diamond_r$
- $(\textbf{let}\ x = \textbf{rec}(w, w', \lambda i.\lambda \vec{y}.t)\ \textbf{in}\ u)^\diamond_r = \ \textbf{var}\ z := w';\ \textbf{for}\ i := 0\ \textbf{until}\ w^\diamond\ \{\textbf{cst}\ y = z;\ t^\diamond_z\}_z;\ \textbf{cst}\ x = z;\ (u)^\diamond_r$
- $(\textbf{let}\ \vec{x} = t\ \textbf{in}\ u)^\diamond_r \ = \ \textbf{var}\ z;\ \{(t)^\diamond_z\}_z;\ \textbf{cst}\ \vec{x} = z;\ (u)^\diamond_r$

**Remark 2.17.** Note that all identifiers of the source term are mapped to read-only variables. Indeed, mutable are introduced locally, assigned and then only used to initialize local read-only variables. This property ensures that mutable variables do not occur in the body of procedures in the resulting LOOP$^\omega$ program: the only mutable variables are fresh variables introduced during the translation.

### 2.4.1 Retraction

We prove that for any term $t$ of $\mathcal{L}$, the term $(t^\diamond_{\vec{r}})^\star_{\vec{r}}$ is convertible with $t$. Both terms are not equal in general since some "administrative" redices are introduced by the translations. However, equality holds for integer values.

**Definition 2.18.** *We define the reduction relation $\approx$ as the reflexive, symmetric, transitive and contextual closure of the reduction $\rightsquigarrow$ for arbitrary contexts.*

**Proposition 2.19.**

- *Given a term $t \in \mathcal{L}$ and a fresh mutable variable tuple $\vec{r}$ we have $(t^\diamond_{\vec{r}})^\star_{\vec{r}} \approx t$.*
- *Given a value $w \in \mathcal{W}$, if $w = S^n(0)$ or $w = *$ then $w^{\diamond\star} = w$ else $w^{\diamond\star} \approx w$.*

**Proof.** See Appendix-A. ☐

**Proposition 2.20.** *For any value $w$, if $w = \bar{q}$ or $w = *$ then $w^{\star\diamond} = w$.*

**Proof.** By Proposition 2.19, if $w = \bar{q}$ or $w = *$ then $w^{\star\diamond\star} = w^\star$. ☐

# 3 Pseudo-dynamic Type System

In this section, we present the simple type system for language **F** and the pseudo-dynamic type system for language **I**. Then we show that both translation $\star$ and $\diamond$ preserve typability and that the transition semantics of **I** enjoys the usual "type preservation" and "progress" properties.

## 3.1 Functional simple type system FS

The functional simple type system **FS** is defined as usual for a simply typed $\lambda$-calculus extended with tuples, natural numbers and with primitive recursion at all types. The set $\Sigma_{\textbf{FS}}$ of simple functional types is defined by the following grammar:

$$\sigma \ ::= \ \textbf{nat} \ | \ \textbf{unit} \ | \ \sigma_1 \to \sigma_2 \ | \ \sigma_1 \times ... \times \sigma_n$$

$$\frac{x\colon \tau \in \Gamma}{\Gamma \vdash x\colon \tau} \quad \text{(IDENT)}$$

$$\Gamma \vdash 0\colon \mathbf{nat} \quad \text{(ZERO)}$$

$$\frac{\Gamma \vdash t\colon \mathbf{nat}}{\Gamma \vdash S(t)\colon \mathbf{nat}} \quad \text{(SUCC)}$$

$$\frac{\Gamma \vdash t\colon \mathbf{nat}}{\Gamma \vdash \mathbf{pred}(t)\colon \mathbf{nat}} \quad \text{(PRED)}$$

$$\frac{\Gamma \vdash t_1\colon \tau_1 \quad ... \quad \Gamma \vdash t_n\colon \tau_n}{\Gamma \vdash (t_1, ..., t_n)\colon \tau_1 \times ... \times \tau_n} \quad \text{(TUPLE)}$$

$$\frac{}{\Gamma \vdash ()\colon \mathbf{unit}} \quad \text{(UNIT)}$$

$$\frac{\Gamma, x_1\colon \tau_1, ..., x_n\colon \tau_n \vdash t\colon \tau \qquad \Gamma \vdash u\colon \tau_1 \times ... \times \tau_n}{\Gamma \vdash \mathbf{let}\ (x_1, ..., x_n) = u\ \mathbf{in}\ t\colon \tau} \quad \text{(LET)}$$

$$\frac{\Gamma, x\colon \tau \vdash t\colon \sigma}{\Gamma \vdash \lambda x.t\ \colon \tau \rightarrow \sigma} \quad \text{(ABS)}$$

$$\frac{\Gamma \vdash t_1\colon \sigma \rightarrow \tau \quad \Gamma \vdash t_2\colon \sigma}{\Gamma \vdash t_1\ t_2\colon \tau} \quad \text{(APP)}$$

$$\frac{\Gamma \vdash t_1\colon \mathbf{nat} \quad \Gamma \vdash t_2\colon \tau \quad \Gamma, x\colon \mathbf{nat}, y\colon \tau \vdash t_3\colon \tau}{\Gamma \vdash \mathbf{rec}(t_1, t_2, \lambda x.\lambda y.t_3)\colon \tau} \quad \text{(REC)}$$

**Figure 3.1.** Functional type system **FS**

The type system is summarized in Figure 3.1.

## 3.2 Pseudo-dynamic imperative type system IS

The static type system described in this section is called “pseudo-dynamic” since the type of a mutable variable is allowed to change during execution. It is however fully static in the sense that it guarantees statically that no type error can occur at run-time. As a side benefit, we obtain a convenient way to address the issue of uninitialized variables: any mutable variable can be initialized with the $*$ (which denotes the single value of type **unit**) and its type shall change later (when assigned its first relevant value).

The pseudo-dynamic type system may also be seen as a simple effect system [28, 71] since it is able to guarantee the absence of side-effects, aliasing and fix-points in well-typed programs. Its key feature which enable this property is the distinction between mutable variables and read-only variables. More formally, the set $\Sigma_{\mathbf{IS}}$ of imperative types is defined by the following grammar:

$$\sigma, \tau \ ::= \ \mathbf{nat} \mid \mathbf{proc}\ (\mathbf{in}\ \vec{\tau}; \mathbf{out}\ \vec{\sigma}) \mid (\tau_1, ..., \tau_n) \mid \mathbf{unit}$$

A typing environment has the form $\Gamma; \Omega$ where $\Gamma$ and $\Omega$ are (possibly empty) lists of pairs $x\colon \tau$ ($x$ ranges over variables and $\tau$ over types). $\Gamma$ stands for read-only variables (constants and **in** parameters) and $\Omega$ stands for mutable variables (local variables and **out** parameters). We use two typing judgments, one for expressions and one for sequences: $\Gamma; \Omega \vdash e\colon \tau$ has the usual meaning, whereas in $\Gamma; \Omega \vdash s \ \triangleright \Omega'$, the environment $\Omega'$ contains the types of the mutable variables at the end of the sequence $s$. The type system is given in Figure 3.2. As usual, we consider programs up to renaming of bound variables, where the notion of free variable of a command is defined in the standard way.

**Remark 3.1.** Let us recall important features of this pseudo-dynamic type system shared with the static type system described in [15]:

- *(scoping rules)*. As usual for $C$-like languages, the scope of a constant (rule T.CST) or a variable (rule T.VAR) extends from the point of declaration to the end of the block containing the declaration.

$$\frac{x:\tau \in \Gamma;\Omega}{\Gamma;\Omega \vdash x:\tau} \quad (\text{T.ENV})$$

$$\frac{}{\Gamma;\Omega \vdash \bar{q}:\mathbf{nat}} \quad (\text{T.NUM})$$

$$\frac{}{\Gamma;\Omega \vdash *:\mathbf{unit}} \quad (\text{T.UNIT})$$

$$\frac{\Gamma;\Omega \vdash e_1:\tau_1 \quad \ldots \quad \Gamma;\Omega \vdash e_n:\tau_n}{\Gamma;\Omega \vdash (e_1,\ldots,e_n):(\tau_1,\ldots,\tau_n)} \quad (\text{T.TUPLE})$$

$$\frac{\vec{z} \neq \emptyset \qquad \Gamma, \vec{y}:\vec{\sigma}; \vec{z}:\overrightarrow{\mathbf{unit}} \vdash s \triangleright \vec{z}:\vec{\tau}}{\Gamma;\Omega \vdash \mathbf{proc}\ (\mathbf{in}\ \vec{y}; \mathbf{out}\ \vec{z})\{s\}_{\vec{z}}:\mathbf{proc}\ (\mathbf{in}\ \vec{\sigma}; \mathbf{out}\ \vec{\tau})} \quad (\text{T.PROC})$$

$$\frac{}{\Gamma;\Omega,\Omega' \vdash \varepsilon \triangleright \Omega'} \quad (\text{T.EMPTY})$$

$$\frac{\Gamma;\Omega,\vec{x}:\vec{\sigma} \vdash c \triangleright \vec{x}:\vec{\tau} \qquad \Gamma;\Omega,\vec{x}:\vec{\tau} \vdash s \triangleright \Omega'}{\Gamma;\Omega,\vec{x}:\vec{\sigma} \vdash c;\ s \triangleright \Omega'} \quad (\text{T.SEQ})$$

$$\frac{\Gamma;\Omega \vdash e:\tau \qquad \Gamma, y:\tau;\Omega \vdash s \triangleright \Omega'}{\Gamma;\Omega \vdash \mathbf{cst}\ y = e;\ s \triangleright \Omega'} \quad (\text{T.CST})$$

$$\frac{\Gamma;\Omega \vdash e:\tau \qquad \Gamma;\Omega, y:\tau \vdash s \triangleright \Omega' \qquad y \notin \Omega'}{\Gamma;\Omega \vdash \mathbf{var}\ y := e;\ s \triangleright \Omega'} \quad (\text{T.VAR})$$

$$\frac{\Gamma;\Omega,\vec{y}:\vec{\sigma} \vdash e:(\vec{\tau}) \qquad \Gamma;\Omega,\vec{y}:\vec{\tau} \vdash s \triangleright \Omega'}{\Gamma;\Omega,\vec{y}:\vec{\sigma} \vdash \vec{y} := e;\ s \triangleright \Omega'} \quad (\text{T.ASSIGN})$$

$$\frac{\Gamma;\vec{x}:\vec{\tau} \vdash s \triangleright \vec{x}:\vec{\sigma}}{\Gamma;\Omega,\vec{x}:\vec{\tau} \vdash \{s\}_{\vec{x}} \triangleright \vec{x}:\vec{\sigma}} \quad (\text{T.BLOCK})$$

$$\frac{}{\Gamma;\Omega, y:\mathbf{nat} \vdash \mathbf{inc}(y) \triangleright y:\mathbf{nat}} \quad (\text{T.INC})$$

$$\frac{}{\Gamma;\Omega, y:\mathbf{nat} \vdash \mathbf{dec}(y) \triangleright y:\mathbf{nat}} \quad (\text{T.DEC})$$

$$\frac{\Gamma;\Omega,\vec{x}:\vec{\sigma} \vdash e:\mathbf{nat} \quad \Gamma, y:\mathbf{nat};\vec{x}:\vec{\sigma} \vdash s \triangleright \vec{x}:\vec{\sigma}}{\Gamma;\Omega,\vec{x}:\vec{\sigma} \vdash \mathbf{for}\ y := 0\ \mathbf{until}\ e\ \{s\}_{\vec{x}} \triangleright \vec{x}:\vec{\sigma}} \quad (\text{T.FOR})$$

$$\frac{\Gamma;\Omega,\vec{r}:\vec{\omega} \vdash p:\mathbf{proc}\ (\mathbf{in}\ \vec{\sigma}; \mathbf{out}\ \vec{\tau}) \quad \Gamma;\Omega,\vec{r}:\vec{\omega} \vdash \vec{e}:\vec{\sigma}}{\Gamma;\Omega,\vec{r}:\vec{\omega} \vdash p(\vec{e};\vec{r}) \triangleright \vec{r}:\vec{\tau}} \quad (\text{T.CALL})$$

**Figure 3.2.** Imperative type system

- *(no side-effects).* Rule (T.PROC) implies that the only mutable variables which may occur inside the body of a procedure are its **out** parameters and its local mutable variables. This is enough to guarantee the absence of side-effects. However, side-effects can still be simulated by passing the non-local variable as an explicit **in out** parameter (see section-3.5).
- *(no fix-points).* Rule (T.PROC) also forbids the reading of non-local mutable variables: this is necessary to prevents the definition of fix-points in the language.

Let us define formally the notions of well-typed stores and states.

**Definition 3.2.** (store typing). *We say that a store $\mu$ is typable of output typing environment $\Omega = z_1:\tau_1, \ldots, z_n:\tau_n$, denoted $\mu \triangleright \Omega$ if and only if $\vec{z} \in dom(\mu)$ and for all $(z_i, w_i) \in \mu$ we have $\emptyset;\emptyset \vdash w_i:\tau_i$.*

**Definition 3.3.** (state typing). *We say that a state $(s, \mu)$ is typable of output typing environment $\Omega'$ for a restriction of the store to the variables $\vec{z}:\vec{\tau}$, which we write as $\vec{z}:\vec{\tau} \vdash (s,\mu) \triangleright \Omega'$, if and only if $\mu \triangleright \vec{z}:\vec{\tau}$ and $\emptyset;\vec{z}:\vec{\tau} \vdash s \triangleright \Omega'$.*

## 3.3 Translations between IS and FS

We define translations $^\star$ and $\diamond$ for simple types (which also form a retraction at the type level) and we show that both translations preserve typing.

**Definition 3.4.** *For any type $\tau \in \Sigma_{\mathbf{IS}}$, the corresponding type $\tau^\star \in \Sigma_{\mathbf{FS}}$ is defined inductively as follows:*

- $\mathbf{unit}^\star = \mathbf{unit}$
- $\mathbf{nat}^\star = \mathbf{nat}$
- $(\tau_1, ..., \tau_n)^\star = (\tau_1^\star \times ... \times \tau_n^\star)$
- $\mathbf{proc}\ (\mathbf{in}\ \vec{\tau}; \mathbf{out}\ \vec{\sigma})^\star = \vec{\tau}^\star \to \vec{\sigma}^\star$

**Definition 3.5.** *For any type $\sigma \in \Sigma_{\mathbf{FS}}$ the translation $\sigma^\diamond \in \Sigma_{\mathbf{IS}}$ is defined as follows:*

- $\mathbf{unit}^\diamond = \mathbf{unit}$
- $\mathbf{nat}^\diamond = \mathbf{nat}$
- $(\sigma_1 \times ... \times \sigma_n)^\diamond = (\sigma_1^\diamond, ..., \sigma_n^\diamond)$
- $(\vec{\sigma} \to \vec{\tau})^\diamond = \mathbf{proc}\ (\mathbf{in}\ \vec{\sigma}^\diamond; \mathbf{out}\ \vec{\tau}^\diamond)$

**Proposition 3.6.** *(retraction at the type level).*

1. *For any type $\sigma \in \Sigma_{\mathbf{IS}}$, we have $\sigma^{\star\diamond} = \sigma$.*
2. *For any type $\sigma \in \Sigma_{\mathbf{FS}}$, we have $\sigma^{\diamond\star} = \sigma$.*

**Proof.** Straightforward induction on the translations $\sigma^\diamond$ and $\sigma^\star$. □

**Theorem 3.7.** *For any environments $\Gamma$ and $\Omega$, any expression $e$, any sequence $s$ we have:*

- $\Gamma; \Omega \vdash e: \tau$ in **IS** implies $\Gamma^\star, \Omega^\star \vdash e^\star: \tau^\star$ in **FS**.
- $\Gamma; \Omega \vdash s \rhd \vec{z}: \vec{\sigma}$ in **IS** implies $\Gamma^\star, \Omega^\star \vdash (s)^\star_{\vec{z}}: \vec{\sigma}^\star$ in **FS**.

**Proof.** By induction on the typing derivation. □

**Theorem 3.8.** *For any state $(s, \mu)$, if $\vec{z}: \vec{\tau} \vdash (s, \mu) \rhd \Omega$ in* **IS** *with $\vec{z}: \vec{\sigma} \subset \Omega$, then $\vdash (s)^\star_{\vec{z}}[\mu(\vec{x})^\star / \vec{x}]: \vec{\sigma}^\star$ in* **FS**.

**Proof.** By definition of state typing, $\vec{z}: \vec{\tau} \vdash (s, \mu) \rhd \Omega$ implies $\emptyset; \vec{z}: \vec{\tau} \vdash s \rhd \Omega$ and for all $(z_i, \mu(z_i)) \in \mu$, $\emptyset; \emptyset \vdash \mu(z_i): \tau_i$. By Theorem 3.7, on one hand $\emptyset; \vec{z}: \vec{\tau} \vdash s \rhd \Omega$ implies $\vec{z}: \vec{\tau}^\star \vdash (s)^\star_{\vec{z}}: \vec{\sigma}^\star$, and on the other hand $\emptyset; \emptyset \vdash \mu(z_i): \tau_i$ implies $\vdash \mu(z_i)^\star: \tau_i^\star$. Since $(s)^\star_{\vec{z}}$ is well typed in the environment $\vec{z}: \vec{\tau}^\star$, the variables in $\vec{x}$ which are not in $\vec{z}$ are not free in $(s)^\star_{\vec{z}}$. Hence, by the substitution lemma, $\vdash (s)^\star_{\vec{z}}[\mu(\vec{x})^\star / \vec{x}]: \vec{\sigma}^\star$. □

**Theorem 3.9.**

- *Given a term $t \in \mathcal{L}$ such that $\Gamma \vdash t: \vec{\sigma}$ in* **FS** *with $\Gamma, \vec{\sigma} \in \Sigma_{\mathbf{FS}}$ and a fresh mutable variable tuple $\vec{r}$ of any type $\vec{\sigma}\,' \in \Sigma_{\mathbf{ID}}$ we have $\Gamma^\diamond; \vec{r}: \vec{\sigma}\,' \vdash t^\diamond_{\vec{r}} \rhd \vec{r}: \vec{\sigma}^\diamond$ in* **IS**.
- *Given a value $v \in \mathcal{V}$ such that $\Gamma \vdash v: \sigma$ in* **FS** *with $\Gamma, \sigma \in \Sigma_{\mathbf{FS}}$, we have $\Gamma^\diamond; \vdash v^\diamond: \sigma^\diamond$ in* **IS**.

**Proof.** By induction on the typing derivation. □

## 3.4 Properties of the pseudo-dynamic type system

As expected, the transition semantics preserves typing and the usual "progress" property holds.

**Theorem 3.10.** *(preservation). For any state $(s, \mu)$, if $\vec{z}: \vec{\tau} \vdash (s, \mu) \rhd \Omega$ in* **IS** *and $(s, \mu) \mapsto (s', \mu')$ then there exists $\vec{\tau}\,'$ such that $\vec{z}: \vec{\tau}\,' \vdash (s', \mu') \rhd \Omega$, in the simple type system.*

**Proof.** By induction on the transition, and by case analysis on the typing derivation (see Appendix B.3). □

**Lemma 3.11.** *(progress). For any state* $(s,\mu)$*, if* $\vec{z}:\vec{\tau}\vdash(s,\mu)\rhd\Omega$ *in* **IS** *then either* $s=\varepsilon$ *and no more evaluation step can occur, or there is a unique state* $(s',\mu')$ *such that* $(s,\mu)\mapsto(s',\mu')$*.*

**Proof.** By induction on the typing derivation (see Appendix B.4). □

**Lemma 3.12.** *(termination). For any state* $(s,\ \mu)$*, if* $\vec{z}:\vec{\tau}\vdash(s,\ \mu)\rhd\Omega$ *in* **IS** *then the evaluation of* $(s,\ \mu)$ *terminates.*

**Proof.** By contradiction, let us assume that there is an infinite sequence of evaluation steps of $(s,\ \mu)$. By Proposition 2.12, with the fact that there cannot be an infinite sequence of evaluation steps using only rule (S.VAR-I), we have an infinite sequence of evaluation steps of $(s)^{\star}_{\vec{z}}[\mu(\vec{x})^{\star}/\vec{x}]$. By Theorem 3.8, $\vec{z}:\vec{\tau}\vdash(s,\mu)\rhd\Omega$ implies $\vdash(s)^{\star}_{\vec{z}}[\mu(\vec{x})^{\star}/\vec{x}]:\vec{\sigma}^{\star}$ and since typable terms of system T are strongly normalizing, we have a contradiction. □

**Proposition 3.13.** *For any* $(s,\ \mu)$*,* $\Omega$ *and* $\vec{z}$*, if* $\vec{z}:\vec{\tau}\vdash(s,\ \mu)\rhd\Omega$ *in* **IS***, then there is a unique store* $\mu'$ *such that* $(s,\mu)\mapsto^{n}(\varepsilon,\mu')$ *for some* $n$*.*

**Proof.** Since, by Lemma 3.12, no infinite evaluation of $(s,\ \mu)$ can occur, we prove the property by induction on the length $n$ of the longest sequence of evaluation steps from $(s,\ \mu)$, using appropriately Theorem 3.10 in the induction step. □

## 3.5 Global variables

Recall that the imperative type systems **IS** (and also **ID**, in the next section) forbids any access to global mutable variables. It is straightforward to address this restriction by passing the global variable as an explicit **in out** parameter to each procedure declaration. The same variable is then given as argument for each procedure call. Moreover, an **in out** parameter can be encoded with one **in** parameter and one **out** parameter, where each procedure initialize the variable with its input value before executing its body. To handle more conveniently a list of global variables $\vec{z}$ we introduce the following abbreviations:

$$\begin{aligned}\mathbf{proc}(\mathbf{in}\ \vec{x};\mathbf{out}\ \vec{y})_{\vec{z}}\{s\}_{\vec{y},\vec{z}} &= \mathbf{proc}(\mathbf{in}\ \vec{x},\vec{z}\,';\mathbf{out}\ \vec{y},\vec{z})\{\vec{z}:=\vec{z}\,';s\}_{\vec{y},\vec{z}}\\ p(\vec{e};\vec{y})_{\vec{z}} &= p(\vec{e},\vec{z};\vec{y},\vec{z})\end{aligned}$$

This transformation corresponds to the usual state-passing style transform in functional programming. Up to curryfication, we also obtain a state monad [50]. At the type level, however, since the type of a mutable variable can be changed by an assignment, this transform do not correspond to the usual state monad $\tau\ ST=\sigma\to(\tau\times\sigma)$ where $\sigma$ is the fixed type of the global state. We obtain instead a parameterized state monad [5], $(\sigma,\tau,\sigma')\ ST=\sigma\to(\tau\times\sigma')$ where $\sigma$ is the input type of the global state and $\sigma'$ is its output state.

This remark shows that the pseudo-dynamic type system is quite expressive and enables to type programs which would usually require an ad-hoc effect system [71].

# 4 Dependent Type Systems

In this section, we present the dependently-typed systems for languages **F** and **I**. As in the non-dependent case, we show that both translation $\star$ and $\diamond$ preserve typability. As a corollary, we obtain a soundness result (theorem 4.8) and a representation theorem (proposition 4.10) for dependently-typed imperative programs.

## 4.1 Functional dependent type system FD

Following the definition of **ML1P** [48] (or similarly **IT**($\mathbb{N}$) in [49]), we enrich language **F** with dependent types. The type system is parameterized by a first-order signature and an equational system $\mathcal{E}$ which defines a set of functions in the style of Herbrand-Gödel. We consider only the sort **nat** (with constructors 0 and **s**), and we assume that $\mathcal{E}$ contains at least the usual defining equations for addition, multiplication and a predecessor function **p** (which is essential to derive all axioms of Peano's arithmetic [49]). The syntax of formulas is the following (where $n,m$ are first-order terms):

$$\tau\ ::=\ \mathbf{nat}(n)\quad|\quad(n=m)\quad|\quad\forall\vec{\imath}\,(\tau_1\Rightarrow\tau_2)\quad|\quad\exists\vec{\imath}\,(\tau_1\wedge\ldots\wedge\tau_k)$$

Note that first-order quantifiers are provided in the form of dependent products and dependent sums. As usual, implication and conjunction are recovered as special non-dependent cases (when $\vec{\imath}$ is empty). Similarly, relativized quantification $\forall x(\mathbf{nat}(x) \Rightarrow \varphi)$ and $\exists x(\mathbf{nat}(x) \wedge \varphi)$ are also obtained as special cases.

The functional dependent type system is summarized in Figure 4.1 (where $\top$ denotes $n = n$ for some $n$ and $\vdash_{\mathcal{E}} n = m$ means that either $n = m$ or $m = n$ is an instance of $\mathcal{E}$).

The main difference between this type system and the deduction system **ML1P** described in [48] comes from the fact that a derived sequent is directly annotated by a realizer (a functional term), whereas in [48] an extraction function (or forgetful map) $\kappa$ needs to be applied to the derivation to obtain the realizer. In other words, if $\Pi$ is a derivation of a sequent $\Gamma \vdash \sigma$ in **ML1P**, then $\Gamma \vdash \kappa(\Pi) \colon \sigma$ is derivable in **FD**. Conversely, if $\Pi$ is a derivation of $\Gamma \vdash t \colon \sigma$ in **FD**, then $\Pi$ is also derivation of $\Gamma \vdash \sigma$ in **ML1P** (just remove the realizers from the derivation). Let us recall the subject reduction property of **ML1P** [48] and derive the same property for **FD** as a corollary.

**Theorem 4.1.** *(subject reduction for* **ML1P***).*

- *If* $\Pi$ *Prawitz-reduces to* $\Pi'$*, then* $\kappa\Pi$ *reduces to* $\kappa\Pi'$*.*
- *If* $t = \kappa\Pi$ *reduces to* $t'$ *then* $t' = \kappa\Pi'$ *for some* $\Pi'$ *such that* $\Pi$ *Prawitz-reduces to* $\Pi'$*.*

**Corollary 4.2.** *(subject reduction for* **FD***). If* $\Gamma \vdash t \colon \sigma$ *in* **FD** *and* $t \rightsquigarrow t'$ *then* $\Gamma \vdash t' \colon \sigma$*.*

**Proof.** Let $\Pi$ be a derivation of $\Gamma \vdash t \colon \sigma$ in **FD**, then $\kappa(\Pi) = t$ and $\Pi$ is also a derivation of $\Gamma \vdash \sigma$ in **ML1P**. By the above theorem, if $t \rightsquigarrow t'$ then $t' = \kappa\Pi'$ for some derivation $\Pi'$ of the same sequent $\Gamma \vdash \sigma$ in **ML1P**. Consequently, $\Gamma \vdash t' \colon \sigma$ is derivable since $t' = \kappa\Pi'$. □

Similarly, we obtain the representation theorem for **FD** as a corollary of the same property for **ML1P** [48, 49].

**Proposition 4.3.** *(representation theorem for* **FD***) Given an equational system* $\mathcal{E}$ *and an n-ary function symbol f, if*

$$\vdash_{\mathcal{E}} t \colon \forall \vec{n}.\mathbf{nat}(\vec{n}) \Rightarrow \mathbf{nat}(f(\vec{n}))$$

*is derivable in* **FD** *then t represents f.*

**Definition 4.4.** *(forgetful map). For any functional dependent type* $\tau$*, the computational content* $\kappa\tau$ *of* $\tau$ *is defined inductively as follows:*

- $\kappa(n = m) = \mathbf{unit}$
- $\kappa(\mathbf{nat}(n)) = \mathbf{nat}$
- $\kappa(\forall \vec{\imath}\,(\sigma \Rightarrow \tau)) = \kappa\sigma \rightarrow \kappa\tau$
- $\kappa(\exists \vec{\imath}\,(\tau_1 \wedge \ldots \wedge \tau_n) = \kappa\tau_1 \times \ldots \times \kappa\tau_n$

### 4.1.1 Example: the addition function

Recall the usual Peano's axiom for addition (see in appendix D the conventions we use in the examples):

$$\begin{array}{lrcl} (1) & x + 0 & = & x \\ (2) & x + \mathbf{s}(i) & = & \mathbf{s}(x + i) \end{array}$$

The proof of $\forall n(\mathbf{nat}(n) \Rightarrow \forall m(\mathbf{nat}(m) \Rightarrow \mathbf{nat}(n + m)))$ gives us a term of **F** that computes the addition of two natural numbers. Here follows, in a "pure" natural deduction style, the proof annotated by the terms of **F**.

$$\dfrac{\dfrac{\dfrac{y \colon \mathbf{nat}(m) \qquad \dfrac{x \colon \mathbf{nat}(n)}{x \colon \mathbf{nat}(n+0)}\text{by (1)} \qquad \dfrac{\dfrac{z \colon \mathbf{nat}(n+u)}{S(z) \colon \mathbf{nat}(\mathbf{s}(n+u))}}{S(z) \colon \mathbf{nat}(n+\mathbf{s}(u))}\text{by (2)}}{\mathbf{rec}(y, x, \lambda i.\lambda z.S(z)) \colon \mathbf{nat}(n+m)}}{\lambda y.\mathbf{rec}(y, x, \lambda i.\lambda z.S(z)) \colon \forall m(\mathbf{nat}(m) \Rightarrow \mathbf{nat}(n+m))}}{\lambda x.\lambda y.\mathbf{rec}(y, x, \lambda i.\lambda z.S(z)) \colon \forall n(\mathbf{nat}(n) \Rightarrow \forall m(\mathbf{nat}(m) \Rightarrow \mathbf{nat}(n+m)))}$$

$$\frac{x\colon\tau\in\Gamma}{\Gamma\vdash x\colon\tau} \quad (\textsc{ident})$$

$$\Gamma\vdash 0\colon \mathbf{nat}(0) \quad (\textsc{zero})$$

$$\frac{\Gamma\vdash t\colon \mathbf{nat}(n)}{\Gamma\vdash S(t)\colon \mathbf{nat}(\mathbf{s}(n))} \quad (\textsc{succ})$$

$$\frac{\Gamma\vdash t\colon \mathbf{nat}(n)}{\Gamma\vdash \mathbf{pred}(t)\colon \mathbf{nat}(\mathbf{p}(n))} \quad (\textsc{pred})$$

$$\frac{\Gamma\vdash t_1\colon \tau_1[\vec{m}/\vec{\imath}] \quad \dots \quad \Gamma\vdash t_k\colon \tau_k[\vec{m}/\vec{\imath}]}{\Gamma\vdash (t_1,\dots,t_k)\colon \exists\vec{\imath}\,(\tau_1\wedge\dots\wedge\tau_k)} \quad (\textsc{tuple})$$

$$\frac{\Gamma, x_1\colon\tau_1,\dots,x_k\colon\tau_k\vdash t\colon\tau \qquad \Gamma\vdash u\colon \exists\vec{\imath}\,(\tau_1\wedge\dots\wedge\tau_k)}{\Gamma\vdash \mathbf{let}\ (x_1,\dots,x_k)=u\ \mathbf{in}\ t\colon\tau} \quad (\textsc{let})^\star$$

$$\frac{\Gamma, x\colon\tau\vdash t\colon\sigma}{\Gamma\vdash \lambda x.t\ \colon \forall\vec{\imath}\,(\tau\Rightarrow\sigma)} \quad (\textsc{abs})^\star$$

$$\frac{\Gamma\vdash t_1\colon \forall\vec{\imath}\,(\sigma\Rightarrow\tau) \quad \Gamma\vdash t_2\colon \sigma[\vec{n}/\vec{\imath}]}{\Gamma\vdash t_1\ t_2\colon \tau[\vec{n}/\vec{\imath}]} \quad (\textsc{app})$$

$$\frac{\Gamma\vdash t_1\colon \mathbf{nat}(n) \quad \Gamma\vdash t_2\colon \tau[\mathbf{0}/i] \quad \Gamma, x\colon\mathbf{nat}(i), y\colon\tau\vdash t_3\colon\tau[\mathbf{s}(i)/i]}{\Gamma\vdash \mathbf{rec}(t_1,t_2,\lambda x.\lambda y.t_3)\colon\tau[n/i]} \quad (\textsc{rec})^\star$$

$$\frac{\vdash_{\mathcal{E}} n=m}{\Gamma\vdash ()\colon (n=m)} \quad (\textsc{equal})$$

$$\frac{\Gamma\vdash t\colon\tau[n/i] \quad \Gamma\vdash v\colon (n=m)}{\Gamma\vdash t\colon\tau[m/i]} \quad (\textsc{subst})$$

*where $\vec{\imath}\notin\mathcal{FV}(\Gamma)$ in (ABS), $\vec{\imath}\notin\mathcal{FV}(\Gamma,\tau)$ in (LET) and $i\notin\mathcal{FV}(\Gamma)$ in (REC)

**Figure 4.1.** Functional dependent type system

## 4.2 Imperative dependent type system ID

As in the functional case, the type system is parameterized by equational system $\mathcal{E}$. The syntax of imperative dependent types is the following:

$$\sigma,\tau \ ::=\ \mathbf{nat}(n) \mid \mathbf{proc}\ \forall\vec{\imath}\,(\mathbf{in}\ \vec{\tau};\mathbf{out}\ \vec{\sigma}) \mid \exists\vec{\jmath}\,(\tau_1,\dots,\tau_n) \mid n=m$$

The dependent type system is summarized in Figure 4.2 (where $\top$ denotes $n=n$ for some $n$ and $\vdash_{\mathcal{E}} n=m$ means that either $n=m$ or $m=n$ is an instance of $\mathcal{E}$).

The store typing and the state typing are defined in the same way as for the pseudo-dynamic type system.

**Definition.** (store typing). *We say that a store $\mu$ is typable of output typing environment $\Omega=z_1\colon\tau_1,\dots,z_n\colon\tau_n$, denoted $\mu\rhd\Omega$ if and only if $\vec{z}\in dom(\mu)$ and for all $(z_i,w_i)\in\mu$ we have $\emptyset;\emptyset\vdash w_i\colon\tau_i$.*

**Definition.** (state typing). *We say that a state $(s,\ \mu)$ is typable of output typing environment $\Omega'$ for a restriction of the store to the variables $\vec{z}\colon\vec{\tau}$, which we write as $\vec{z}\colon\vec{\tau}\vdash(s,\mu)\rhd\Omega'$, if and only if $\mu\rhd\vec{z}\colon\vec{\tau}$ and $\emptyset;\vec{z}\colon\vec{\tau}\vdash s\rhd\Omega'$.*

**Definition 4.5.** *(forgetful map). For any imperative dependent type $\tau$, the computational content $\kappa\tau$ of $\tau$ is defined inductively as follows:*

- $\kappa(n=m)=\mathbf{unit}$
- $\kappa(\mathbf{nat}(n))=\mathbf{nat}$

$$\frac{x\colon\tau \in \Gamma;\Omega}{\Gamma;\Omega \vdash x\colon\tau} \quad (\text{T.ENV})$$

$$\frac{}{\Gamma;\Omega \vdash \bar{q}\colon \mathbf{nat}(\mathbf{s}^q(\mathbf{0}))} \quad (\text{T.NUM})$$

$$\frac{\vdash_{\mathcal{E}} n = m}{\Gamma;\Omega \vdash *: n = m} \quad (\text{T.EQUAL})$$

$$\frac{\Gamma;\Omega \vdash e_1\colon \tau_1[\vec{m}/\vec{\imath}] \quad \dots \quad \Gamma;\Omega \vdash e_n\colon \tau_n[\vec{m}/\vec{\imath}]}{\Gamma;\Omega \vdash (e_1,\dots,e_n)\colon \exists\vec{\imath}\,(\tau_1,\dots,\tau_n)} \quad (\text{T.TUPLE})$$

$$\frac{\vec{z} \neq \emptyset \qquad \Gamma,\vec{y}:\vec{\sigma};\vec{z}:\vec{\top} \vdash s \rhd \vec{z}:\vec{\tau}}{\Gamma;\Omega \vdash \mathbf{proc}\ (\mathbf{in}\ \vec{y};\mathbf{out}\ \vec{z})\{s\}_{\vec{z}}\colon \mathbf{proc}\ \forall\vec{\imath}\,(\mathbf{in}\ \vec{\sigma};\mathbf{out}\ \vec{\tau})} \quad (\text{T.PROC})^*$$

$$\frac{\Gamma;\Omega \vdash e'\colon \tau[n/i] \qquad \Gamma;\Omega \vdash e\colon n = m}{\Gamma;\Omega \vdash e'\colon \tau[m/i]} \quad (\text{T.SUBST-I})$$

$$\frac{\Gamma;\Omega \vdash s \rhd \Omega'[n/i] \qquad \Gamma;\Omega \vdash e\colon n = m}{\Gamma;\Omega \vdash s \rhd \Omega'[m/i]} \quad (\text{T.SUBST-II})$$

$$\frac{}{\Gamma;\Omega,\Omega' \vdash \varepsilon \rhd \Omega'} \quad (\text{T.EMPTY})$$

$$\frac{\Gamma;\Omega,\vec{x}:\vec{\sigma} \vdash c \rhd \vec{x}:\vec{\tau} \qquad \Gamma;\Omega,\vec{x}:\vec{\tau} \vdash s \rhd \Omega'}{\Gamma;\Omega,\vec{x}:\vec{\sigma} \vdash c;\ s \rhd \Omega'} \quad (\text{T.SEQ})$$

$$\frac{\Gamma;\Omega \vdash e\colon\tau \qquad \Gamma, y\colon\tau;\Omega \vdash s \rhd \Omega'}{\Gamma;\Omega \vdash \mathbf{cst}\ y = e;\ s \rhd \Omega'} \quad (\text{T.CST})$$

$$\frac{\Gamma;\Omega \vdash e\colon\tau \qquad \Gamma;\Omega, y\colon\tau \vdash s \rhd \Omega' \qquad y \notin \Omega'}{\Gamma;\Omega \vdash \mathbf{var}\ y := e;\ s \rhd \Omega'} \quad (\text{T.VAR})$$

$$\frac{\Gamma;\Omega,\vec{y}:\vec{\sigma} \vdash e\colon \exists\vec{\imath}\,(\tau_1,\dots,\tau_n) \qquad \Gamma;\Omega, y_1\colon\tau_1,\dots,y_n\colon\tau_n \vdash s \rhd \Omega'}{\Gamma;\Omega,\vec{y}:\vec{\sigma} \vdash \vec{y} := e;\ s \rhd \Omega'} \quad (\text{T.ASSIGN})^*$$

$$\frac{\Gamma;\vec{x}:\vec{\tau} \vdash s \rhd \vec{x}:\vec{\sigma}}{\Gamma;\Omega,\vec{x}:\vec{\tau} \vdash \{s\}_{\vec{x}} \rhd \vec{x}:\vec{\sigma}} \quad (\text{T.BLOCK})$$

$$\frac{}{\Gamma;\Omega, y\colon \mathbf{nat}(n) \vdash \mathbf{inc}(y) \rhd y\colon \mathbf{nat}(\mathbf{s}(n))} \quad (\text{T.INC})$$

$$\frac{}{\Gamma;\Omega, y\colon \mathbf{nat}(n) \vdash \mathbf{dec}(y) \rhd y\colon \mathbf{nat}(\mathbf{p}(n))} \quad (\text{T.DEC})$$

$$\frac{\Gamma;\Omega,\vec{x}:\vec{\sigma}[\mathbf{0}/i] \vdash e\colon \mathbf{nat}(n) \qquad \Gamma, y\colon \mathbf{nat}(i);\vec{x}:\vec{\sigma} \vdash s \rhd \vec{x}:\vec{\sigma}[\mathbf{s}(i)/i]}{\Gamma;\Omega,\vec{x}:\vec{\sigma}[\mathbf{0}/i] \vdash \mathbf{for}\ y := 0\ \mathbf{until}\ e\ \{s\}_{\vec{x}} \rhd \vec{x}:\vec{\sigma}[n/i]} \quad (\text{T.FOR})^*$$

$$\frac{\Gamma;\Omega,\vec{r}:\vec{\omega} \vdash p\colon \mathbf{proc}\ \forall\vec{\imath}\,(\mathbf{in}\ \vec{\sigma};\mathbf{out}\ \vec{\tau}) \qquad \Gamma;\Omega,\vec{r}:\vec{\omega} \vdash \vec{e}:\vec{\sigma}[\vec{m}/\vec{\imath}]}{\Gamma;\Omega,\vec{r}:\vec{\omega} \vdash p(\vec{e};\vec{r}) \rhd \vec{r}:\vec{\tau}[\vec{m}/\vec{\imath}]} \quad (\text{T.CALL})$$

*where $\vec{\imath} \notin \mathcal{FV}(\Gamma)$ in (T.PROC) and $i \notin \mathcal{FV}(\Gamma)$ in (T.FOR) and $\vec{\jmath} \notin \mathcal{FV}(\Gamma,\Omega,\Omega')$ in (T.ASSIGN)

**Figure 4.2.** Imperative dependent type system

- $\kappa(\exists\vec{\jmath}\,(\tau_1,\dots,\tau_n)) = (\kappa\tau_1,\dots,\kappa\tau_n)$
- $\kappa(\mathbf{proc}\ \forall\vec{\imath}\,(\mathbf{in}\ \vec{\sigma};\mathbf{out}\ \vec{\tau})) = \mathbf{proc}\ (\mathbf{in}\ \kappa\vec{\tau};\mathbf{out}\ \kappa\vec{\sigma})$

**Proposition 4.6.** *(erasure). If $\Gamma;\Omega \vdash s \rhd \Omega'$ is derivable in* **ID** *then $\kappa\Gamma;\kappa\Omega \vdash s \rhd \kappa\Omega'$ is derivable in* **IS**.

**Proof.** By induction on the typing derivation of $\Gamma;\Omega \vdash s \rhd \Omega'$. □

### 4.2.1 Example: the addition procedure

Complete type derivations in **ID** are tedious. In the following examples, we prefer instead to provide only some type annotations on the right-hand side of the program. Although we did not formally define this syntax, we believe that it is self-explanatory and contains enough information to reconstruct the complete type derivation in **ID**. For instance, here is the procedure *add* which computes the addition together with the sketch of its type derivation:

| | |
|---|---|
| **cst** $add =$ **proc** (**in** $X, Y$; **out** $Z$) { | $-\,(X\colon \mathbf{nat}(x), Y\colon \mathbf{nat}(y))[Z\colon \top]$ |
| $\quad Z := X$; | $\mid\quad [Z\colon \mathbf{nat}(x+0)]\quad$ by (1) |
| $\quad$ **for** $I := 0$ **until** $Y$ { | $\mid\quad -\,(I\colon \mathbf{nat}(i))[Z\colon \mathbf{nat}(x+i)]$ |
| $\qquad$ **inc**($Z$); | $\mid\quad \mid\quad [Z\colon \mathbf{nat}(x+\mathbf{s}(i))]\quad$ by (2) |
| $\quad \}_Z$; | $\mid\quad [Z\colon \mathbf{nat}(x+y)]$ |
| $\}_Z$ | $(add\colon \mathbf{proc}\ \forall x, y(\mathbf{in}\ \mathbf{nat}(x), \mathbf{nat}(y); \mathbf{out}\ \mathbf{nat}(x+y)))$ |

### 4.2.2 Example: the Ackermann procedure

We recall the equations which define a variant the Ackermann function [49]:

$$
\begin{array}{lll}
(1)\ \mathbf{a}(0, n) & = & \mathbf{s}(n) \\
(2)\ \mathbf{a}(\mathbf{s}(z), 0) & = & \mathbf{s}(\mathbf{s}(0)) \\
(3)\ \mathbf{a}(\mathbf{s}(z), \mathbf{s}(u)) & = & \mathbf{a}(z, \mathbf{a}(\mathbf{s}(z), u))
\end{array}
$$

Similarly, from a proof of $\forall m, n(\mathbf{nat}(m) \wedge \mathbf{nat}(n) \Rightarrow \mathbf{nat}(\mathbf{a}(m, n)))$ in **FD** in monadic normal form, by applying translation $^\diamond$ by hand, we obtain a procedure which computes $\mathbf{a}(m, n)$ (the functional typing derivation is in Appendix D.3.1). Here is the definition of the procedure *ack* with its typing annotations.

| | |
|---|---|
| **cst** $ack =$ **proc** (**in** $M, N$; **out** $Z$) { | $-\,(M\colon \mathbf{nat}(m), N\colon \mathbf{nat}(n))[Z\colon \top]$ |
| | $\mid$ |
| $\quad$ **var** $G :=$ **proc** (**in** $Y$; **out** $P$) { | $\mid\quad -\,(Y\colon \mathbf{nat}(y))[P\colon \top]$ |
| $\qquad P := Y$; | $\mid\quad \mid\quad [P\colon \mathbf{nat}(y)]$ |
| $\qquad$ **inc**($P$); | $\mid\quad \mid\quad [P\colon \mathbf{nat}(\mathbf{s}(y))]$ |
| $\quad \}_P$; | $\mid\quad [G\colon \mathbf{proc}\ \forall y(\mathbf{in}\ \mathbf{nat}(y); \mathbf{out}\ \mathbf{nat}(\mathbf{a}(0, y)))]\quad$ by (1) |
| | $\mid$ |
| $\quad$ **for** $I := 0$ **until** $M$ { | $\mid\quad -\,(I\colon \mathbf{nat}(i))[G\colon \mathbf{proc}\ \forall y(\mathbf{in}\ \mathbf{nat}(y); \mathbf{out}\ \mathbf{nat}(\mathbf{a}(i, y)))]$ |
| $\qquad$ **cst** $H = G$; | $\mid\quad \mid\quad (H\colon \mathbf{proc}\ \forall y(\mathbf{in}\ \mathbf{nat}(y); \mathbf{out}\ \mathbf{nat}(\mathbf{a}(i, y))))$ |
| | $\mid\quad \mid$ |
| $\qquad G :=$ **proc** (**in** $Y$; **out** $P$) { | $\mid\quad \mid\quad -\,(Y\colon \mathbf{nat}(y))[P\colon \top]$ |
| $\qquad\quad P := 2$; | $\mid\quad \mid\quad \mid\quad [P\colon \mathbf{nat}(\mathbf{a}(\mathbf{s}(i), 0))]\quad$ by (2) |
| $\qquad\quad$ **for** $J := 0$ **until** $Y$ { | $\mid\quad \mid\quad \mid\quad -\,(J\colon \mathbf{nat}(j))[P\colon \mathbf{nat}(\mathbf{a}(\mathbf{s}(i), j))]$ |
| $\qquad\qquad H(P; P)$; | $\mid\quad \mid\quad \mid\quad \mid\quad [P\colon \mathbf{nat}(\mathbf{a}(\mathbf{s}(i), \mathbf{s}(j)))]\quad$ by (3) |
| $\qquad\quad \}_P$; | $\mid\quad \mid\quad \mid\quad [P\colon \mathbf{nat}(\mathbf{a}(\mathbf{s}(i), y)))]$ |
| $\qquad \}_P$; | $\mid\quad \mid\quad [G\colon \mathbf{proc}\ \forall y(\mathbf{in}\ \mathbf{nat}(y); \mathbf{out}\ \mathbf{nat}(\mathbf{a}(\mathbf{s}(i), y))))]$ |
| | $\mid\quad \mid$ |
| $\quad \}_G$; | $\mid\quad [G\colon \mathbf{proc}\ \forall y(\mathbf{in}\ \mathbf{nat}(y); \mathbf{out}\ \mathbf{nat}(\mathbf{a}(m, y)))]$ |
| | $\mid$ |
| $\quad G(N; Z)$; | $\mid\quad [Z\colon \mathbf{a}(m, n)]$ |
| $\}_Z$ | $(ack\colon \mathbf{proc}\ \forall m, n(\mathbf{in}\ \mathbf{nat}(m), \mathbf{nat}(n); \mathbf{out}\ \mathbf{nat}(\mathbf{a}(m, n))))$ |

## 4.3 Translation from ID to FD

We show that translation $^\star$ preserves dependent types.

**Definition 4.7.** *(translation of dependent types). For any imperative dependent type $\tau$, the corresponding functional dependent type $\tau^\star$ is defined inductively as follows:*

- $(t = u)^\star = (t = u)$

- $(\mathbf{nat}(u))^\star = \mathbf{nat}(u)$
- $(\exists\vec{\imath}\,(\tau_1, \ldots, \tau_n))^\star = \exists\vec{\imath}\,(\tau_1^\star \wedge \ldots \wedge \tau_n^\star)$
- $(\mathbf{proc}\ \forall\vec{\imath}\,(\mathbf{in}\ \vec{\tau}; \mathbf{out}\ \vec{\sigma}))^\star = \forall\vec{\imath}\,(\vec{\tau}^\star \Rightarrow \vec{\sigma}^\star)$

**Theorem 4.8.** *(Soundness for* **ID***). For any environments* $\Gamma$ *and* $\Omega$*, any expression* $e$*, any sequence* $s$ *we have:*

– $\Gamma; \Omega \vdash e{:}\,\tau$ *in* **ID** *implies* $\Gamma^\star, \Omega^\star \vdash e^\star{:}\,\tau^\star$ *in* **FD***.*

– $\Gamma; \Omega \vdash s \rhd \vec{z}: \vec{\sigma}$ *in* **ID** *implies* $\Gamma^\star, \Omega^\star \vdash (s)^\star_{\vec{z}}{:}\,\vec{\sigma}^\star$ *in* **FD***.*

**Proof.** See Appendix C.3. ☐

**Theorem 4.9.** *For any state* $(s, \mu)$*, if* $\vec{z}: \vec{\tau} \vdash (s, \mu) \rhd \vec{z}: \vec{\sigma}$ *in* **ID** *then* $\vdash (s)^\star_{\vec{z}}[\mu(\vec{x})^\star/\vec{x}]{:}\,\vec{\sigma}^\star$ *in* **FD***.*

**Proof.** By definition of state typing, $\vec{z}: \vec{\tau} \vdash (s, \mu) \rhd \vec{z}: \vec{\sigma}$ implies $\emptyset; \vec{z}: \vec{\tau} \vdash s \rhd \vec{z}: \vec{\sigma}$ and for all $(z_i, \mu(z_i)) \in \mu$, $\emptyset$; $\emptyset \vdash \mu(z_i){:}\,\tau_i$. By theorem 4.8, on one hand $\emptyset; \vec{z}: \vec{\tau} \vdash s \rhd \vec{z}: \vec{\sigma}$ implies $\vec{z}: \vec{\tau}^\star \vdash (s)^\star_{\vec{z}}{:}\,\vec{\sigma}^\star$, and on the other hand $\emptyset$; $\emptyset \vdash \mu(z_i){:}\,\tau_i$ implies $\vdash \mu(z_i)^\star{:}\,\tau_i^\star$. Since $(s)^\star_{\vec{z}}$ is well typed in the environment $\vec{z}: \vec{\tau}^\star$, the variables in $\vec{x}$ which are not in $\vec{z}$ are not free in $(s)^\star_{\vec{z}}$. Hence, by the substitution lemma, $\vdash (s)^\star_{\vec{z}}[\mu(\vec{x})^\star/\vec{x}]{:}\,\vec{\sigma}^\star$. ☐

## 4.4 Properties of dependently-typed imperative programs

We are now ready to state and prove the representation theorem for dependently-typed imperative programs. This theorem is a corollary of the representation theorem for **FD** and the simulation theorem.

**Corollary 4.10.** *(representation theorem for* **ID***). Given an equational system* $\mathcal{E}$ *and an n-ary function symbol f, if*

$$\vdash p{:}\,\mathbf{proc}\ \forall\vec{n}(\mathbf{in}\ \mathbf{nat}(\vec{n}); \mathbf{out}\ \mathbf{nat}(f(\vec{n})))$$

*is derivable in* **ID** *then p represents f.*

**Proof.** Indeed, $\vdash p^\star{:}\ \forall\vec{n}.\mathbf{nat}(\vec{n}) \Rightarrow \mathbf{nat}(f(\vec{n}))$ is derivable in **FD**, and thus $p^\star$ represents $f$ by proposition 4.3. Since by Proposition 4.6, $\vdash p{:}\,\mathbf{proc}(\mathbf{in}\ \mathbf{nat}; \mathbf{out}\ \mathbf{nat})$ is derivable in **IS**, we know that $p$ always terminates by lemma 3.12 and computes $p^\star$ by proposition 2.12. ☐

## 4.5 Translation from FD to ID

We close this section by some properties of translation $^\diamond$.

**Definition 4.11.** *For any type* $\sigma \in \Sigma_{\mathbf{FD}}$ *the translation* $\sigma^\diamond$ *is defined as follows:*

– $(n = m)^\diamond = (n = m)$

– $(\mathbf{nat}(n))^\diamond = \mathbf{nat}(n)$

– $(\exists\vec{\jmath}\,(\sigma_1 \wedge \ldots \wedge \sigma_n))^\diamond = \exists\vec{\jmath}\,(\sigma_1^\diamond, \ldots, \sigma_n^\diamond)$

– $(\forall\vec{\imath}\,(\vec{\tau} \Rightarrow \vec{\sigma}))^\diamond = \mathbf{proc}\ \forall\vec{\imath}\,(\mathbf{in}\ \vec{\tau}^\diamond; \mathbf{out}\ \vec{\sigma}^\diamond)$

As expected, Proposition 3.6 is extended as follows.

**Proposition 4.12.** *(retraction).*

1. *For any type* $\sigma \in \Sigma_{\mathbf{ID}}$*, we have* $\sigma^{\star\diamond} = \sigma$*.*
2. *For any type* $\sigma \in \Sigma_{\mathbf{FD}}$*, we have* $\sigma^{\diamond\star} = \sigma$*.*

**Proof.** Straightforward induction on translations $\sigma^\diamond$ and $\sigma^\star$. ☐

**Proposition 4.13.** *(erasure and translation commute). For any imperative dependent type* $\sigma$ *we have* $\kappa(\sigma^\star) = (\kappa\sigma)^\star$*.*

**Proof.** Straightforward induction on types. □

**Theorem 4.14.**

- *Given a term* $t \in \mathcal{L}$ *such that* $\Gamma \vdash t\colon \vec{\sigma}$ *in* **FD** *with* $\Gamma, \vec{\sigma} \in \Sigma_{\mathbf{FD}}$ *and a fresh mutable variable tuple* $\vec{r}$ *of any type* $\vec{\sigma}\,' \in \Sigma_{\mathbf{ID}}$ *we have* $\Gamma^{\diamond}; \vec{r} : \vec{\sigma}\,' \vdash t^{\diamond}_{\vec{r}} \rhd \vec{r} : \vec{\sigma}^{\diamond}$ *in* **ID**.
- *Given a value* $v \in \mathcal{V}$ *such that* $\Gamma \vdash v\colon \sigma$ *in* **FD** *with* $\Gamma, \sigma \in \Sigma_{\mathbf{FD}}$, *we have* $\Gamma^{\diamond}; \vdash v^{\diamond}\colon \sigma^{\diamond}$ *in* **ID**.

**Proof.** By induction on the typing derivation (see Appendix C.4). □

**Remark 4.15.** Translation $^{\diamond}$ is only defined above for terms of $\mathcal{L}$. Translating an arbitrary term (typable in **FD**) into an imperative program (typable in **ID**), just requires to put the term in monadic normal form. More details are given in Appendix C.5.

# 5 Control operators

In order to extend the imperative language **I** with non-local jumps, we first extend the functional language **F** with control operators. The resulting dependent type system $\mathbf{FD}^c$ corresponds thus to classical logic [31] (Peano's arithmetic in fact). In this section, we rephrase known results from [55, 57] in our setting. However, since **FD** is based on Leivant's **ML1P**, our variant may seem closer to Parigot's type system for the $\lambda\mu$-calculus [61] (albeit in the second-order framework).

## 5.1 Functional dependent type system for control $\mathbf{FD}^c$

In order to extend **FD** to $\mathbf{FD}^c$, we assume the existence of a propositional constant "absurd" written $\bot$, we define the negation $\neg\varphi$ as an abbreviation for $\varphi \Rightarrow \bot$ and we add two constants **callcc** and **throw** with the following types:

$$\begin{aligned} \mathbf{callcc} &: (\neg\varphi \Rightarrow \varphi) \Rightarrow \varphi \\ \mathbf{throw} &: (\neg\varphi \wedge \varphi) \Rightarrow \psi \end{aligned}$$

This choice of control operators is taken from [32] but it would be equivalent to take for instance $\mathcal{A}$ and $\mathcal{C}$ from [21] as in [55, 57]. Note that we do not consider any direct style semantics of these operators in this paper. Instead, we give an indirect semantics as a CPS-transformation [65].

## 5.2 CPS translation

As is well-known [33], it is natural to factor a CPS-transformation through Moggi's computational meta-language [53, 54]. Since we are interested in providing a semantics for imperative programs and since the output of translation $^{\star}$ is already a term in monadic normal form, the CPS-transformation needed is almost straightforward. We still have to be careful since in a dependent type system a monad is actually a modality [12, 8], and we have to deal with first-order quantifiers.

Following [12], we write $\neg_o\varphi$ for $\varphi \Rightarrow o$ where $o$ is a fixed propositional variable. The continuation monad $\nabla$ is then defined as $\nabla\varphi = \neg_o\neg_o\varphi$ together with the following two abbreviations (which corresponds to *unit* and *bind*):

$$\begin{aligned} \mathbf{val}\ u &= \lambda z.(z\ u) \\ \mathbf{let\ val}\ x = u\ \mathbf{in}\ t &= \lambda z.(u\ \lambda x.(t\ z)) \end{aligned}$$

Moreover, in the continuation monad, control operators *callcc* and *throw* are definable as the following abbreviations [59]:

$$\begin{aligned} \mathit{callcc} &= \lambda h.\lambda k.(h\ k\ k) \\ \mathit{throw} &= \lambda(k,a).\lambda k'.(k\ a) \end{aligned}$$

Let us now prove that for any monadic normal form (possibly containing **callcc** and **throw**) typable in $\mathbf{FD}^c$, its call-by-value CPS-transform is typable in **FD**. The translation of dependent types is defined as follows:

**Definition 5.1.** *(translation of dependent types from* $\mathbf{FD}^c$ *to* $\mathbf{FD}$*)*

$$\begin{array}{rcl}
\mathbf{nat}(n)^\circ &=& \mathbf{nat}(n)\\
(n=m)^\circ &=& (n=m)\\
(\exists\vec{n}\,(\varphi_1\wedge\ldots\wedge\varphi_n))^\circ &=& \exists\vec{n}\,(\varphi_1^\circ\wedge\ldots\wedge\varphi_n^\circ)\\
(\forall\vec{n}\,(\varphi\Rightarrow\psi))^\circ &=& \forall\vec{n}\,(\varphi^\circ\Rightarrow\nabla\psi^\circ)\\
(\neg\varphi)^\circ &=& \neg_o\varphi^\circ\\
\bot^\circ &=& o
\end{array}$$

**Remark 5.2.** If we instantiate the monad, and restrict ourselves to relativized quantifiers we obtain as expected Murthy's variant [55, 57] of Kuroda's translation [43].

**Definition 5.3.** *For any value* $v\in\mathcal{V}$ *and any term* $t\in\mathcal{L}$ *possibly containing* **callcc** *and* **throw**, *the call-by-value CPS-transform* $v^\bullet$ *and* $t^\circ$ *are defined by mutual induction as follows:*

$$\begin{array}{rcl}
()^\bullet &=& ()\\
x^\bullet &=& x\\
0^\bullet &=& 0\\
S(v)^\bullet &=& S(v^\bullet)\\
(\lambda x.u)^\bullet &=& (\lambda x.u^\circ)\\
(v_1,\ldots,v_k)^\bullet &=& (v_1^\bullet,\ldots,v_k^\bullet)\\
(\mathbf{callcc})^\bullet &=& \mathit{callcc}\\
(\mathbf{throw})^\bullet &=& \mathit{throw}
\end{array}$$

$$\begin{array}{rcl}
(v)^\circ &=& \mathbf{val}\ (v^\bullet)\\
(v_1\ v_2)^\circ &=& (v_1^\bullet\ v_2^\bullet)\\
(\mathbf{let}\ (x_1,\ldots,x_n)=t\ \mathbf{in}\ u)^\circ &=& \mathbf{let\ val}\ y=t^\circ\ \mathbf{in\ let}\ (x_1,\ldots,x_n)=y\ \mathbf{in}\ u^\circ\\
\mathbf{rec}(v,u,\lambda x.\lambda y.t)^\circ &=& \mathbf{rec}(v^\bullet,u^\circ,\lambda x.\lambda r.\mathbf{let\ val}\ y=r\ \mathbf{in}\ t^\circ)\\
\mathbf{pred}(v)^\circ &=& \mathbf{val\ pred}(v^\bullet)
\end{array}$$

**Remark 5.4.** The translation above is defined for a syntax slightly more general than $\mathcal{L}$ since we only need here to distinguish values from computations. It is however straightforward to check that any term of $\mathcal{L}$ belongs to $dom(^\circ)$ and any value of $\mathcal{V}$ belongs to $dom(^\bullet)$.

**Lemma 5.5.** *The following typing rules are derivable in* $\mathbf{FD}$*:*

$$\frac{\Gamma\vdash u:\varphi}{\Gamma\vdash\mathbf{val}\ u:\nabla\varphi}\qquad\qquad\frac{\Gamma\vdash u:\nabla\varphi\qquad\Gamma,x:\varphi\vdash t:\nabla\psi}{\Gamma\vdash\mathbf{let\ val}\ x=u\ \mathbf{in}\ t:\nabla\psi}$$

**Proof.** Straightforward (see Appendix B). □

**Lemma 5.6.** *Abbreviations callcc and throw are typable in* $\mathbf{FD}$ *as follows:*

$$\begin{array}{rcl}
\mathit{callcc} &:& ((\varphi^\circ\Rightarrow o)\Rightarrow\nabla\varphi^\circ)\Rightarrow\nabla\varphi^\circ\\
\mathit{throw} &:& ((\varphi^\circ\Rightarrow o)\wedge\varphi^\circ)\Rightarrow\nabla\psi^\circ
\end{array}$$

**Proof.** Straightforward (see Appendix B). □

**Lemma 5.7.** *For any term* $t$ *of* $\mathcal{L}$ *(resp. any value* $v$ *of* $\mathcal{V}$*) possibly containing* **callcc** *and* **throw**, *if* $\Gamma\vdash t:\varphi$ *(resp.* $\Gamma\vdash v:\varphi$*) is derivable in* $\mathbf{FD}^c$ *then* $\Gamma^\circ\vdash t^\circ:\nabla\varphi^\circ$ *(resp.* $\Gamma^\circ\vdash v^\bullet:\varphi^\circ$*) is derivable in* $\mathbf{FD}$.

**Proof.** By induction on the typing derivation where the basic cases for *callcc* and *throw* are obtained by Lemma 5.6:

- (IDENT)

$$\Gamma,x:\varphi\vdash x:\varphi$$

Indeed,

$$\Gamma^\circ, x\colon \varphi^\circ \vdash x\colon \varphi^\circ$$

- (EQUAL)

$$\frac{\vdash_{\mathcal{E}} n = m}{\Gamma \vdash ()\colon (n = m)}$$

Indeed,

$$\frac{\vdash_{\mathcal{E}} n = m}{\Gamma^\circ \vdash ()\colon (n = m)}$$

- (SUBST)

$$\frac{\Gamma \vdash t\colon \varphi[n/i] \qquad \Gamma \vdash v\colon (n = m)}{\Gamma \vdash t\colon \varphi[m/i]}$$

Indeed,

$$\frac{\Gamma^\circ \vdash t^\circ\colon \nabla \varphi^\circ[n/i] \qquad \Gamma^\circ \vdash v^\bullet\colon (n = m)}{\Gamma^\circ \vdash t^\circ\colon \nabla \varphi^\circ[m/i]}$$

- (ZERO)

$$\Gamma \vdash 0\colon \mathbf{nat}(0)$$

Indeed,

$$\Gamma^\circ \vdash 0\colon \mathbf{nat}(0)$$

- (SUCC)

$$\frac{\Gamma \vdash v\colon \mathbf{nat}(n)}{\Gamma \vdash S(v)\colon \mathbf{nat}(\mathbf{s}n)}$$

Indeed,

$$\frac{\Gamma^\circ \vdash v^\bullet\colon \mathbf{nat}(n)}{\Gamma^\circ \vdash S(v^\bullet)\colon \mathbf{nat}(\mathbf{s}n)}$$

- (ABS) where $\vec{\imath} \notin \mathcal{FV}(\Gamma)$

$$\frac{\Gamma, x\colon \varphi \vdash u\colon \psi}{\Gamma \vdash \lambda x.u\colon \forall \vec{\imath}\,(\varphi \Rightarrow \psi)}$$

Indeed,

$$\frac{\Gamma^\circ, x\colon \varphi^\circ \vdash u^\circ\colon \nabla \psi^\circ}{\Gamma^\circ \vdash \lambda x.u^\circ\colon \forall \vec{\imath}\,(\varphi^\circ \Rightarrow \nabla \psi^\circ)}$$

- (APP)

$$\frac{\Gamma \vdash v_1\colon \forall \vec{\imath}\,(\varphi \Rightarrow \psi) \qquad \Gamma \vdash v_2\colon \varphi[\vec{n}/\vec{\imath}]}{\Gamma \vdash (v_1\ v_2)\colon \psi[\vec{n}/\vec{\imath}]}$$

Indeed,

$$\frac{\Gamma \vdash v_1^\bullet\colon \forall \vec{\imath}\,(\varphi^\circ \Rightarrow \nabla \psi^\circ) \qquad \Gamma^\circ \vdash v_2^\bullet\colon \varphi^\circ[\vec{n}/\vec{\imath}]}{\Gamma \vdash (v_1^\bullet\ v_2^\bullet)\colon \nabla \psi^\circ[\vec{n}/\vec{\imath}]}$$

- (TUPLE)

$$\frac{\Gamma \vdash v_1\colon \varphi_1[\vec{n}/\vec{\imath}] \quad \dots \quad \Gamma \vdash v_k\colon \varphi_k[\vec{n}/\vec{\imath}]}{\Gamma \vdash (v_1, \dots, v_k)\colon \exists \vec{\imath}\,(\varphi_1 \wedge \dots \wedge \varphi_k)}$$

Indeed,

$$\frac{\Gamma^\circ \vdash v_1^\bullet\colon \varphi_1^\circ[\vec{n}/\vec{\imath}] \quad \dots \quad \Gamma^\circ \vdash v_k^\bullet\colon \varphi_k^\circ[\vec{n}/\vec{\imath}]}{\Gamma^\circ \vdash (v_1^\bullet, \dots, v_k^\bullet)\colon \exists \vec{\imath}\,(\varphi_1^\circ \wedge \dots \wedge \varphi_k^\circ)}$$

- (LET) where $\vec{n} \notin \mathcal{FV}(\Gamma, \psi)$

$$\frac{\Gamma \vdash t\colon \exists \vec{\imath}\,(\varphi_1 \wedge \dots \wedge \varphi_k) \quad \Gamma, x_1\colon \varphi_1[\vec{n}/\vec{\imath}], \dots, x_k\colon \varphi_k[\vec{n}/\vec{\imath}] \vdash u\colon \psi}{\Gamma \vdash \mathbf{let}\ (x_1, \dots, x_k) = t\ \mathbf{in}\ u\colon \psi}$$

Indeed, since $\vec{n} \notin \mathcal{FV}(\Gamma^\circ, \psi^\circ)$

$$\frac{\Gamma^\circ \vdash t\colon \nabla(\exists \vec{\imath}\,(\varphi_1^\circ \wedge \dots \wedge \varphi_k^\circ)) \qquad \dfrac{\Gamma^\circ \vdash t\colon \exists \vec{\imath}\,(\varphi_1^\circ \wedge \dots \wedge \varphi_k^\circ) \quad \Gamma^\circ, x_1\colon \varphi_1^\circ[\vec{n}/\vec{\imath}], \dots, x_k\colon \varphi_k^\circ[\vec{n}/\vec{\imath}] \vdash u^\circ\colon \nabla \psi^\circ}{\Gamma^\circ \vdash \mathbf{let}\ (x_1, \dots, x_k) = y\ \mathbf{in}\ u^\circ\colon \nabla \psi^\circ}}{\Gamma^\circ \vdash \mathbf{let\ val}\ y = t^\circ\ \mathbf{in\ let}\ (x_1, \dots, x_k) = y\ \mathbf{in}\ u^\circ\colon \nabla \psi^\circ}$$

- (REC) where $i \notin \mathcal{FV}(\Gamma)$

$$\frac{\Gamma \vdash v\colon \mathbf{nat}(n) \quad \Gamma \vdash u\colon \varphi[0/i] \quad \Gamma, x\colon \mathbf{nat}(i), y\colon \varphi \vdash t\colon \varphi[\mathbf{s}(i)/i]}{\Gamma \vdash \mathbf{rec}(v, u, \lambda x.\lambda y.t)\colon \varphi[n/i]}$$

  Indeed, since $i \notin \mathcal{FV}(\Gamma^\circ)$

$$\frac{\Gamma^\circ \vdash v^\bullet\colon \mathbf{nat}(n) \quad \Gamma^\circ \vdash u^\circ\colon \nabla\varphi^\circ[0/i] \quad \dfrac{\dfrac{\Gamma^\circ, r\colon \nabla\varphi^\circ \vdash r\colon \nabla\varphi^\circ \quad \Gamma^\circ, x\colon \mathbf{nat}(i), y\colon \varphi^\circ \vdash t^\circ\colon \nabla\varphi^\circ[\mathbf{s}(i)/i]}{\Gamma^\circ, x\colon \mathbf{nat}(i), r\colon \nabla\varphi^\circ \vdash \mathbf{let\ val}\ y = r\ \mathbf{in}\ t^\circ\colon \nabla\varphi^\circ[\mathbf{s}(i)/i]}}{\Gamma^\circ, x\colon \mathbf{nat}(i) \vdash \lambda r.\mathbf{let\ val}\ y = r\ \mathbf{in}\ t^\circ\colon \nabla\varphi^\circ \Rightarrow \nabla\varphi^\circ[\mathbf{s}(i)/i]}}{\Gamma^\circ \vdash \mathbf{rec}(v^\bullet, u^\circ, \lambda x.\lambda r.\mathbf{let\ val}\ y = r\ \mathbf{in}\ t^\circ)\colon \nabla\varphi^\circ[n/i]}$$

- (PRED)

$$\frac{\Gamma \vdash v\colon \mathbf{nat}(n)}{\Gamma \vdash \mathbf{pred}(v)\colon \mathbf{nat}(\mathbf{p}n)}$$

  Indeed,

$$\frac{\dfrac{\Gamma^\circ \vdash v^\bullet\colon \mathbf{nat}(n)}{\Gamma^\circ \vdash \mathbf{pred}(v^\bullet)\colon \mathbf{nat}(\mathbf{p}n)}}{\Gamma^\circ \vdash \mathbf{val\ pred}(v^\bullet)\colon \nabla\mathbf{nat}(\mathbf{p}n)}$$

□

As a corollary of Lemma 5.7, we obtain a representation theorem for $\mathbf{FD}^c$.

**Theorem 5.8.** *(representation theorem for $\mathbf{FD}^c$). Given an equational system $\mathcal{E}$ and an n-ary function symbol f, if $\vdash t\colon \forall\vec{n}.\mathbf{nat}(\vec{n}) \Rightarrow \mathbf{nat}(f(\vec{n})))$ is derivable in $\mathbf{FD}^c$ then t represents f.*

**Proof.** By Lemma 5.7 $\vdash t^\circ\colon \forall\vec{n}.\mathbf{nat}(\vec{n}) \Rightarrow \nabla\mathbf{nat}(f(\vec{n})))$ is derivable in **FD**. Then, using Friedman's top level trick [27, 55], we replace $o$ by $\mathbf{nat}(f(\vec{n}))$ in the derivation, we obtain that $\vdash \lambda\vec{x}.(t^\circ\ \vec{x}\ id)\colon \forall\vec{n}.\mathbf{nat}(\vec{n}) \Rightarrow \mathbf{nat}(f(\vec{n}))$ is also derivable in **FD**, and thus $\lambda\vec{x}.(t^\circ\ \vec{x}\ id)$ represents $f$. □

# 6 Non-local jumps

In this section we extend language **I** with control. Since control in imperative language are usually given in the form of several ad-hoc statements (such as exits from loops, exception handling, generators), there is no natural primitive statements. Consequently, we chose to retrofit operators **callcc** and **throw** to language **I**. We do not claim that these are natural control statement in an imperative language, but they are merely primitive constructs which can be used to encode other statements as derived forms. This main advantage of this approach is that we derive immediately a sound program logic for imperative programs with control.

## 6.1 Dependent imperative type system with control $\mathbf{ID}^c$

Similarly to the functional case, we extend type system **ID** with a propositional type constant $\perp$, we define $\neg\vec{\sigma}$ as an abbreviation for **proc** (**in** $\vec{\sigma}$; **out** $\perp$), and we add to **ID** two primitive procedures **callcc** and **throw** with the following types:

$$\begin{array}{rcl}\mathbf{callcc} & : & \mathbf{proc}\ (\mathbf{in\ proc}\ (\mathbf{in}\ \neg\vec{\sigma}; \mathbf{out}\ \vec{\sigma}); \mathbf{out}\ \vec{\sigma}) \\ \mathbf{throw} & : & \mathbf{proc}\ (\mathbf{in}\ \neg\vec{\sigma}, \vec{\sigma}; \mathbf{out}\ \vec{\tau})\end{array}$$

Note that the type of **callcc** is exactly $((\neg\vec{\sigma} \Rightarrow \vec{\sigma}) \Rightarrow \vec{\sigma})^\diamond$ and the type of **throw** is exactly $((\neg\vec{\sigma} \wedge \vec{\sigma}) \Rightarrow \vec{\tau})^\diamond$. If we assume that **callcc** and **throw** are mapped by $^\star$ to their functional counterpart, we have the following properties by construction:

**Proposition 6.1.** *For any environments $\Gamma$ and $\Omega$, any expression e, any sequence s, possibly containing procedures* **callcc** *and* **throw***, we have:*

- $\Gamma; \Omega \vdash e\colon \tau$ *in* $\mathbf{ID}^c$ *implies* $\Gamma^\star, \Omega^\star \vdash e^\star\colon \tau^\star$ *in* $\mathbf{FD}^c$.
- $\Gamma; \Omega \vdash s \rhd \vec{z}\colon \vec{\sigma}$ *in* $\mathbf{ID}^c$ *implies* $\Gamma^\star, \Omega^\star \vdash (s)^\star_{\vec{z}}\colon \vec{\sigma}^\star$ *in* $\mathbf{FD}^c$.

**Proposition 6.2.**

- *Given a term $t \in \mathcal{L}$ possibly containing* **callcc** *and* **throw** *such that $\Gamma \vdash t\colon \vec{\sigma}$ in $\mathbf{FD}^c$ and a fresh mutable variable tuple $\vec{r}$ of any type $\vec{\sigma}\,' \in \Sigma_{\mathbf{ID}}$ we have $\Gamma^\diamond; \vec{r}:\vec{\sigma}\,' \vdash t^\diamond_{\vec{r}} \rhd \vec{r}:\vec{\sigma}^\diamond$ in $\mathbf{ID}^c$.*
- *Given a value $v \in \mathcal{V}$ possibly containing* **callcc** *and* **throw** *such that $\Gamma \vdash v\colon \sigma$ in $\mathbf{FD}^c$ for any environment $\Omega$ we have $\Gamma^\diamond; \Omega \vdash v^\diamond\colon \sigma^\diamond$ in $\mathbf{ID}^c$.*

Since our semantics of $\mathbf{ID^c}$ is indirect, no representation theorem for $\mathbf{ID^c}$ can be claimed. However, we still have the following corollary:

**Corollary 6.3.** *Given an equational system $\mathcal{E}$ and an n-ary function symbol $f$, if $\vdash p\colon \mathbf{proc}(\{\vec{n}\}\mathbf{in}\ \mathbf{nat}(\vec{n});\ \mathbf{out}\ \mathbf{nat}(f(\vec{n})))$ is derivable in $\mathbf{ID^c}$ then $p^\star$ represents $f$.*

**Proof.** Since $\vdash p^\star\colon \forall \vec{n}.\mathbf{nat}(\vec{n}) \Rightarrow \mathbf{nat}(f(\vec{n})))$ is derivable in $\mathbf{FD}^c$ and by Theorem 5.8, $p^\star$ represents $f$. ☐

## 6.2 Syntax and typing extensions with control operators

In order to get closer to some usual syntax for jumps in imperative language, we introduce the following two abbreviations:

$$
\begin{aligned}
k\colon \{s\}_{\vec{z}} &= \mathbf{cst}\ \vec{z}\,' = \vec{z};\ \mathbf{callcc}(\mathbf{proc}(\mathbf{in}\ k; \mathbf{out}\ \vec{z})\{\vec{z} := \vec{z}\,';\ s\}_{\vec{z}}; \vec{z}) \\
\mathbf{jump}(k, \vec{e}\,)_{\vec{z}} &= \mathbf{throw}(k, \vec{e}\,; \vec{z}\,)
\end{aligned}
$$

The first abbreviation corresponds to the declaration of a (first-class) label. Recall that our type systems requires that the current mutable variables be explicitly passed inside the body of the procedure, hence the constants declaration. The second abbreviation is a "jump with parameters" to *the end* of the block annotated with the label given as argument. Note that the output variables are important only for typing purpose (since the **jump** never returns), they are thus written as a subscript.

**Proposition 6.4.** *The following typing rules are derivable in $\mathbf{ID}^c$.*

$$
\frac{\Gamma, k\colon \neg\vec{\sigma}; \vec{z}:\vec{\tau} \vdash s \rhd \vec{z}:\vec{\sigma} \qquad \Gamma; \Omega, \vec{z}:\vec{\sigma} \vdash s' \rhd \Omega'}{\Gamma; \Omega, \vec{z}:\vec{\tau} \vdash k\colon \{s\}_{\vec{z}};\ s' \rhd \Omega'}
$$

$$
\frac{\Gamma; \Omega, \vec{z}:\vec{\tau} \vdash k\colon \neg\vec{\sigma} \qquad \Gamma; \Omega, \vec{z}:\vec{\tau} \vdash \vec{e}:\vec{\sigma}}{\Gamma; \Omega, \vec{z}:\vec{\tau} \vdash \mathbf{jump}(k, \vec{e}\,)_{\vec{z}} \rhd \vec{z}:\vec{\tau}\,'}
$$

**Proof.** See Appendix C.7. ☐

## 6.3 Imperative delimited continuations

As a concluding example we show how to encode delimited continuation operators **shift** and **reset** [18] in $\mathbf{ID}^c$. This example is generic since it was shown by Filinski [23, 24] that any representable monad can be encoded using **shift** and **reset**. We also refer the reader to [75] for a detailed analysis of various type systems for **shift** and **reset** in the monadic framework, to [3] for a type-theoretic study of delimited continuations and to [4] for a generalization of Danvy and Filinski's type system to allow for polymorphic delimited continuations.

Our encoding follows [23] which contains the proof that **shift** and **reset** can themselves be implemented using **callcc**, **throw** and one global mutable variable storing the meta-continuation. The idea behind this encoding is best understood at the type level. First recall that the orignal semantics of these operators was given in terms of a double CPS-transform [18] (indeed, a single CPS transform is not enough to obtain a term whose semantics is independent of the evaluation strategy). The first transform corresponds to a parameterized continuation monad [6]:

$$
M(\alpha, \beta, \gamma) = (\gamma \to \beta) \to \alpha
$$

The second transform corresponds to the usual continuation monad, with a fixed *output type o* :

$$
\nabla\sigma = (\sigma \to o) \to o
$$

Composing both transforms [50] yields the following parameterized monad:

$$\begin{aligned}(\gamma \to \nabla\beta) \to \nabla\alpha &\cong ((\gamma \times (\beta \to o)) \to o) \to ((\alpha \to o) \to o)\\ &\cong (\alpha \to o) \to (((\gamma \times (\beta \to o)) \to o) \to o)\\ &= (\alpha \to o) \to \nabla(\gamma \times (\beta \to o))\end{aligned}$$

Up to simple type isomorphisms, we recognize the parameterized state monad transformer applied to the continuation monad. This monad correspond thus exactly to composing the state passing style transform (where the state is a continuation) with a CPS transform. This is the type isomorphism which exploited in [23] to encode **shift** et **reset** in direct style with a global state (always containing a continuation, called the meta-continuation) and **callcc**/**throw**.

Relying on higher-order mutable variables and the abbreviations for global variables from section 3.5, Filinski's implementation can thus be almost mechanically translated in $\mathbf{ID}^c$ (the type derivations are given in Appendix D.4):

$$\begin{array}{lcl}\mathbf{reset} & : & \mathbf{proc}(\mathbf{in}\ \mathbf{proc}(\mathbf{in}\ \neg\alpha; \mathbf{out}\ \beta, \neg\beta), \neg\gamma; \mathbf{out}\ \alpha, \neg\gamma)\\ \mathbf{reset} & = & \mathbf{proc}(\mathbf{in}\ p; \mathbf{out}\ r)_{mk}\{\\ & & \quad k\colon\{\\ & & \qquad \mathbf{cst}\ m = mk;\\ & & \qquad mk := \mathbf{proc}(\mathbf{in}\ r; \mathbf{out}\ z)\{\mathbf{jump}\ (k, r, m)_z;\}_z;\\ & & \qquad \mathbf{var}\ y;\ \ p(;y)_{mk};\\ & & \qquad \mathbf{jump}\ (mk, y)_{r,mk};\\ & & \quad \}_{r,mk};\\ & & \}_{r,mk};\end{array}$$

$$\begin{array}{lcl}\mathbf{shift} & : & \mathbf{proc}(\mathbf{in}\ \mathbf{proc}(\mathbf{in}\ \mathbf{proc}(\mathbf{in}\ \alpha, \neg\beta; \mathbf{out}\ \gamma, \neg\beta), \neg\delta; \mathbf{out}\ \epsilon, \neg\epsilon), \neg\delta; \mathbf{out}\ \alpha, \neg\gamma)\\ \mathbf{shift} & = & \mathbf{proc}(\mathbf{in}\ p; \mathbf{out}\ r)_{mk}\{\\ & & \quad k\colon\{\\ & & \qquad \mathbf{proc}\ q(\mathbf{in}\ v; \mathbf{out}\ r)_{mk}\{\\ & & \qquad\quad \mathbf{reset}\ (\mathbf{proc}(\mathbf{out}\ z)_{mk}\{\mathbf{jump}\ (k, v, mk)_{z,mk};\}_{z,mk}; r)_{mk};\\ & & \qquad \}_{r,mk};\\ & & \qquad \mathbf{var}\ y;\ \ p(q;y)_{mk};\\ & & \qquad \mathbf{jump}\ (mk, y)_{r,mk};\\ & & \quad \}_{r,mk};\\ & & \}_{r,mk}\end{array}$$

Of course, the image of those procedures by translation $\star$ yields functional terms typable in $\mathbf{FD}^c$. Those terms are given in Appendix E in Standard ML syntax [52]. The SML signature *CONT* is slightly different from [32] but they are equivalent (see [23] for an implementation of a similar signature in SML/NJ [1]). Their functional types are reproduced here:

$$\begin{array}{lcl}\mathit{reset} & : & (\neg\alpha \Rightarrow \beta \wedge \neg\beta) \wedge \neg\gamma \Rightarrow \alpha \wedge \neg\gamma\\ \mathit{shift} & : & ((\alpha \wedge \neg\beta \Rightarrow \gamma \wedge \neg\beta) \wedge \neg\delta \Rightarrow \varepsilon \wedge \neg\varepsilon) \wedge \neg\delta \Rightarrow \alpha \wedge \neg\gamma\end{array}$$

These types could be made a little more readable by using a parameterized state monad. However, we recognize the type of **shift** and **reset** from [18] where $(\alpha \wedge \neg\sigma) \Rightarrow (\beta \wedge \neg\tau)$ is written in the form $\alpha/\tau \to \beta/\sigma$. Our encoding thus provides a formulas-as-types interpretation of the full type system from [18] in a dependently-typed framework.

### 6.3.1 Example

In [75], Wadler presents several simple examples using **shift** and **reset**, and its third example, which requires the full type system from [18] to type check, is the following:

$$\begin{array}{l}\mathbf{let}\ g = (\mathbf{reset}\ (\mathbf{if}\ (\mathbf{shift}\ \lambda f.f)\ \mathbf{then}\ 2\ \mathbf{else}\ 3))\\ \mathbf{in}\ (g\ \mathit{True}) + (g\ \mathit{False})\end{array}$$

Walder explains informally the semantics of his example as follows: "*Here f (and hence g) is bound to the function that returns 2 if passed True, and 3 if passed False, hence the value of the given term is 5.*"

Now the question is "how to prove formally the correctness of this program?". The solution we propose consists in first translating the expression into an imperative program (with **shift**/**reset** defined as above) and then proving its correctness by deriving the expected specification in $\mathbf{ID}^c$. We thus obtain following imperative program (where the conditional is simulated by a for-loop):

$$
\begin{array}{l}
\mathbf{cst}\ q\ =\ \mathbf{proc}(;\ \mathbf{out}\ r)_{mk}\ \{ \\
\quad \mathbf{cst}\ p\ =\ \mathbf{proc}(\mathbf{in}\ f;\ \mathbf{out}\ h)_{mk}\ \{\ h := f; \} \\
\quad \mathbf{var}\ b;\ \mathit{shift}(p;\ b)_{mk}; \\
\quad r := 3; \\
\quad \mathbf{for}\ i := 0\ \mathbf{until}\ b\ \{ \\
\quad\quad r := 2; \\
\quad \}_{r,mk}; \\
\}; \\
\mathbf{var}\ g;\ \mathit{reset}(q;\ g)_{mk}; \\
\mathbf{var}\ x;\ g(0;\ x)_{mk}; \\
\mathbf{var}\ y;\ g(1;\ y)_{mk}; \\
\mathit{add}(x,\ y;\ z)_{mk};
\end{array}
$$

It is then possible to show that $z\colon \top \vdash s \rhd z\colon \mathbf{nat}(f_{32}(0) + f_{32}(1))$ is derivable in $\mathbf{ID}^c$ where $f_{32}$ is defined by the equations:

$$
\begin{array}{l}
f_{32}(0) = 3 \\
f_{32}(S(i)) = 2
\end{array}
$$

We shall not detail the type derivation since it is rather technical. However, we have formally specified $\mathbf{ID}^c$, $\mathbf{FD}^c$, the translation $^\star$ and our encoding of **shift** and **reset** in Twelf [64]. Moreover, thanks to Twelf logic programming engine, those specifications are executable and we have mechanically checked the correctness of the above example (together with a few others from [75]). The interested reader is referred to [14] for more details.

# Appendix A Properties of I and F

## A.1 Basic properties of I

**Definition.** *The sets $\mathcal{FI}(s)$, $\mathcal{FI}(c)$ and $\mathcal{FI}(e)$ of free identifiers (including both variable and constant identifiers) of a sequence, a command and an expression are defined by mutual induction as follows:*

- $\mathcal{FI}(y) = \{y\}$
- $\mathcal{FI}(*) = \mathcal{FI}(\bar{q}) = \emptyset$
- $\mathcal{FI}(e_1, ..., e_n) = \mathcal{FI}(e_1) \cup ... \cup \mathcal{FI}(e_n)$
- $\mathcal{FI}(\mathbf{proc}\ (\mathbf{in}\ \vec{y}\,; \mathbf{out}\ \vec{z}\,)\ \{s\}_{\vec{z}}) = \mathcal{FI}(s) \setminus (\vec{y} \cup \vec{z}\,)$

- $\mathcal{FI}(\mathbf{inc}(y)) = \mathcal{FI}(\mathbf{dec}(y)) = \{y\}$
- $\mathcal{FI}(\{s\}_{\vec{x}}) = \mathcal{FI}(s) \cup \vec{x}$
- $\mathcal{FI}(\vec{y} := e) = \{\vec{y}\,\} \cup \mathcal{FI}(e)$
- $\mathcal{FI}(p(\vec{e}\,; \vec{y}\,)) = \vec{y} \cup \mathcal{FI}(\vec{e}\,) \cup \mathcal{FI}(p)$
- $\mathcal{FI}(\mathbf{for}\ y := 0\ \mathbf{until}\ e\ \{s\}_{\vec{x}}) = \mathcal{FI}(\vec{e}\,) \cup (\mathcal{FI}(s) \setminus \{y\}) \cup \vec{x}$

- $\mathcal{FI}(\varepsilon) = \emptyset$
- $\mathcal{FI}(c\,;\ s) = \mathcal{FI}(c) \cup \mathcal{FI}(s)$
- $\mathcal{FI}(\mathbf{cst}\ y = e;\ s) = \mathcal{FI}(\mathbf{var}\ y := e;\ s) = \mathcal{FI}(\vec{e}\,) \cup (\mathcal{FI}(s) \setminus \{y\})$

## A.2 Translation from F to I and retraction

**Definition A.1.** *We define the translation of any term $t$ of* **F** *into a term $t^\natural$ of $\mathcal{L}$ by the following equations:*

$$\begin{array}{rcl}
x^\natural &=& x \\
()^\natural &=& () \\
S^n(0)^\natural &=& S^n(0) \\
S^n(t)^\natural &=& \mathbf{let}\ x = t^\natural\ \mathbf{in}\ \mathbf{let}\ x = \mathbf{succ}(x)\ \mathbf{in}\ ...\ \mathbf{let}\ x = \mathbf{succ}(x)\ \mathbf{in}\ x \\
(\lambda x.t)^\natural &=& \lambda x.t^\natural \\
\\
\mathbf{pred}(t)^\natural &=& \mathbf{let}\ x = t^\natural\ \mathbf{in}\ \mathbf{let}\ x = \mathbf{pred}(x)\ \mathbf{in}\ x \\
\mathbf{rec}(t_1, t_2, t_3)^\natural &=& \mathbf{let}\ a = t_1^\natural\ \mathbf{in}\ \mathbf{let}\ b = t_2^\natural\ \mathbf{in}\ \mathbf{let}\ c = t_3^\natural\ \mathbf{in} \\
&& \mathbf{let}\ z = \mathbf{rec}(a, b, \lambda x.\lambda y.\mathbf{let}\ d = c\ x\ \mathbf{in}\ \mathbf{let}\ e = d\ y\ \mathbf{in}\ e)\ \mathbf{in}\ z \\
(t\ u)^\natural &=& \mathbf{let}\ x = t^\natural\ \mathbf{in}\ \mathbf{let}\ y = u^\natural\ \mathbf{in}\ \mathbf{let}\ r = x\ y\ \mathbf{in}\ r \\
(\mathbf{let}\ \vec{x} = u\ \mathbf{in}\ t)^\natural &=& \mathbf{let}\ y = u^\natural\ \mathbf{in}\ \mathbf{let}\ \vec{x} = y\ \mathbf{in}\ t^\natural \\
(t_1, ..., t_n)^\natural &=& \mathbf{let}\ x_1 = t_1^\natural\ \mathbf{in}\ ...\ \mathbf{let}\ x_n = t_n^\natural\ \mathbf{in}\ (x_1, ..., x_n)
\end{array}$$

**Proposition A.2.** *For any term $t$ of* **F**, *we have $t^\natural \in \mathcal{L}$.*

**Proof.** Straightforward induction on $t$. □

**Lemma A.3.** *Given a term $t \in \mathcal{L}$ and a fresh mutable variable tuple $\vec{r}$ we have $\vec{r} \notin \mathcal{FV}(((t)^\diamond_{\vec{r}})^\star_{\vec{r}})$.*

**Proof.** By induction on $t$.

- $((\vec{w}\,)^\diamond_{r})^\star_{r}$
  $= (\vec{r} := \vec{w}\,;)^\star_{r}$

$= \textbf{let } r_1 = (w_1^\diamond)^\star \textbf{ in } \dots \textbf{ let } r_n = (w_n^\diamond)^\star \textbf{ in } \vec{r}$

We easily conclude since $\vec{r}$ does not occur in $(\vec{w}^\diamond)^\star$.

- $((\textbf{let } y = w \textbf{ in } u)_{\vec{r}}^\diamond)_{\vec{r}}^\star$
  $= (\textbf{cst } y = w^\diamond;\ (u)_{\vec{r}}^\diamond)_{\vec{r}}^\star$
  $= \textbf{let } y = (w^\diamond)^\star \textbf{ in } ((u)_{\vec{r}}^\diamond)_{\vec{r}}^\star$
  By induction hypothesis, $\vec{r} \notin \mathcal{FV}(((u)_{\vec{r}}^\diamond)_{\vec{r}}^\star)$, and $\vec{r}$ does not occur in $(w^\diamond)^\star$.
- $((\textbf{let } y = \textbf{succ}(w) \textbf{ in } u)_{\vec{r}}^\diamond)_{\vec{r}}^\star$
  $= (\textbf{var } z := w^\diamond;\ \textbf{inc}(z);\ \textbf{cst } y = z;\ (u)_{\vec{r}}^\diamond)_{\vec{r}}^\star$
  $= (\textbf{let } z = \textbf{succ}(z) \textbf{ in let } y = z \textbf{ in } ((u)_{\vec{r}}^\diamond)_{\vec{r}}^\star)[(w^\diamond)^\star/z]$
  $= (\textbf{let } z = \textbf{succ}((w^\diamond)^\star) \textbf{ in let } y = z \textbf{ in } ((u)_{\vec{r}}^\diamond)_{\vec{r}}^\star)$
  By induction hypothesis, $\vec{r} \notin \mathcal{FV}(((u)_{\vec{r}}^\diamond)_{\vec{r}}^\star)$, and $\vec{r}$ does not occur in $(w^\diamond)^\star$.
- The case of **pred** is similar to **succ**.
- $((\textbf{let } \vec{x} = \textbf{rec}(w, \vec{w}, \lambda i.\lambda \vec{y}.t) \textbf{ in } u)_{\vec{r}}^\diamond)_{\vec{r}}^\star$
  $= (\textbf{var } \vec{z} := \vec{w};\ \textbf{for } i := 0 \textbf{ until } w^\diamond\ \{\textbf{cst } \vec{y} = \vec{z};\ (t)_{\vec{z}}^\diamond\}_{\vec{z}};\ \textbf{cst } \vec{x} = \vec{z};\ (u)_{\vec{r}}^\diamond)_{\vec{r}}^\star$
  $= \textbf{let } \vec{z} = \textbf{rec}((w^\diamond)^\star, \vec{z}, \lambda i.\lambda \vec{z}.\textbf{let } \overline{y = z} \textbf{ in } ((t)_{\vec{z}}^\diamond)_{\vec{z}}^\star) \textbf{ in let } \overline{x = z} \textbf{ in } ((u)_{\vec{r}}^\diamond)_{\vec{r}}^\star)[(\vec{w}^\diamond)^\star/\vec{z}]$
  $= \textbf{let } \vec{z} = \textbf{rec}((w^\diamond)^\star, (\vec{w}^\diamond)^\star, \lambda i.\lambda \vec{z}.\textbf{let } \overline{y = z} \textbf{ in } ((t)_{\vec{z}}^\diamond)_{\vec{z}}^\star) \textbf{ in let } \overline{x = z} \textbf{ in } ((u)_{\vec{r}}^\diamond)_{\vec{r}}^\star)$
  By induction hypothesis, $\vec{r} \notin \mathcal{FV}(((u)_{\vec{r}}^\diamond)_{\vec{r}}^\star)$, and $\vec{r}$ does not occur in $(w^\diamond)^\star$, $(\vec{w}^\diamond)^\star$ and $((t)_{\vec{z}}^\diamond)_{\vec{z}}^\star)$.
- $((\textbf{let } \vec{x} = w\ \vec{w} \textbf{ in } u)_{\vec{r}}^\diamond)_{\vec{r}}^\star$
  $= (\textbf{var } \vec{z};\ w^\diamond(\vec{w}^\diamond; \vec{z});\ \textbf{cst } \vec{x} = \vec{z};\ (u)_{\vec{r}}^\diamond)_{\vec{r}}^\star$
  $= (\textbf{let } \vec{z} = (w^\diamond)^\star\ (\vec{w}^\diamond)^\star \textbf{ in let } x_1 = z_1 \textbf{ in } \dots \textbf{ let } x_n = z_n \textbf{ in } ((u)_{\vec{r}}^\diamond)_{\vec{r}}^\star)[\vec{()}/\vec{z}]$
  $= \textbf{let } \vec{z} = (w^\diamond)^\star\ (\vec{w}^\diamond)^\star \textbf{ in let } x_1 = z_1 \textbf{ in } \dots \textbf{ let } x_n = z_n \textbf{ in } ((u)_{\vec{r}}^\diamond)_{\vec{r}}^\star$
  By induction hypothesis, $\vec{r} \notin \mathcal{FV}(((u)_{\vec{r}}^\diamond)_{\vec{r}}^\star)$, and $\vec{r}$ does not occur in $(w^\diamond)^\star$ and $(\vec{w}^\diamond)^\star$.
- $((\textbf{let } \vec{x} = t \textbf{ in } u)_{\vec{r}}^\diamond)_{\vec{r}}^\star$
  $= (\textbf{var } \vec{z};\ \{(t)_{\vec{z}}^\diamond\}_{\vec{z}};\ \textbf{cst } \vec{x} = \vec{z};\ (u)_{\vec{r}}^\diamond))_{\vec{r}}^\star$
  $= (\textbf{let } \vec{z} = ((t)_{\vec{z}}^\diamond)_{\vec{z}}^\star \textbf{ in let } x_1 = z_1 \textbf{ in } \dots \textbf{ let } x_n = z_n \textbf{ in } ((u)_{\vec{r}}^\diamond)_{\vec{r}}^\star)[\vec{()}/\vec{z}]$
  $= \textbf{let } \vec{z} = ((t)_{\vec{z}}^\diamond)_{\vec{z}}^\star \textbf{ in let } x_1 = z_1 \textbf{ in } \dots \textbf{ let } x_n = z_n \textbf{ in } ((u)_{\vec{r}}^\diamond)_{\vec{r}}^\star$
  By induction hypothesis, $\vec{r} \notin \mathcal{FV}(((u)_{\vec{r}}^\diamond)_{\vec{r}}^\star)$, and $\vec{r}$ does not occur in $((t)_{\vec{z}}^\diamond)_{\vec{z}}^\star)$.

□

**Definition A.4.** *We define the reduction relation $\twoheadrightarrow$ as the reflexive, transitive and contextual closure of the reduction $\rightsquigarrow$ for arbitrary contexts.*

**Proposition.** *We prove the following properties, which clearly implies $((t)_{\vec{r}}^\diamond)_{\vec{r}}^\star \approx t$ and if $w = S^n(0)$ or $w = *$ then $w^{\star\diamond} = w$ else $w^{\star\diamond} \approx w$.*

- *Given a term $t \in \mathcal{L}$ and a fresh mutable variable $r$ we have $((t)_{\vec{r}}^\diamond)_{\vec{r}}^\star \twoheadrightarrow t$.*
- *Given a value $v \in \mathcal{W}$, if $w = S^n(0)$ or $w = *$ then $w^{\star\diamond} = w$ else $w^{\star\diamond} \twoheadrightarrow w$.*

**Proof.** By mutual induction.

- $(S^n(0)^\diamond)^\star = \bar{n}^\star = S^n(0)$.
- $(y^\diamond)^\star = y^\star = y$.
- $(()^\diamond)^\star = *^\star = ()$.
- $((\lambda \vec{x}.t)^\diamond)^\star$
  $= (\textbf{proc}(\textbf{in } \vec{x}; \textbf{out } \vec{z})\ \{(t)_{\vec{z}}^\diamond\}_{\vec{z}})^\star$
  $= \lambda \vec{x}.((t)_{\vec{z}}^\diamond)_{\vec{z}}^\star[()/\vec{z}]$
  $= \lambda \vec{x}.((t)_{\vec{z}}^\diamond)_{\vec{z}}^\star$ since $\vec{z} \notin \mathcal{FV}(((t)_{\vec{z}}^\diamond)_{\vec{z}}^\star)$ by Lemma A.3
  $\twoheadrightarrow \lambda \vec{x}.t$ by induction hypothesis.
- $((\vec{w})_{\vec{r}}^\diamond)_{\vec{r}}^\star$
  $= (\vec{r} := \vec{w}^\diamond;)_{\vec{r}}^\star$
  $= \textbf{let } r_1 = (w_1^\diamond)^\star \textbf{ in } \dots \textbf{ let } r_n = (w_n^\diamond)^\star \textbf{ in } \vec{r}$
  $\rightsquigarrow^n (\vec{w}^\diamond)^\star$
  $\twoheadrightarrow \vec{w}$ by induction hypothesis.

- $((\mathbf{let}\ y = w\ \mathbf{in}\ u)^{\diamond}_{\vec{r}})^{\star}_{\vec{r}}$
  $= (\mathbf{cst}\ y = w^{\diamond};\ (u)^{\diamond}_{\vec{r}})^{\star}_{\vec{r}}$
  $= \mathbf{let}\ y = (w^{\diamond})^{\star}\ \mathbf{in}\ ((u)^{\diamond}_{\vec{r}})^{\star}_{\vec{r}}$
  $\twoheadrightarrow \mathbf{let}\ y = w\ \mathbf{in}\ u$ by induction hypothesis.

- $((\mathbf{let}\ y = \mathbf{succ}(w)\ \mathbf{in}\ u)^{\diamond}_{\vec{r}})^{\star}_{\vec{r}}$
  $= (\mathbf{var}\ z := w^{\diamond};\ \mathbf{inc}(z);\ \mathbf{cst}\ y = z;\ (u)^{\diamond}_{\vec{r}})^{\star}_{\vec{r}}$
  $= (\mathbf{let}\ z = \mathbf{succ}(z)\ \mathbf{in}\ \mathbf{let}\ y = z\ \mathbf{in}\ ((u)^{\diamond}_{\vec{r}})^{\star}_{\vec{r}})[(w^{\diamond})^{\star}/z]$
  $= (\mathbf{let}\ z = \mathbf{succ}((w^{\diamond})^{\star})\ \mathbf{in}\ \mathbf{let}\ y = z\ \mathbf{in}\ ((u)^{\diamond}_{\vec{r}})^{\star}_{\vec{r}})$
  $\rightsquigarrow (\mathbf{let}\ y = \mathbf{succ}((w^{\diamond})^{\star})\ \mathbf{in}\ ((u)^{\diamond}_{\vec{r}})^{\star}_{\vec{r}})$ since $z \notin \mathcal{FV}((u)^{\diamond}_{\vec{r}})^{\star}_{\vec{r}})$
  $\twoheadrightarrow \mathbf{let}\ y = \mathbf{succ}(w)\ \mathbf{in}\ u$ by induction hypothesis.

- The case of **pred** is similar to **succ**.

- $((\mathbf{let}\ \vec{x} = \mathbf{rec}(w, \vec{w}, \lambda i.\lambda \vec{y}.t)\ \mathbf{in}\ u)^{\diamond}_{\vec{r}})^{\star}_{\vec{r}}$
  $= (\mathbf{var}\ \vec{z} := \vec{w};\ \mathbf{for}\ i := 0\ \mathbf{until}\ w^{\diamond}\ \{\mathbf{cst}\ \vec{y} = \vec{z};\ (t)^{\diamond}_{\vec{z}}\}_{\vec{z}};\ \mathbf{cst}\ \vec{x} = \vec{z};\ (u)^{\diamond}_{\vec{r}})^{\star}_{\vec{r}}$
  $= \mathbf{let}\ \vec{z} = \mathbf{rec}((w^{\diamond})^{\star}, \vec{z}, \lambda i.\lambda \vec{z}.\mathbf{let}\ \overrightarrow{y = z}\ \mathbf{in}\ ((t)^{\diamond}_{\vec{z}})^{\star}_{\vec{z}})\ \mathbf{in}\ \mathbf{let}\ \overrightarrow{x = z}\ \mathbf{in}\ ((u)^{\diamond}_{\vec{r}})^{\star}_{\vec{r}})[(\vec{w}^{\diamond})^{\star}/\vec{z}]$
  $= \mathbf{let}\ \vec{z} = \mathbf{rec}((w^{\diamond})^{\star}, (\vec{w}^{\diamond})^{\star}, \lambda i.\lambda \vec{z}.\mathbf{let}\ \overrightarrow{y = z}\ \mathbf{in}\ ((t)^{\diamond}_{\vec{z}})^{\star}_{\vec{z}})\ \mathbf{in}\ \mathbf{let}\ \overrightarrow{x = z}\ \mathbf{in}\ ((u)^{\diamond}_{\vec{r}})^{\star}_{\vec{r}})$
  $\twoheadrightarrow \mathbf{let}\ \vec{z} = \mathbf{rec}((w^{\diamond})^{\star}, (\vec{w}^{\diamond})^{\star}, \lambda i.\lambda \vec{z}.((t)^{\diamond}_{\vec{z}})^{\star}_{\vec{z}}[\vec{z}/\vec{y}])\ \mathbf{in}\ ((u)^{\diamond}_{\vec{r}})^{\star}_{\vec{r}})[\vec{z}/\vec{x}]$
  $\twoheadrightarrow \mathbf{let}\ \vec{z} = \mathbf{rec}(w, \vec{w}, \lambda i.\lambda \vec{z}.t[\vec{z}/\vec{y}])\ \mathbf{in}\ u[\vec{z}/\vec{x}]$ by induction hypothesis
  $= \mathbf{let}\ \vec{x} = \mathbf{rec}(w, \vec{w}, \lambda i.\lambda \vec{y}.t)\ \mathbf{in}\ u$ modulo $\alpha$-conversion.

- $((\mathbf{let}\ \vec{x} = w\ \vec{w}\ \mathbf{in}\ u)^{\diamond}_{\vec{r}})^{\star}_{\vec{r}}$
  $= (\mathbf{var}\ \vec{z};\ w^{\diamond}(\vec{w}^{\diamond}; \vec{z});\ \mathbf{cst}\ \vec{x} = \vec{z};\ (u)^{\diamond}_{\vec{r}})^{\star}_{\vec{r}}$
  $= (\mathbf{let}\ \vec{z} = (w^{\diamond})^{\star}\ (\vec{w}^{\diamond})^{\star}\ \mathbf{in}\ \mathbf{let}\ x_1 = z_1\ \mathbf{in}\ \dots\ \mathbf{let}\ x_n = z_n\ \mathbf{in}\ ((u)^{\diamond}_{\vec{r}})^{\star}_{\vec{r}})[\vec{()}/\vec{z}]$
  $= \mathbf{let}\ \vec{z} = (w^{\diamond})^{\star}\ (\vec{w}^{\diamond})^{\star}\ \mathbf{in}\ \mathbf{let}\ x_1 = z_1\ \mathbf{in}\ \dots\ \mathbf{let}\ x_n = z_n\ \mathbf{in}\ ((u)^{\diamond}_{\vec{r}})^{\star}_{\vec{r}}$
  $\twoheadrightarrow \mathbf{let}\ \vec{z} = (w^{\diamond})^{\star}\ (\vec{w}^{\diamond})^{\star}\ \mathbf{in}\ ((u)^{\diamond}_{\vec{r}})^{\star}_{\vec{r}}[\vec{z}/\vec{x}]$
  $\twoheadrightarrow \mathbf{let}\ \vec{z} = w\ \vec{w}\ \mathbf{in}\ u[\vec{z}/\vec{x}]$ by induction hypothesis
  $= \mathbf{let}\ \vec{x} = w\ \vec{w}\ \mathbf{in}\ u$ modulo $\alpha$-conversion.

- $((\mathbf{let}\ \vec{x} = t\ \mathbf{in}\ u)^{\diamond}_{\vec{r}})^{\star}_{\vec{r}}$
  $= (\mathbf{var}\ \vec{z};\ \{(t)^{\diamond}_{\vec{z}}\}_{\vec{z}};\ \mathbf{cst}\ \vec{x} = \vec{z};\ (u)^{\diamond}_{\vec{r}}))^{\star}_{\vec{r}}$
  $= (\mathbf{let}\ \vec{z} = ((t)^{\diamond}_{\vec{z}})^{\star}_{\vec{z}}\ \mathbf{in}\ \mathbf{let}\ x_1 = z_1\ \mathbf{in}\ \dots\ \mathbf{let}\ x_n = z_n\ \mathbf{in}\ ((u)^{\diamond}_{\vec{r}})^{\star}_{\vec{r}})[\vec{()}/\vec{z}]$
  $= (\mathbf{let}\ \vec{z} = ((t)^{\diamond}_{\vec{z}})^{\star}_{\vec{z}}\ \mathbf{in}\ \mathbf{let}\ x_1 = z_1\ \mathbf{in}\ \dots\ \mathbf{let}\ x_n = z_n\ \mathbf{in}\ ((u)^{\diamond}_{\vec{r}})^{\star}_{\vec{r}})$
  $\twoheadrightarrow \mathbf{let}\ \vec{z} = ((t)^{\diamond}_{\vec{z}})^{\star}_{\vec{z}}\ \mathbf{in}\ ((u)^{\diamond}_{\vec{r}})^{\star}_{\vec{r}}[\vec{z}/\vec{x}]$
  $\twoheadrightarrow \mathbf{let}\ \vec{z} = t\ \mathbf{in}\ u[\vec{z}/\vec{x}]$ by induction hypothesis
  $= \mathbf{let}\ \vec{x} = t\ \mathbf{in}\ u$ modulo $\alpha$-conversion.

□

$$\frac{x\colon\tau \in \Gamma;\Omega}{\Gamma;\Omega \vdash x\colon\tau} \quad \text{(T.ENV)}$$

$$\frac{}{\Gamma;\Omega \vdash \bar{q}\colon \mathbf{nat}} \quad \text{(T.NUM)}$$

$$\frac{}{\Gamma;\Omega \vdash *\colon \mathbf{unit}} \quad \text{(T.UNIT)}$$

$$\frac{\Gamma;\Omega \vdash \vec{e}\colon\vec{\tau}}{\Gamma;\Omega \vdash (\vec{e})\colon(\vec{\tau})} \quad \text{(T.TUPLE)}$$

$$\frac{\vec{z} \neq \emptyset \qquad \Gamma, \vec{y}\colon\vec{\sigma}; \vec{z}\colon\overrightarrow{\mathbf{unit}} \vdash s \rhd \vec{z}\colon\vec{\tau}}{\Gamma;\Omega \vdash \mathbf{proc}\ (\mathbf{in}\ \vec{y}; \mathbf{out}\ \vec{z})\{s\}_{\vec{z}}\colon \mathbf{proc}\ (\mathbf{in}\ \vec{\sigma}; \mathbf{out}\ \vec{\tau})} \quad \text{(T.PROC)}$$

$$\frac{}{\Gamma;\Omega,\Omega' \vdash \varepsilon \rhd \Omega'} \quad \text{(T.EMPTY)}$$

$$\frac{\Gamma;\Omega \vdash e\colon\tau \qquad \Gamma, y\colon\tau;\Omega \vdash s \rhd \Omega'}{\Gamma;\Omega \vdash \mathbf{cst}\ y = e;\ s \rhd \Omega'} \quad \text{(T.CST)}$$

$$\frac{\Gamma;\Omega \vdash e\colon\tau \qquad \Gamma;\Omega, y\colon\tau \vdash s \rhd \Omega' \qquad y \notin \Omega'}{\Gamma;\Omega \vdash \mathbf{var}\ y := e;\ s \rhd \Omega'} \quad \text{(T.VAR)}$$

$$\frac{\Gamma;\vec{x}\colon\vec{\sigma} \vdash s \rhd \vec{x}\colon\vec{\tau} \qquad \Gamma;\Omega,\vec{x}\colon\vec{\tau} \vdash s' \rhd \Omega'}{\Gamma;\Omega,\vec{x}\colon\vec{\sigma} \vdash \{s\}_{\vec{x}};\ s' \rhd \Omega'} \quad \text{(T.BLOCK)}$$

$$\frac{\Gamma;\Omega, y\colon\mathbf{nat} \vdash s \rhd \Omega'}{\Gamma;\Omega, y\colon\mathbf{nat} \vdash \mathbf{inc}(y);\ s \rhd \Omega'} \quad \text{(T.INC)}$$

$$\frac{\Gamma;\Omega, y\colon\mathbf{nat} \vdash s \rhd \Omega'}{\Gamma;\Omega, y\colon\mathbf{nat} \vdash \mathbf{dec}(y);\ s \rhd \Omega'} \quad \text{(T.DEC)}$$

$$\frac{\Gamma;\Omega,\vec{y}\colon\vec{\sigma} \vdash e\colon(\vec{\tau}) \qquad \Gamma;\Omega,\vec{y}\colon\vec{\tau} \vdash s \rhd \Omega'}{\Gamma;\Omega,\vec{y}\colon\vec{\sigma} \vdash \vec{y} := e;\ s \rhd \Omega'} \quad \text{(T.ASSIGN)}$$

$$\frac{\Gamma;\Omega,\vec{x}\colon\vec{\sigma} \vdash e\colon\mathbf{nat} \quad \Gamma, y\colon\mathbf{nat};\vec{x}\colon\vec{\sigma} \vdash s \rhd \vec{x}\colon\vec{\sigma} \quad \Gamma;\Omega,\vec{x}\colon\vec{\sigma} \vdash s' \rhd \Omega'}{\Gamma;\Omega,\vec{x}\colon\vec{\sigma} \vdash \mathbf{for}\ y := 0\ \mathbf{until}\ e\ \{s\}_{\vec{x}};\ s' \rhd \Omega'} \quad \text{(T.FOR)}$$

$$\frac{\Gamma;\Omega,\vec{r}\colon\vec{\omega} \vdash p\colon \mathbf{proc}\ (\mathbf{in}\ \vec{\tau}; \mathbf{out}\ \vec{\sigma}) \quad \Gamma;\Omega,\vec{r}\colon\vec{\omega} \vdash \vec{e}\colon\vec{\tau} \quad \Gamma;\Omega,\vec{r}\colon\vec{\sigma} \vdash s \rhd \Omega'}{\Gamma;\Omega,\vec{r}\colon\vec{\omega} \vdash p(\vec{e};\vec{r});\ s \rhd \Omega'} \quad \text{(T.CALL)}$$

**Figure B.1.** Alternative imperative pseudo-dynamic type system

# Appendix B Properties of IS and FS

## B.1 Alternative pseudo-dynamic type system

We first present in Figure B.1 a different (but equivalent) formulation of the pseudo-dynamic type system which is easier to deal with when proving properties by induction on sequences.

## B.2 Preliminary properties

**Lemma B.1.** *If* $\Gamma, x\colon\tau;\Omega \vdash s \rhd \Omega'$ *and* $\emptyset;\emptyset \vdash e\colon\tau$ *in* **IS** *then* $\Gamma;\Omega \vdash s[x \leftarrow e] \rhd \Omega'$ *in* **IS**.

**Proof.** Straightforward induction on $s$. ☐

**Lemma B.2.** *If* $\Gamma; x\colon\tau,\Omega \vdash s \rhd x\colon\sigma,\Omega'$ *in* **IS** *then* $\Gamma; y\colon\tau,\Omega \vdash s[x \hookleftarrow y] \rhd y\colon\sigma,\Omega'$ *in* **IS**.

**Proof.** Straightforward induction on $s$. ☐

**Lemma B.3.** *If* $\Gamma;\Omega\vdash s\rhd\Omega'$ *in* **IS** *then for any* $x\!:\!\sigma$, $\Gamma,x\!:\!\sigma;\Omega\vdash s\rhd\Omega'$ *and* $\Gamma;\Omega,x\!:\!\sigma\vdash s\rhd\Omega'$ *in* **IS**.

**Proof.** Straightforward by induction on the typing derivation. ☐

**Lemma B.4.** *If* $\mu\rhd\Omega$ *and* $\emptyset;\Omega\vdash e\!:\tau$ *in* **IS** *and* $e=_\mu w$, *then we have* $\emptyset;\emptyset\vdash w\!:\tau$ *in* **IS**.

**Proof.** The case $e=w$ is trivial and if $e$ is some variable $x\in\Omega$ then by definition of $\mu\rhd\Omega$, we have $\emptyset;\emptyset\vdash\mu(x)=w\!:\tau$. ☐

## B.3 Reduction preserves typing

**Theorem.** *For any state* $(s,\mu)$, $\Omega$ *and* $\vec{z}$, *if* $\vec{z}\!:\!\vec{\tau}\vdash(s,\mu)\rhd\Omega$ *in* **IS** *and* $(s,\mu)\mapsto(s',\mu')$ *then there exists* $\vec{\tau}\,'$ *such that* $\vec{z}\!:\!\vec{\tau}\,'\vdash(s',\mu')\rhd\Omega$ *in* **IS**.

**Proof.** By induction on the derivation of $(s,\mu)\mapsto(s',\mu')$, and then by analysis of the typing derivation.

- (S.BLOCK-I): we have $\mu\rhd\Delta,\vec{x}\!:\!\vec{\tau}$ and

  $$\frac{\overline{\emptyset;\vec{x}\!:\!\vec{\tau}\vdash\varepsilon\rhd\vec{x}\!:\!\vec{\tau}}\qquad\emptyset;\Delta,\vec{x}\!:\!\vec{\tau}\vdash s\ \rhd\Delta'}{\emptyset;\Delta,\vec{x}\!:\!\vec{\tau}\vdash\{\}_{\vec{x}};\ s\ \rhd\Delta'}$$

  then we get $\emptyset;\Delta,\vec{x}\!:\!\vec{\tau}\vdash s\ \rhd\Delta'$ hence $\mu\rhd\Delta,\vec{x}\!:\!\vec{\tau}$ and $\Delta,\vec{x}\!:\!\vec{\tau}\vdash(s,\mu)\ \rhd\Delta'$.

- (S.BLOCK-II): we have $\mu\rhd\Delta,\vec{x}\!:\!\vec{\sigma}$

  $$\frac{\emptyset;\vec{x}\!:\!\vec{\sigma}\vdash s_1\rhd\vec{x}\!:\!\vec{\tau}\qquad\emptyset;\Delta,\vec{x}\!:\!\vec{\tau}\vdash s_2\ \rhd\Delta'}{\emptyset;\Delta,\vec{x}\!:\!\vec{\sigma}\vdash\{s_1\}_{\vec{x}};\ s_2\ \rhd\Delta'}$$

  By induction hypothesis on $\vec{x}\!:\!\vec{\sigma}\vdash(s_1,\mu)\rhd\vec{x}\!:\!\vec{\tau}$, we obtain $\vec{x}\!:\!\vec{\sigma}\,'\vdash(s_1',\mu')\rhd\vec{x}\!:\!\vec{\tau}$ which gives us $\emptyset;\vec{x}\!:\!\vec{\sigma}\,'\vdash s_1'\rhd\vec{x}\!:\!\vec{\tau}$ and $\mu'\rhd\Delta,\vec{x}\!:\!\vec{\sigma}\,'$. We can build the following typing derivation to conclude:

  $$\frac{\emptyset;\vec{x}\!:\!\vec{\sigma}\,'\vdash s_1'\rhd\vec{x}\!:\!\vec{\tau}\qquad\emptyset;\Delta,\vec{x}\!:\!\vec{\tau}\vdash s\ \rhd\Delta'}{\emptyset;\Delta,\vec{x}\!:\!\vec{\sigma}\,'\vdash\{s_1'\}_{\vec{x}};\ s_2\ \rhd\Delta'}$$

- (S.VAR-I): we have $\mu\rhd\Omega$

  $$\frac{\emptyset;\Omega\vdash e\!:\tau\qquad\overline{\emptyset;\Omega,y\!:\tau\vdash\varepsilon\rhd\Omega}}{\emptyset;\Omega\vdash\textbf{var}\ y:=e;\ \varepsilon\rhd\Omega}$$

  then we get $\emptyset;\Omega\vdash\varepsilon\rhd\Omega$.

- (S.VAR-II): we have $\mu\rhd\Delta$ and

  $$\frac{\emptyset;\Delta\vdash e\!:\tau\qquad\emptyset;\Delta,y\!:\tau\vdash s\rhd\Omega\qquad y\notin\Omega}{\emptyset;\Delta\vdash\textbf{var}\ y:=e;\ s\rhd\Omega}$$

  By Lemma B.4, $\mu\rhd\Delta$ and $\emptyset;\Delta\vdash e\!:\tau$ and $e=_\mu w$ implies $\emptyset;\emptyset\vdash w\!:\tau$. By definition of store typing, $(\mu, y\leftarrow w)\rhd\Delta,y\!:\tau$. By induction hypothesis, since $\Delta,y\!:\tau\vdash(s,(\mu,y\leftarrow w))\rhd\Omega$ is derivable, we obtain $\Gamma, y\!:\sigma\vdash(s',(\mu',y\leftarrow w'))\rhd\Omega$ which implies $\emptyset;\Gamma,y\!:\sigma\vdash s'\rhd\Omega$ with $(\mu',y\leftarrow w')=\Gamma,y\!:\sigma$. This last assertion trivially implies $\emptyset;\Gamma\vdash w'\!:\sigma$ by definition of store typing. We can then build the following typing derivation to conclude:

  $$\frac{\emptyset;\Gamma\vdash w'\!:\sigma\qquad\emptyset;\Gamma,y\!:\sigma\vdash s'\rhd\Omega\qquad y\notin\Omega}{\emptyset;\Gamma\vdash\textbf{var}\ y:=w';\ s'\rhd\Omega}$$

- (S.ASSIGN): we have $\mu\rhd\Delta,\vec{y}\!:\!\vec{\sigma}$ and

  $$\frac{\emptyset;\Delta,\vec{y}\!:\!\vec{\sigma}\vdash e\!:(\vec{\tau})\qquad\emptyset;\Delta,\vec{y}\!:\!\vec{\tau}\vdash s\ \rhd\Delta'}{\emptyset;\Delta,\vec{y}\!:\!\vec{\sigma}\vdash\vec{y}:=e;\ s\ \rhd\Delta'}$$

then we get $\emptyset; \Delta, \vec{y} : \vec{\tau} \vdash s \rhd \Delta'$. By Lemma B.4, $\mu \rhd \Delta, \vec{y} : \vec{\sigma}$ and $\emptyset; \Delta, \vec{y} : \vec{\sigma} \vdash e : (\vec{\tau})$ and $e =_\mu (\vec{w})$ implies $\emptyset; \emptyset \vdash \vec{w} : \vec{\tau}$. Then, by definition of store typing, we obtain $\mu[\vec{y} \leftarrow \vec{w}] \rhd \Delta, \vec{y} : \vec{\tau}$.

- (S.INC): we have $\mu \rhd \Delta, y : \mathbf{nat}$ and
$$\frac{\emptyset; \Delta, y : \mathbf{nat} \vdash s \rhd \Delta'}{\emptyset; \Delta, y : \mathbf{nat} \vdash \mathbf{inc}(y);\ s \rhd \Delta'}$$
then
$$\frac{\dfrac{}{\emptyset; \Delta, y : \mathbf{nat} \vdash \overline{q+1} : \mathbf{nat}} \quad \emptyset; \Delta, y : \mathbf{nat} \vdash s \rhd \Delta'}{\emptyset; \Delta, y : \mathbf{nat} \vdash y := \overline{q+1};\ s \rhd \Delta'}$$

- (S.DEC): similar to above.
- (S.CALL): we have $\mu \rhd \Delta, \vec{r} : \vec{\omega}$ and
$$\frac{\emptyset; \Delta, \vec{r} : \vec{\omega} \vdash p : \mathbf{proc}\ (\mathbf{in}\ \vec{\tau}; \mathbf{out}\ \vec{\sigma}) \quad \emptyset; \Delta, \vec{r} : \vec{\omega} \vdash \vec{e} : \vec{\tau} \quad \emptyset; \Delta, \vec{r} : \vec{\sigma} \vdash s \rhd \Delta'}{\emptyset; \Delta, \vec{r} : \vec{\omega} \vdash p(\vec{e}, \vec{r});\ s \rhd \Delta'}$$
By Lemma B.4, $\mu \rhd \Delta, \vec{r} : \vec{\omega}$ and $\emptyset; \Delta, \vec{r} : \vec{\omega} \vdash \vec{e} : \vec{\tau}$ and $\vec{e} =_\mu \vec{w}$ implies $\emptyset; \emptyset \vdash \vec{w} : \vec{\tau}$. Still by Lemma B.4, $\mu \rhd \Delta, \vec{r} : \vec{\omega}$ and $\emptyset; \Delta, \vec{r} : \vec{\omega} \vdash p : \mathbf{proc}(\mathbf{in}\ \vec{\tau};\ \mathbf{out}\ \vec{\sigma})$ and $p =_\mu \mathbf{proc}(\mathbf{in}\ \vec{y};\ \mathbf{out}\ \vec{x})\{s'\}_{\vec{x}}$ implies $\emptyset; \emptyset \vdash \mathbf{proc}(\mathbf{in}\ \vec{y}; \mathbf{out}\ \vec{x})\{s\}_{\vec{x}} : \mathbf{proc}(\mathbf{in}\ \vec{\tau}; \mathbf{out}\ \vec{\sigma})$, that is
$$\frac{\vec{z} \neq \emptyset \qquad \emptyset; \vec{y} : \vec{\sigma}; \vec{x} : \overrightarrow{\mathbf{unit}} \vdash s' \rhd \vec{x} : \vec{\sigma}}{\emptyset; \emptyset \vdash \mathbf{proc}\ (\mathbf{in}\ \vec{y}; \mathbf{out}\ \vec{x})\{s'\}_{\vec{x}} : \mathbf{proc}(\mathbf{in}\ \vec{\tau}; \mathbf{out}\ \vec{\sigma})}$$
By Lemmas B.1 and B.2, $\emptyset; \vec{y} : \vec{\sigma}; \vec{x} : \overrightarrow{\mathbf{unit}} \vdash s' \rhd \vec{x} : \vec{\sigma}$ and $\emptyset; \emptyset \vdash \vec{w} : \vec{\tau}$ implies $\emptyset; \vec{r} : \overrightarrow{\mathbf{unit}} \vdash s'[\vec{y} \leftarrow \vec{w}][\vec{x} \hookleftarrow \vec{r}] \rhd \vec{r} : \vec{\sigma}$. By definition of store typing, we have $\mu[\vec{r} \leftarrow *] \rhd \Delta, \vec{r} : \overrightarrow{\mathbf{unit}}$ and we can then build the following typing derivation to conclude:
$$\frac{\emptyset; \vec{r} : \overrightarrow{\mathbf{unit}} \vdash s'[\vec{y} \leftarrow \vec{w}][\vec{x} \hookleftarrow \vec{r}] \rhd \vec{r} : \vec{\sigma} \quad \emptyset; \Delta, \vec{r} : \vec{\sigma} \vdash s \rhd \Delta'}{\emptyset; \Delta, \vec{r} : \overrightarrow{\mathbf{unit}} \vdash \{s'[\vec{y} \leftarrow \vec{w}][\vec{x} \hookleftarrow \vec{r}]\}_{\vec{x}};\ s \rhd \Delta'}$$

- (S.CST): we have $\mu \rhd \Delta$ and
$$\frac{\emptyset; \Delta \vdash e : \tau \quad y : \tau; \Delta \vdash s \rhd \Omega}{\emptyset; \Delta \vdash \mathbf{cst}\ y = e;\ s \rhd \Omega}$$
By Lemma B.4, $\mu \rhd \Delta$ and $\emptyset; \Delta \vdash e : \tau$ and $e =_\mu w$ implies $\emptyset; \emptyset \vdash w : \tau$. By Lemma B.1, $y : \tau; \Delta \vdash s \rhd \Omega$ and $\emptyset; \emptyset \vdash w : \tau$ implies $\emptyset; \Delta \vdash s[y \leftarrow w] \rhd \Omega$.
- (S.FOR-I): we have $\mu \rhd \Delta, \vec{x} : \vec{\sigma}$ and
$$\frac{\emptyset; \Delta, \vec{x} : \vec{\sigma} \vdash e : \mathbf{nat} \quad y : \mathbf{nat}; \vec{x} : \vec{\sigma} \vdash s \rhd \vec{x} : \vec{\sigma} \quad \emptyset; \Delta, \vec{x} : \vec{\sigma} \vdash s' \rhd \Delta'}{\emptyset; \Delta, \vec{x} : \vec{\sigma} \vdash \mathbf{for}\ y := 0\ \mathbf{until}\ e\ \{s\}_{\vec{x}};\ s' \rhd \Delta'}$$
We have immediately $\emptyset; \Delta, \vec{x} : \vec{\sigma} \vdash s' \rhd \Delta'$.
- (S.FOR-II): we have $\mu \rhd \Delta, \vec{x} : \vec{\sigma}$ and
$$\frac{\emptyset; \Delta, \vec{x} : \vec{\sigma} \vdash e : \mathbf{nat} \quad y : \mathbf{nat}; \vec{x} : \vec{\sigma} \vdash s \rhd \vec{x} : \vec{\sigma} \quad \emptyset; \Delta, \vec{x} : \vec{\sigma} \vdash s' \rhd \Delta'}{\emptyset; \Delta, \vec{x} : \vec{\sigma} \vdash \mathbf{for}\ y := 0\ \mathbf{until}\ e\ \{s\}_{\vec{x}};\ s' \rhd \Delta'}$$
By Lemma B.1, $y : \mathbf{nat}; \vec{x} : \vec{\sigma} \vdash s \rhd \vec{x} : \vec{\sigma}$ and $\emptyset; \emptyset \vdash \bar{q} : \mathbf{nat}$ implies $\emptyset; \vec{x} : \vec{\sigma} \vdash s[y \leftarrow \bar{q}] \rhd \vec{x} : \vec{\sigma}$. We can then build the following typing derivation to conclude:
$$\frac{\dfrac{\dfrac{}{\emptyset; \Delta, \vec{x} : \vec{\sigma} \vdash \bar{q} : \mathbf{nat}} \quad y : \mathbf{nat}; \vec{x} : \vec{\sigma} \vdash s \rhd \vec{x} : \vec{\sigma} \quad \emptyset; \vec{x} : \vec{\sigma} \vdash s[y \leftarrow \bar{q}] \rhd \vec{x} : \vec{\sigma}}{\emptyset; \vec{x} : \vec{\sigma} \vdash \mathbf{for}\ y := 0\ \mathbf{until}\ \bar{q}\ \{s\}_{\vec{x}};\ s[y \leftarrow \bar{q}] \rhd \vec{x} : \vec{\sigma}} \quad \emptyset; \Delta, \vec{x} : \vec{\sigma} \vdash s' \rhd \Delta'}{\emptyset; \Delta, \vec{x} : \vec{\sigma} \vdash \{\mathbf{for}\ y := 0\ \mathbf{until}\ \bar{q}\ \{s\}_{\vec{x}};\ s[y \leftarrow \bar{q}]\}_{\vec{x}};\ s' \rhd \Delta'}$$

□

## B.4 Progress

**Proposition.** *For any state* $(s, \mu)$*,* $\Omega$ *and* $\vec{z}$*, if* $\vec{z}:\vec{\tau} \vdash (s, \mu) \rhd \Omega$ *in* **IS** *then either* $s = \varepsilon$ *and no more reduction can occur, or there is a state* $(s', \mu')$ *such that* $(s, \mu) \mapsto (s', \mu')$*.*

**Proof.** By induction on $\vec{z}:\vec{\tau} \vdash (s, \mu) \rhd \Omega$.

- $s \equiv \varepsilon$: then we are in the first case.
- $s \equiv (\mathbf{cst}\ y = e;\ s_1)$: if $e \equiv x$ then by definition of state typing, $x \in \mathit{dom}(\mu)$ ; we then have $((\mathbf{cst}\ y = e;\ s_1), \mu) \mapsto (s_1[y \leftarrow \varphi_\mu(e)], \mu))$.
- $s \equiv (\mathbf{var}\ y := e;\ s_1)$: if $e \equiv x$ then by definition of state typing, $x \in \mathit{dom}(\mu)$ ; by induction hypothesis on $\vec{z}:\vec{\tau}, y{:}\,\tau \vdash (s_1, (\mu, y \leftarrow \varphi_\mu(e))) \rhd \Omega, y{:}\,\tau'$, we have either $s_1 \equiv \varepsilon$ or $(s_1, (\mu, y \leftarrow \varphi_\mu(e))) \mapsto (s_1', (\mu', y \leftarrow w'))$ ; in the first case, we have $((\mathbf{var}\ y := e;\ \varepsilon), \mu) \mapsto (\varepsilon, \mu)$, and in the second case we have $((\mathbf{var}\ y := e;\ s_1), \mu) \mapsto ((\mathbf{var}\ y := w';\ s_1'), \mu')$.
- $s \equiv (\{s_1\}_{\vec{z}\,'};\ s_2)$: by induction hypothesis on $\vec{z}\,'{:}\,\vec{\sigma} \vdash (s_1, \mu) \rhd \vec{z}\,'{:}\,\vec{\sigma}\,'$, we have either $s_1 \equiv \varepsilon$ or $(s_1, \mu) \mapsto (s_1', \mu')$ ; in the first case, we have $((\{\}_{\vec{z}\,'};\ s_2), \mu) \mapsto (s_2, \mu)$, and in the second case we have $((\{s_1\}_{\vec{z}\,'};\ s_2), \mu) \mapsto ((\{s_1'\}_{\vec{z}\,'};\ s_2), \mu')$.
- $s \equiv (\mathbf{inc}(y);\ s_1)$: by definition of state typing, $y \in \mathit{dom}(\mu)$ ; we have $((\mathbf{inc}(y);\ s_1), \mu) \mapsto ((y := \overline{q+1};\ s_1), \mu)$.
- the case for **dec** is similar to **inc**.
- $s \equiv (\vec{y} := e;\ s_1)$: if $e \equiv x$ then by definition of state typing, $x \in \mathit{dom}(\mu)$, hence $e =_\mu (\vec{w})$ can always be derived ; we have $((\vec{y} := e;\ s_1), \mu) \mapsto (s_1, \mu[\vec{y} \leftarrow \vec{w}])$.
- $s \equiv (p(\vec{e}\,; \vec{r}\,);\ s_1)$: if $e_i \equiv x$ then by definition of state typing, $x \in \mathit{dom}(\mu)$, similarly for $p$ ; we have $((p(\vec{e}\,; \vec{r}\,);\ s), \mu) \mapsto ((\{s'[\vec{y} \leftarrow \vec{w}][\vec{z} \hookleftarrow \vec{r}\,]\}_{\vec{r}};\ s), \mu[\vec{r} \leftarrow *])$.
- $s \equiv (\mathbf{for}\ y := 0\ \mathbf{until}\ e\ \{s_1\}_{\vec{z}\,'};\ s_2)$: if $e \equiv x$ then by definition of state typing, $x \in \mathit{dom}(\mu)$ ; either $e =_\mu \bar{0}$ and $((\mathbf{for}\ y := 0\ \mathbf{until}\ e\ \{s_1\}_{\vec{z}\,'};\ s_2), \mu) \mapsto (s_2, \mu)$, or $e \neq_\mu \bar{0}$ and $((\mathbf{for}\ y := 0\ \mathbf{until}\ e\ \{s_1\}_{\vec{z}\,'};\ s_2), \mu) \mapsto ((\{\mathbf{for}\ y := 0\ \mathbf{until}\ \bar{q}\ \ \{s_1\}_{\vec{z}\,'};\ s_1[y \leftarrow \bar{q}\,]\}_{\vec{z}\,'};\ s_2), \mu)$. □

## B.5 Expressiveness

**Definition B.5.** *The translation of a type* $\tau \in \Sigma_{\mathbf{FS}}$ *into a type* $\tau^\natural \in \Sigma_{\mathbf{FS}}$ *is defined by the following rules:*

$$\begin{aligned} \mathbf{nat}^\natural &= \mathbf{nat} \\ \mathbf{unit}^\natural &= \mathbf{unit} \\ (\sigma \to \tau)^\natural &= \sigma^\natural \to \tau^\natural \\ (\tau_1 \times \ldots \times \tau_n)^\natural &= (\tau_1^\natural \times \ldots \times \tau_n^\natural) \end{aligned}$$

**Proposition B.6.** *For any functional term* $t$*, if* $\Gamma \vdash t{:}\,\tau$ *in* **FS** *then* $\Gamma^\natural \vdash t^\natural{:}\,\tau^\natural$ *in* **FS***.*

**Proof.** Straightforward induction on $t$. □

# Appendix C Properties of $\mathbf{ID}^{(c)}$ and $\mathbf{FD}^{(c)}$

## C.1 Alternative dependent type system

We first present in Figure C.1 a different (but equivalent) formulation of the imperative dependent type system which is easier to deal with when proving properties by induction on sequences.

## C.2 Preliminary properties

**Lemma C.1.** *If* $\Gamma, x{:}\,\tau; \Omega \vdash s \rhd \Omega'$ *and* $\emptyset; \emptyset \vdash e{:}\,\tau$ *in* **ID** *then* $\Gamma; \Omega \vdash s[x \leftarrow e] \rhd \Omega'$ *in* **ID**.

**Proof.** Straightforward induction on $s$. ☐

**Lemma C.2.** *If* $\Gamma; x{:}\,\tau, \Omega \vdash s \rhd x{:}\,\sigma, \Omega'$ *in* **ID** *then* $\Gamma; y{:}\,\tau, \Omega \vdash s[x \hookleftarrow y] \rhd y{:}\,\sigma, \Omega'$ *in* **ID**.

**Proof.** Straightforward induction on $s$. ☐

**Lemma C.3.** *If* $\Gamma; \Omega \vdash s \rhd \Omega'$ *in* **ID** *then for any* $x{:}\,\sigma$, $\Gamma, x{:}\,\sigma; \Omega \vdash s \rhd \Omega'$ *and* $\Gamma; \Omega, x{:}\,\sigma \vdash s \rhd \Omega'$ *in* **ID**.

**Proof.** Straightforward induction on the typing derivation. ☐

**Lemma C.4.** *If* $\mu \rhd \Omega$ *and* $\emptyset; \Omega \vdash e{:}\,\tau$ *in* **ID** *and* $e =_\mu w$, *then we have* $\emptyset; \emptyset \vdash w{:}\,\tau$ *in* **ID**.

**Proof.** The case $e = w$ is trivial and if $e$ is some variable $x \in \Omega$ then by definition of $\mu \rhd \Omega$, we have $\emptyset; \emptyset \vdash \mu(x) = w{:}\,\tau$. ☐

## C.3 Translation from ID to FD

**Theorem.** *(Soundness for* **ID***). For any environments* $\Gamma$ *and* $\Omega$*, any expression* $e$*, any sequence* $s$ *we have:*

- $\Gamma; \Omega \vdash e{:}\,\tau$ *in* **ID** *implies* $\Gamma^\star, \Omega^\star \vdash e^\star{:}\,\tau^\star$ *in* **FD**.
- $\Gamma; \Omega \vdash s \rhd \vec{z}: \vec{\sigma}$ *in* **ID** *implies* $\Gamma^\star, \Omega^\star \vdash (s)^\star_{\vec{z}}{:}\,\vec{\sigma}^\star$ *in* **FD**.

**Proof.** We proceed by induction on the typing derivation:

- (T.ENV)

$$\frac{y{:}\,\tau \in \Gamma, \Omega}{\Gamma; \Omega \vdash y{:}\,\tau}$$

Indeed,

$$\frac{y{:}\,\tau^\star \in \Gamma^\star, \Omega^\star}{\Gamma^\star, \Omega^\star \vdash y{:}\,\tau^\star}$$

- (T.NUM)

$$\frac{}{\Gamma; \Omega \vdash \bar{q} : \mathbf{nat}(\mathbf{s}^q(0))}$$

Indeed,

$$\frac{\dfrac{\Gamma^\star, \Omega^\star \vdash 0{:}\,\mathbf{nat}(0)}{\cdots}}{\Gamma^\star, \Omega^\star \vdash S^q(0){:}\,\mathbf{nat}(\mathbf{s}^q(0))}$$

- (T.TUPLE)

$$\frac{\Gamma; \Omega \vdash \vec{e} : \vec{\tau}[\vec{u}/\vec{\imath}]}{\Gamma; \Omega \vdash (\vec{e}) : \exists \vec{\jmath}\,(\vec{\tau})}$$

$$\frac{x\colon\tau\in\Gamma;\Omega}{\Gamma;\Omega\vdash x\colon\tau} \quad (\textsc{t.env})$$

$$\frac{}{\Gamma;\Omega\vdash\bar{q}\colon\mathbf{nat}(\mathbf{s}^q(\mathbf{0}))} \quad (\textsc{t.num})$$

$$\frac{\vdash_{\mathcal{E}} n=m}{\Gamma;\Omega\vdash *\colon n=m} \quad (\textsc{t.equal})$$

$$\frac{\Gamma;\Omega\vdash\vec{e}\colon\vec{\tau}[\vec{u}/\vec{\imath}]}{\Gamma;\Omega\vdash(\vec{e})\colon\exists\vec{\jmath}(\vec{\tau})} \quad (\textsc{t.tuple})$$

$$\frac{\vec{z}\neq\emptyset \qquad \Gamma,\vec{y}\colon\vec{\sigma};\vec{z}\colon\vec{\top}\vdash s\rhd\vec{z}\colon\vec{\tau}}{\Gamma;\Omega\vdash\mathbf{proc}\ (\mathbf{in}\ \vec{y};\mathbf{out}\ \vec{z})\{s\}_{\vec{z}}\colon\mathbf{proc}\ \forall\vec{\imath}(\mathbf{in}\ \vec{\sigma};\mathbf{out}\ \vec{\tau})} \quad (\textsc{t.proc})^*$$

$$\frac{\Gamma;\Omega\vdash e'\colon\tau[n/i] \qquad \Gamma;\Omega\vdash e\colon n=m}{\Gamma;\Omega\vdash e'\colon\tau[m/i]} \quad (\textsc{t.subst-i})$$

$$\frac{\Gamma;\Omega\vdash s\rhd\Omega'[n/i] \qquad \Gamma;\Omega\vdash e\colon n=m}{\Gamma;\Omega\vdash s\rhd\Omega'[m/i]} \quad (\textsc{t.subst-ii})$$

$$\frac{}{\Gamma;\Omega,\Omega'\vdash\varepsilon\rhd\Omega'} \quad (\textsc{t.empty})$$

$$\frac{\Gamma;\Omega\vdash e\colon\tau \qquad \Gamma,y\colon\tau;\Omega\vdash s\rhd\Omega'}{\Gamma;\Omega\vdash\mathbf{cst}\ y=e;\ s\rhd\Omega'} \quad (\textsc{t.cst})$$

$$\frac{\Gamma;\Omega\vdash e\colon\tau \qquad \Gamma;\Omega,y\colon\tau\vdash s\rhd\Omega' \qquad y\notin\Omega'}{\Gamma;\Omega\vdash\mathbf{var}\ y:=e;\ s\rhd\Omega'} \quad (\textsc{t.var})$$

$$\frac{\Gamma;\vec{x}\colon\vec{\tau}\vdash s\rhd\vec{x}\colon\vec{\sigma} \qquad \Gamma;\Omega,\vec{x}\colon\vec{\sigma}\vdash s'\rhd\Omega'}{\Gamma;\Omega,\vec{x}\colon\vec{\tau}\vdash\{s\}_{\vec{x}};s'\rhd\Omega'} \quad (\textsc{t.block})$$

$$\frac{\Gamma;\Omega,y\colon\mathbf{nat}(\mathbf{s}(n))\vdash s\rhd\Omega'}{\Gamma;\Omega,y\colon\mathbf{nat}(n)\vdash\mathbf{inc}(y);s\rhd\Omega'} \quad (\textsc{t.inc})$$

$$\frac{\Gamma;\Omega,y\colon\mathbf{nat}(\mathbf{p}(n))\vdash s\rhd\Omega'}{\Gamma;\Omega,y\colon\mathbf{nat}(n)\vdash\mathbf{dec}(y);s\rhd\Omega'} \quad (\textsc{t.dec})$$

$$\frac{\Gamma;\Omega,\vec{y}\colon\vec{\sigma}\vdash e\colon\exists\vec{\imath}(\vec{\tau}) \qquad \Gamma;\Omega,\vec{y}\colon\vec{\tau}\vdash s\rhd\Omega'}{\Gamma;\Omega,\vec{y}\colon\vec{\sigma}\vdash\vec{y}:=e;\ s\rhd\Omega'} \quad (\textsc{t.assign})^*$$

$$\frac{\Gamma;\Omega,\vec{x}\colon\vec{\sigma}[\mathbf{0}/i]\vdash e\colon\mathbf{nat}(n) \quad \Gamma,y\colon\mathbf{nat}(i);\vec{x}\colon\vec{\sigma}\vdash s\rhd\vec{x}\colon\vec{\sigma}[\mathbf{s}(i)/i] \quad \Gamma;\Omega,\vec{x}\colon\vec{\sigma}[n/i]\vdash s'\rhd\Omega'}{\Gamma;\Omega,\vec{x}\colon\vec{\sigma}[\mathbf{0}/i]\vdash\mathbf{for}\ y:=0\ \mathbf{until}\ e\ \{s\}_{\vec{x}};s'\rhd\Omega'} \quad (\textsc{t.for})^*$$

$$\frac{\Gamma;\Omega,\vec{r}\colon\vec{\omega}\vdash p\colon\mathbf{proc}\ \forall\vec{\imath}(\mathbf{in}\ \vec{\sigma};\mathbf{out}\ \vec{\tau}) \quad \Gamma;\Omega,\vec{r}\colon\vec{\omega}\vdash\vec{e}\colon\vec{\sigma}[\vec{u}/\vec{\imath}] \quad \Gamma;\Omega,\vec{r}\colon\vec{\tau}[\vec{u}/\vec{\imath}]\vdash s\rhd\Omega'}{\Gamma;\Omega,\vec{r}\colon\vec{\omega}\vdash p(\vec{e};\vec{r});s\rhd\Omega'} \quad (\textsc{t.call})$$

*where $\vec{\imath}\notin\mathcal{FV}(\Gamma)$ in (T.PROC) and $i\notin\mathcal{FV}(\Gamma)$ in (T.FOR) and $\vec{\imath}\notin\mathcal{FV}(\Gamma,\Omega,\Omega')$ in (T.ASSIGN)

**Figure C.1.** Alternative imperative dependent type system

Indeed,

$$\frac{\Gamma,\Omega\vdash\vec{e}^{\star}\colon\vec{\tau}^{\star}[\vec{u}/\vec{\imath}]}{\Gamma,\Omega\vdash\vec{e}^{\star}\colon\exists\vec{\jmath}(\vec{\tau})}$$

- (T.SUBST-I)

$$\frac{\Gamma;\Omega\vdash e'\colon\tau[n/i] \quad \Gamma;\Omega\vdash e\colon n=m}{\Gamma;\Omega\vdash e'\colon\tau[m/i]}$$

Indeed,

$$\frac{\Gamma^\star, \Omega^\star \vdash e'^\star : \tau[n/i] \quad \Gamma^\star, \Omega^\star \vdash e^\star : n = m}{\Gamma^\star, \Omega^\star \vdash e'^\star : \tau[m/i]}$$

- (T.EQUAL)

$$\frac{\vdash_{\mathcal{E}} n = m}{\Gamma; \Omega \vdash * : n = m}$$

Indeed,

$$\frac{\vdash_{\mathcal{E}} n = m}{\Gamma^\star, \Omega^\star \vdash () : n = m}$$

- (T.SUBST-II)

$$\frac{\Gamma; \Omega \vdash s \rhd \vec{z} : \vec{\sigma}[n/i] \quad \Gamma; \Omega \vdash e : n = m}{\Gamma; \Omega \vdash s \rhd \vec{z} : \vec{\sigma}[m/i]}$$

Indeed,

$$\frac{\Gamma^\star, \Omega^\star \vdash (s)^\star_{\vec{z}} : \vec{\sigma}^\star[n/i] \quad \Gamma; \Omega \vdash e^\star : n = m}{\Gamma^\star, \Omega^\star \vdash (s)^\star_{\vec{z}} : \vec{\sigma}^\star[m/i]}$$

- (T.EMPTY)

$$\frac{}{\Gamma; \Omega, \vec{z} : \vec{\sigma} \vdash \varepsilon \rhd \vec{z} : \vec{\sigma}}$$

Indeed,

$$\frac{}{\Gamma, \Omega, \vec{z} : \vec{\sigma}^\star \vdash \vec{z} : \vec{\sigma}^\star}$$

- (T.CST)

$$\frac{\Gamma; \Omega \vdash e : \tau \qquad \Gamma, y : \tau; \Omega \vdash s \rhd \vec{z} : \vec{\sigma}}{\Gamma; \Omega \vdash \textbf{cst } y = e;\ s \rhd \vec{z} : \vec{\sigma}}$$

Indeed,

$$\frac{\Gamma^\star, \Omega^\star \vdash e^\star : \tau^\star \qquad \Gamma^\star, y : \tau^\star, \Omega^\star \vdash (s)^\star_{\vec{z}} : \vec{\sigma}^\star}{\Gamma^\star, \Omega^\star \vdash \textbf{let } y = e^\star \textbf{ in } (s)^\star_{\vec{z}} : \vec{\sigma}^\star}$$

- (T.VAR)

$$\frac{\Gamma; \Omega \vdash e : \tau \qquad \Gamma; \Omega, y : \tau \vdash s \rhd \vec{z} : \vec{\sigma} \qquad y \notin \vec{z}}{\Gamma; \Omega \vdash \textbf{var } y := e;\ s \rhd \vec{z} : \vec{\sigma}}$$

Indeed, by the substitution lemma,

$$\frac{\Gamma^\star, \Omega^\star \vdash e^\star : \tau^\star \qquad \Gamma^\star, y : \tau^\star, \Omega^\star \vdash (s)^\star_{\vec{z}} : \vec{\sigma}^\star}{\Gamma^\star, \Omega^\star \vdash (s)^\star_{\vec{z}}[e^\star/y] : \vec{\sigma}^\star}$$

- (T.BLOCK)

$$\frac{\Gamma; \vec{x} : \vec{\tau} \vdash s \rhd \vec{x} : \vec{\sigma}' \qquad \Gamma; \Omega, \vec{x} : \vec{\sigma}' \vdash s' \rhd \vec{z} : \vec{\sigma}}{\Gamma; \Omega, \vec{x} : \vec{\tau} \vdash \{s\}_{\vec{x}}; s' \rhd \vec{z} : \vec{\sigma}}$$

Indeed,

$$\frac{\Gamma^\star, \vec{x} : \vec{\tau}^\star \vdash (s)^\star_{\vec{x}} : \vec{\sigma}'^\star \qquad \Gamma^\star, \Omega^\star, \vec{x} : \vec{\sigma}'^\star \vdash (s')^\star_{\vec{z}} : \vec{\sigma}^\star}{\Gamma^\star, \Omega^\star, \vec{x} : \vec{\tau}^\star \vdash \textbf{let } \vec{x} = (s)^\star_{\vec{x}} \textbf{ in } (s')^\star_{\vec{z}} : \vec{\sigma}^\star}$$

- (T.INC)

$$\frac{\Gamma; \Omega, y : \textbf{nat}(\textbf{s}(n)) \vdash s \rhd \vec{z} : \vec{\sigma}}{\Gamma; \Omega, y : \textbf{nat}(n) \vdash \textbf{inc}(y);\ s \rhd \vec{z} : \vec{\sigma}}$$

Indeed,

$$\frac{\Gamma^\star, \Omega^\star, y : \textbf{nat}(n) \vdash \textbf{succ} : \forall x(\textbf{nat}(x) \Rightarrow \textbf{nat}(\textbf{s}(x))) \quad \Gamma^\star, \Omega^\star, y : \textbf{nat}(n) \vdash y : \textbf{nat}(n)}{\Gamma^\star, \Omega^\star, y : \textbf{nat}(n) \vdash \textbf{succ}(y) : \textbf{nat}(\textbf{s}(n))}$$

and

$$\frac{\Gamma^\star, \Omega^\star, y : \textbf{nat}(n) \vdash \textbf{succ}(y) : \textbf{nat}(\textbf{s}(n)) \qquad \Gamma^\star, \Omega^\star, y : \textbf{nat}(\textbf{s}(n)) \vdash (s)^\star_{\vec{z}} : \vec{\sigma}^\star}{\Gamma^\star, \Omega^\star, y : \textbf{nat}(n) \vdash \textbf{let } y = \textbf{succ}(y) \textbf{ in } (s)^\star_{\vec{z}} : \vec{\sigma}^\star}$$

- (T.DEC)

$$\frac{\Gamma; \Omega, y : \textbf{nat}(\textbf{p}(n)) \vdash s \rhd \vec{z} : \vec{\sigma}}{\Gamma; \Omega, y : \textbf{nat}(n) \vdash \textbf{dec}(y);\ s \rhd \vec{z} : \vec{\sigma}}$$

Indeed,

$$\dfrac{\dfrac{\Gamma^\star, \Omega^\star, y\colon \mathbf{nat}(n) \vdash y\colon \mathbf{nat}(n)}{\Gamma^\star, \Omega^\star, y\colon \mathbf{nat}(n) \vdash \mathbf{pred}(y)\colon \mathbf{nat}(\mathbf{p}(n))} \qquad \Gamma^\star, \Omega^\star, y\colon \mathbf{nat}(\mathbf{p}(n)) \vdash (s)^\star_{\vec{z}}\colon \vec{\sigma}^\star}{\Gamma^\star, \Omega^\star, y\colon \mathbf{nat}(n) \vdash \mathbf{let}\ y = \mathbf{pred}(y)\ \mathbf{in}\ (s)^\star_{\vec{z}}\colon \vec{\sigma}^\star}$$

- (T.ASSIGN)

$$\dfrac{\Gamma; \Omega, \vec{y} : \vec{\sigma}\,' \vdash e\colon \exists \vec{\jmath}\,(\vec{\tau}\,) \qquad \Gamma; \Omega, \vec{y} : \vec{\tau} \vdash s \rhd \vec{z} : \vec{\sigma}}{\Gamma; \Omega, \vec{y} : \vec{\sigma}\,' \vdash \vec{y} := e;\ \ s \rhd \vec{z} : \vec{\sigma}}$$

with $\vec{\jmath} \notin \mathcal{FV}(\Gamma, \Omega, \vec{\sigma}\,)$. Indeed,

$$\dfrac{\Gamma^\star, \Omega^\star, \vec{y} : \vec{\sigma}\,'^\star \vdash e^\star\colon \exists \vec{\jmath}\,(\vec{\tau}^\star) \qquad \Gamma^\star, \Omega^\star, \vec{y} : \vec{\tau}^\star \vdash (s)^\star_{\vec{z}}\colon \vec{\sigma}^\star}{\Gamma^\star, \Omega^\star, \vec{y} : \vec{\sigma}\,'^\star \vdash \mathbf{let}\ \vec{y} = e^\star\ \mathbf{in}\ (s)^\star_{\vec{z}}\colon \vec{\sigma}^\star}$$

since $\vec{\jmath} \notin \mathcal{FV}(\Gamma^\star, \Omega^\star, \vec{\sigma}^\star)$.

- (T.FOR)

$$\dfrac{\Gamma; \Omega, \vec{x} : \vec{\sigma}\,[\mathbf{0}/i] \vdash e\colon \mathbf{nat}(n) \quad \Gamma, y\colon \mathbf{nat}(i); \vec{x} : \vec{\sigma} \vdash s \rhd \vec{x} : \vec{\sigma}\,[\mathbf{s}(i)/i] \quad \Gamma; \Omega, \vec{x} : \vec{\sigma}\,[n/i] \vdash s' \rhd \vec{z} : \vec{\sigma}\,'}{\Gamma; \Omega, \vec{x} : \vec{\sigma}\,[\mathbf{0}/i] \vdash \mathbf{for}\ y := 0\ \mathbf{until}\ e\ \{s\}_{\vec{x}};\ \ s' \rhd \vec{z} : \vec{\sigma}\,'}$$

with $i \notin \mathcal{FV}(\Gamma)$. Indeed,

$$\dfrac{\Gamma^\star, \Omega^\star, \vec{x} : \vec{\sigma}^\star[\mathbf{0}/i] \vdash e^\star\colon \mathbf{nat}(n) \quad \Gamma^\star, \Omega^\star, \vec{x} : \vec{\sigma}^\star[\mathbf{0}/i] \vdash \vec{x} : (\vec{\sigma}^\star[\mathbf{0}/i]) \quad \Gamma^\star, y\colon \mathbf{nat}(i), \vec{x} : \vec{\sigma}^\star \vdash (s)^\star_{\vec{x}}\colon (\vec{\sigma}^\star[\mathbf{s}(i)/i])}{\Gamma^\star, \Omega^\star, \vec{x} : \vec{\sigma}^\star[\mathbf{0}/i] \vdash \mathbf{rec}(e^\star, \vec{x}\,, \lambda y.\lambda \vec{x}\,.(s)^\star_{\vec{x}})\colon (\vec{\sigma}^\star[n/i])}$$

since $i \notin \mathcal{FV}(\Gamma^\star)$, and then

$$\dfrac{\Gamma^\star, \Omega^\star, \vec{x} : \vec{\sigma}^\star[\mathbf{0}/i] \vdash \mathbf{rec}(e^\star, \vec{x}\,, \lambda y.\lambda \vec{x}\,.(s)^\star_{\vec{x}})\colon (\vec{\sigma}^\star[n/i]) \qquad \Gamma^\star, \Omega^\star, \vec{x} : \vec{\sigma}^\star[n/i] \vdash (s')^\star_{\vec{z}}\colon \vec{\sigma}\,'^\star}{\Gamma^\star, \Omega^\star, \vec{x} : \vec{\sigma}^\star[\mathbf{0}/i] \vdash \mathbf{let}\ \vec{x} = \mathbf{rec}(e^\star, \vec{x}\,, \lambda y.\lambda \vec{x}\,.(s)^\star_{\vec{x}})\ \mathbf{in}\ (s')^\star_{\vec{z}}\colon \vec{\sigma}\,'^\star}$$

- (T.PROC)

$$\dfrac{\vec{z} \neq \emptyset \qquad \Gamma, \vec{y} : \vec{\sigma}; \vec{z} : \vec{\top} \vdash s \rhd \vec{z} : \vec{\tau}}{\Gamma; \Omega \vdash \mathbf{proc}\ (\mathbf{in}\ \vec{y}\,; \mathbf{out}\ \vec{z}\,)\{s\}_{\vec{z}}\colon \mathbf{proc}\ \forall \vec{\imath}\,(\mathbf{in}\ \vec{\sigma}\,; \mathbf{out}\ \vec{\tau}\,)}$$

with $\vec{\imath} \notin \mathcal{FV}(\Gamma)$. Indeed,

$$\dfrac{\dfrac{\dfrac{\Gamma^\star, \vec{y} : \vec{\sigma}^\star, \vec{z} : \vec{\top} \vdash (s)^\star_{\vec{z}}\colon \vec{\tau}^\star}{\Gamma^\star, \vec{y} : \vec{\sigma}^\star \vdash (s)^\star_{\vec{z}}[\vec{()}/\vec{z}\,]\colon \vec{\tau}^\star}}{\Gamma^\star \vdash \lambda \vec{y}\,.(s)^\star_{\vec{z}}[\vec{()}/\vec{z}\,]\colon \forall \vec{\imath}\,(\vec{\sigma}^\star \Rightarrow \vec{\tau}^\star)}}{\Gamma^\star, \Omega^\star \vdash \lambda \vec{y}\,.(s)^\star_{\vec{z}}[\vec{()}/\vec{z}\,]\colon \forall \vec{\imath}\,(\vec{\sigma}^\star \Rightarrow \vec{\tau}^\star)}$$

since $\vec{\imath} \notin \mathcal{FV}(\Gamma^\star)$.

- (T.CALL)

$$\dfrac{\Gamma; \Omega, \vec{r} : \vec{\omega} \vdash p\colon \mathbf{proc}\ \forall \vec{\imath}\,(\mathbf{in}\ \vec{\tau}\,; \mathbf{out}\ \vec{\sigma}\,) \quad \Gamma; \Omega, \vec{r} : \vec{\omega} \vdash \vec{e} : \vec{\tau}\,[\vec{u}/\vec{\imath}\,] \quad \Gamma; \Omega, \vec{r} : \vec{\sigma}\,[\vec{u}/\vec{\imath}\,] \vdash s \rhd \vec{z} : \vec{\sigma}\,'}{\Gamma; \Omega, \vec{r} : \vec{\omega} \vdash p(\vec{e}\,; \vec{r}\,)\,; s \rhd \vec{z} : \vec{\sigma}\,'}$$

Indeed,

$$\dfrac{\Gamma^\star, \Omega^\star, \vec{r} : \vec{\omega}^\star \vdash p^\star\colon \forall \vec{\imath}\,(\vec{\tau}^\star \Rightarrow \vec{\sigma}^\star) \quad \Gamma^\star, \Omega^\star, \vec{r} : \vec{\omega}^\star \vdash \vec{e}^\star\colon (\vec{\tau}^\star)[\vec{n}/\vec{\imath}\,]}{\Gamma^\star, \Omega^\star, \vec{r} : \vec{\omega}^\star \vdash (p^\star\ \vec{e}^\star)\colon \vec{\sigma}^\star[\vec{n}/\vec{\imath}\,]}$$

and then

$$\dfrac{\Gamma^\star, \Omega^\star, \vec{r} : \vec{\omega}^\star \vdash (p^\star\ \vec{e}^\star)\colon \vec{\sigma}^\star[\vec{n}/\vec{\imath}\,] \qquad \Gamma^\star, \Omega^\star, \vec{r} : \vec{\sigma}^\star[\vec{n}/\vec{\imath}\,] \vdash (s)^\star_{\vec{z}}\colon \vec{\sigma}\,'^\star}{\Gamma^\star, \Omega^\star, \vec{r} : \vec{\omega}^\star \vdash \mathbf{let}\ \vec{r} = p^\star\ \vec{e}^\star\ \mathbf{in}\ (s)^\star_{\vec{z}}\colon \vec{\sigma}\,'^\star}$$

$\square$

## C.4 Translation from FD to ID

**Notation C.5.** *The following typing rules are derivable.*

$$\frac{\Gamma; \Omega, \vec{y} : \vec{\top} \vdash s \rhd \Omega'}{\Gamma; \Omega \vdash \mathbf{var}\ \vec{y};\ \ s\ \rhd \Omega'}$$

$$\frac{\Gamma; \Omega \vdash \vec{w} : \vec{\tau} \qquad \Gamma; \Omega, \vec{y} : \vec{\tau} \vdash s \rhd \Omega'}{\Gamma; \Omega \vdash \mathbf{var}\ \vec{y} := \vec{w};\ \ s\ \rhd \Omega'}$$

$$\frac{\Gamma, \vec{y} : \vec{\tau}; \Omega, \vec{z} : \vec{\tau} \vdash s \rhd \Omega'}{\Gamma; \Omega, \vec{z} : \vec{\tau} \vdash \mathbf{cst}\ \vec{y} = \vec{z};\ \ s\ \rhd \Omega'}$$

$$\frac{\Gamma; \Omega, \vec{y} : \vec{\sigma} \vdash \vec{w} : \vec{\tau} \qquad \Gamma; \Omega, \vec{y} : \vec{\tau} \vdash s \rhd \Omega'}{\Gamma; \Omega, \vec{y} : \vec{\sigma} \vdash \vec{y} := \vec{w};\ \ s \rhd \Omega'}$$

**Lemma C.6.** *For all $t \in \mathcal{L}_n$, if $\Gamma \vdash t{:}\,\tau$ then $\tau = (\sigma_1 \wedge \ldots \wedge \sigma_n)$ for some $\sigma_1, \ldots, \sigma_n$.*

**Proof.** By induction on $t \in \mathcal{L}_n$.

- $t \equiv v \in \mathcal{L}_n$: by definition of $v \in \mathcal{L}_n$, we have $v = (w_1, \ldots, w_n)$, hence the typing derivation of $\Gamma \vdash v{:}\,\tau$ ends with:

$$\frac{\Gamma \vdash w_1{:}\,\sigma_1[\vec{m}/\vec{\imath}\,] \quad \ldots \quad \Gamma \vdash w_n{:}\,\sigma_n[\vec{m}/\vec{\imath}\,]}{\Gamma \vdash (w_1, \ldots, w_n){:}\,\exists \vec{\imath}\,.(\sigma_1 \wedge \ldots \wedge \sigma_n)}$$

- $t \equiv \mathbf{let}\ \vec{x} = u'\ \mathbf{in}\ u \in \mathcal{L}_n$ for any $u'$: by definition, we have in all cases $u \in \mathcal{L}_n$. By induction hypothesis, we have $\Gamma \vdash u{:}\,(\sigma_1 \wedge \ldots \wedge \sigma_n)$ and the typing derivation of $t$ ends with:

$$\frac{\Gamma \vdash u'{:}\,\exists \vec{\jmath}\,.\vec{\tau} \qquad \Gamma, \vec{x} : \vec{\tau} \vdash u{:}\,(\sigma_1 \wedge \ldots \wedge \sigma_n)}{\Gamma \vdash \mathbf{let}\ \vec{x} = u'\ \mathbf{in}\ u{:}\,(\sigma_1 \wedge \ldots \wedge \sigma_n)}$$

□

**Theorem.**

- *Given a term $t \in \mathcal{L}_n$ such that $\Gamma \vdash t{:}\,(\sigma_1 \wedge \ldots \wedge \sigma_n)$ in* **FD** *with $\Gamma, \vec{\sigma} \in \Sigma_{\mathbf{FD}}$ and a fresh mutable variable tuple $(r_1, \ldots, r_n)$ of any type $\vec{\sigma}\,' \in \Sigma_{\mathbf{ID}}$ we have $\Gamma^\diamond; \vec{r} : \vec{\sigma}\,' \vdash t_{\vec{r}}^\diamond \rhd (r_1{:}\,\sigma_1^\diamond, \ldots, r_n{:}\,\sigma_n{}^\diamond)$ in* **ID**.
- *Given a value $v \in \mathcal{V}$ such that $\Gamma \vdash v{:}\,\sigma$ in* **FD** *with $\Gamma, \sigma \in \Sigma_{\mathbf{FD}}$, for any environment $\Omega$ we have $\Gamma^\diamond$; $\Omega \vdash v^\diamond{:}\,\sigma^\diamond$ in* **ID**.

**Proof.** By mutual induction on $\Gamma \vdash t{:}\,\vec{\sigma}$ and $\Gamma \vdash v{:}\,\sigma$, and by case analysis of the translation.

- $x^\diamond = x$

$$\frac{x{:}\,\sigma \in \Gamma}{\Gamma \vdash x{:}\,\sigma}$$

  Indeed,

$$\frac{x{:}\,\sigma^\diamond \in \Gamma^\diamond}{\Gamma^\diamond; \Omega \vdash x{:}\,\sigma^\diamond}$$

- $(S^n(0))^\diamond = \bar{n}$

$$\frac{\dfrac{\Gamma \vdash 0{:}\,\mathbf{nat}(0)}{\ldots}}{\Gamma \vdash S^n(0){:}\,\mathbf{nat}(\mathbf{s}^n(0))}$$

  Indeed,

$$\frac{}{\Gamma^\diamond; \Omega \vdash \bar{n} : \mathbf{nat}(\mathbf{s}^n(0))}$$

- $()^\diamond = *$

  Indeed,
  $$\frac{}{\Gamma \vdash (): (n=m)}$$
  $$\frac{}{\Gamma^\diamond; \Omega \vdash *: (n=m)}$$

- $(\lambda \vec{x}.t)^\diamond = \mathbf{proc}\ (\mathbf{in}\ \vec{x}; \mathbf{out}\ \vec{z})\ \{t^\diamond_{\vec{z}}\}_{\vec{z}}$ where $\vec{z} = (z_1, ..., z_m)$ and $t \in \mathcal{L}_m$.
  $$\frac{\Gamma, \vec{x}: \vec{\tau} \vdash t: \vec{\sigma}}{\Gamma \vdash \lambda \vec{x}.t: \forall \vec{\imath}\,(\vec{\tau} \Rightarrow \vec{\sigma})}$$

  With $\vec{\imath} \notin \mathcal{FV}(\Gamma)$. By lemma C.6, $t \in \mathcal{L}_m$ and $\Gamma, \vec{x}: \vec{\tau} \vdash t: \vec{\sigma}$ implies $\vec{\sigma} = (\sigma_1 \wedge ... \wedge \sigma_m)$. By induction hypothesis, $\Gamma, \vec{x}: \vec{\tau} \vdash t: \vec{\sigma}$ implies $\Gamma^\diamond, \vec{x}: \vec{\tau}^\diamond; \vec{z}: \vec{\sigma}\,' \vdash t^\diamond_{\vec{z}} \rhd \vec{z}: \vec{\sigma}^\diamond$ for any $\vec{\sigma}\,'$, hence $\Gamma^\diamond, \vec{x}: \vec{\tau}^\diamond; \vec{z}: \vec{\top} \vdash t^\diamond_{\vec{z}} \rhd \vec{z}: \vec{\sigma}^\diamond$. Since $\vec{\imath} \notin \mathcal{FV}(\Gamma)$,
  $$\frac{\Gamma^\diamond, \vec{x}: \vec{\tau}^\diamond; \vec{z}: \vec{\top} \vdash t^\diamond_{\vec{z}} \rhd \vec{z}: \vec{\sigma}^\diamond}{\Gamma^\diamond; \vdash \mathbf{proc}\ (\mathbf{in}\ \vec{x}; \mathbf{out}\ \vec{z})\ \{t^\diamond_{\vec{z}}\}_{\vec{z}}: \mathbf{proc}\ \forall \vec{\imath}\,(\mathbf{in}\ \vec{\tau}^\diamond; \mathbf{out}\ \vec{\sigma}^\diamond)}$$

  and, for any $\Omega$, $\Gamma^\diamond; \Omega \vdash \mathbf{proc}\ (\mathbf{in}\ \vec{x}; \mathbf{out}\ \vec{z})\ \{t^\diamond_{\vec{z}}\}_{\vec{z}}: \mathbf{proc}\ \forall \vec{\imath}\,(\mathbf{in}\ \vec{\tau}^\diamond; \mathbf{out}\ \vec{\sigma}^\diamond)$, by weakening (Lemma C.3).

- $(\vec{w})^\diamond_{\vec{r}} = \vec{r} := \vec{w}^\diamond;$
  $$\frac{\Gamma \vdash w_1: \sigma_1[\vec{n}/\vec{\imath}] \quad ... \quad \Gamma \vdash w_m: \sigma_m[\vec{n}/\vec{\imath}]}{\Gamma \vdash \vec{w}: \exists \vec{\imath}.\vec{\sigma}}$$

  Indeed, by induction hypothesis, $\Gamma \vdash w_i: \sigma_i[\vec{n}/\vec{\imath}]$ implies $\Gamma^\diamond; \Omega \vdash w_i^\diamond: \sigma_i^\diamond[\vec{n}/\vec{\imath}]$ for any $\Omega$, hence $\Gamma^\diamond; \vec{r}: \vec{\sigma}\,' \vdash w_i^\diamond: \sigma_i^\diamond[\vec{n}/\vec{\imath}]$. Then
  $$\frac{\Gamma^\diamond; \vec{r}: \vec{\sigma}\,' \vdash \vec{w}^\diamond: \vec{\sigma}^\diamond[\vec{n}/\vec{\imath}] \quad \Gamma^\diamond; \vec{r}: \vec{\sigma}^\diamond[\vec{n}/\vec{\imath}] \vdash \varepsilon \rhd \vec{r}: \exists \vec{\imath}.\vec{\sigma}^\diamond}{\Gamma^\diamond; \vec{r}: \vec{\sigma}\,' \vdash \vec{r} := \vec{w}^\diamond; \rhd \vec{r}: \exists \vec{\imath}.\vec{\sigma}^\diamond}$$

- $v^\diamond = v^\diamond$
  $$\frac{\Gamma \vdash v: \tau[n/i] \quad \Gamma \vdash v': (n=m)}{\Gamma \vdash v: \tau[m/i]}$$

  Indeed, by induction hypothesis, $\Gamma \vdash v: \tau[n/i]$ implies $\Gamma^\diamond; \Omega \vdash v^\diamond: \tau^\diamond[n/i]$ and $\Gamma \vdash v': (n=m)$ implies $\Gamma^\diamond; \Omega \vdash v'^\diamond: (n=m)$ for any $\Omega$, then
  $$\frac{\Gamma^\diamond; \Omega \vdash v^\diamond: \tau^\diamond[n/i] \quad \Gamma^\diamond; \Omega \vdash v'^\diamond: (n=m)}{\Gamma^\diamond; \Omega \vdash v^\diamond: \tau^\diamond[m/i]}$$

- $(t)^\diamond_{\vec{r}} = (t)^\diamond_{\vec{r}}$
  $$\frac{\Gamma \vdash t: \exists \vec{\jmath}.\vec{\sigma}[n/i] \quad \Gamma \vdash v': (n=m)}{\Gamma \vdash t: \exists \vec{\jmath}.\vec{\sigma}[m/i]}$$

  Indeed, by induction hypothesis, $\Gamma \vdash t: \vec{\sigma}[n/i]$ implies $\Gamma^\diamond; \vec{r}: \vec{\tau}\,' \vdash (t)^\diamond_{\vec{r}} \rhd \vec{r}: \exists \vec{\jmath}.\vec{\sigma}^\diamond[n/i]$ for any $\vec{\tau}\,'$ ; $\Gamma \vdash v'$: $(n=m)$ implies $\Gamma^\diamond; \Omega \vdash v'^\diamond: (n=m)$ for any $\Omega$, hence $\Gamma^\diamond; \vec{r}: \vec{\tau}\,' \vdash v'^\diamond: (n=m)$ ; then
  $$\frac{\Gamma^\diamond; \vec{r}: \vec{\tau}\,' \vdash (t)^\diamond_{\vec{r}} \rhd \vec{r}: \exists \vec{\jmath}.\vec{\sigma}^\diamond[n/i] \quad \Gamma^\diamond; \vec{r}: \vec{\tau}\,' \vdash v'^\diamond: (n=m)}{\Gamma^\diamond; \vec{r}: \vec{\tau}\,' \vdash (t)^\diamond_{\vec{r}} \rhd \vec{r}: \exists \vec{\jmath}.\vec{\sigma}^\diamond[m/i]}$$

- $(\mathbf{let}\ y = w\ \mathbf{in}\ u)^\diamond_{\vec{r}} = \mathbf{cst}\ y = w^\diamond;\ (u)^\diamond_{\vec{r}}$
  $$\frac{\Gamma \vdash w: \tau \quad \Gamma, y: \tau \vdash u: \exists \vec{\imath}.\vec{\sigma}}{\Gamma \vdash \mathbf{let}\ y = w\ \mathbf{in}\ u: \exists \vec{\imath}.\vec{\sigma}}$$
  Indeed, by induction hypothesis:
  - $\Gamma \vdash w: \tau$ implies $\Gamma^\diamond; \Omega \vdash w^\diamond: \tau^\diamond$ for any $\Omega$, hence $\Gamma^\diamond; \vec{r}: \vec{\sigma}\,' \vdash w^\diamond: \tau^\diamond$ ;
  - $\Gamma, y: \tau \vdash u: \exists \vec{\imath}.\vec{\sigma}$ implies $\Gamma^\diamond, y: \tau^\diamond; \vec{r}: \vec{\sigma}\,' \vdash u^\diamond_{\vec{r}} \rhd \exists \vec{\imath}.\vec{r}: \vec{\sigma}^\diamond$.

  Then
  $$\frac{\Gamma^\diamond; \vec{r}: \vec{\sigma}\,' \vdash w^\diamond: \tau^\diamond \quad \Gamma^\diamond, y: \tau^\diamond; \vec{r}: \vec{\sigma}\,' \vdash u^\diamond_{\vec{r}} \rhd \exists \vec{\imath}.\vec{r}: \vec{\sigma}^\diamond}{\Gamma^\diamond; \vec{r}: \vec{\sigma}\,' \vdash \mathbf{cst}\ y = w^\diamond;\ (u)^\diamond_{\vec{r}} \rhd \exists \vec{\imath}.\vec{r}: \vec{\sigma}^\diamond}$$

- $(\textbf{let } y = \textbf{succ}(w) \textbf{ in } u)^{\diamond}_{\vec r} \;=\; \textbf{var } z := w^{\diamond};\ \textbf{inc}(z);\ \textbf{cst } y = z;\ (u)^{\diamond}_{\vec r}$

$$\dfrac{\dfrac{\Gamma \vdash \textbf{succ}\colon \forall x(\textbf{nat}(x) \Rightarrow \textbf{nat}(\textbf{s}(x))) \quad \Gamma \vdash w\colon \textbf{nat}(n)}{\Gamma \vdash \textbf{succ}(w)\colon \textbf{nat}(\textbf{s}(n))} \qquad \Gamma, y\colon \textbf{nat}(\textbf{s}(n))) \vdash u\colon \exists \vec\imath.\vec\sigma}{\Gamma \vdash \textbf{let } y = \textbf{succ}(w) \textbf{ in } u\colon \exists \vec\imath.\vec\sigma}$$

Indeed, by induction hypothesis:

  - $\Gamma \vdash w\colon \textbf{nat}(n)$ implies $\Gamma^{\diamond}; \Omega \vdash w^{\diamond}\colon \textbf{nat}(n)$ for any $\Omega$, hence $\Gamma^{\diamond}; \vec r : \vec\sigma\,' \vdash w^{\diamond}\colon \textbf{nat}(n)$ ;
  - $\Gamma, y\colon \textbf{nat}(\textbf{s}(n)) \vdash u\colon \exists\vec\imath.\vec\sigma$ implies $\Gamma, y\colon \textbf{nat}(\textbf{s}(n)); \vec r : \vec\sigma\,' \vdash u \rhd \exists\vec\imath.\vec r : \vec\sigma^{\diamond}$, and, by Lemma C.3, $\Gamma^{\diamond}, y\colon \textbf{nat}(\textbf{s}(n)); \vec r : \vec\sigma\,', z\colon \textbf{nat}(\textbf{s}(n)) \vdash (u)^{\diamond}_{\vec r} \rhd \exists\vec\imath.\vec r : \vec\sigma^{\diamond}$.

Then

$$\dfrac{\dfrac{\Gamma^{\diamond}, y\colon \textbf{nat}(\textbf{s}(n)); \vec r : \vec\sigma\,', z\colon \textbf{nat}(\textbf{s}(n)) \vdash (u)^{\diamond}_{\vec r} \rhd \exists\vec\imath.\vec r : \vec\sigma^{\diamond}}{\Gamma^{\diamond}; \vec r : \vec\sigma\,', z\colon \textbf{nat}(\textbf{s}(n)) \vdash \textbf{cst } y = z;\ (u)^{\diamond}_{\vec r} \rhd \exists\vec\imath.\vec r : \vec\sigma^{\diamond}}}{\Gamma^{\diamond}; \vec r : \vec\sigma\,', z\colon \textbf{nat}(n) \vdash \textbf{inc}(z);\ \textbf{cst } y = z;\ (u)^{\diamond}_{\vec r} \rhd \exists\vec\imath.\vec r : \vec\sigma^{\diamond}}$$

and

$$\dfrac{\Gamma^{\diamond}; \vec r : \vec\sigma\,' \vdash w^{\diamond}\colon \textbf{nat}(n) \quad \Gamma^{\diamond}; \vec r : \vec\sigma\,', z\colon \textbf{nat}(n) \vdash \textbf{inc}(z);\ \textbf{cst } y = z;\ (u)^{\diamond}_{\vec r} \rhd \exists\vec\imath.\vec r : \vec\sigma^{\diamond}}{\Gamma^{\diamond}; \vec r : \vec\sigma\,' \vdash \textbf{var } z := w^{\diamond};\ \textbf{inc}(z);\ \textbf{cst } y = z;\ (u)^{\diamond}_{\vec r} \rhd \exists\vec\imath.\vec r : \vec\sigma^{\diamond}}$$

- $(\textbf{let } y = \textbf{pred}(w) \textbf{ in } u)^{\diamond}_{\vec r} \;=\; \textbf{var } z := w^{\diamond};\ \textbf{dec}(z);\ \textbf{cst } y = z;\ (u)^{\diamond}_{\vec r}$

$$\dfrac{\dfrac{\Gamma \vdash w\colon \textbf{nat}(n)}{\Gamma \vdash \textbf{pred}(w)\colon \textbf{nat}(\textbf{p}(n))} \qquad \Gamma, y\colon \textbf{nat}(\textbf{p}(n)) \vdash u\colon \exists\vec\imath.\vec\sigma}{\Gamma \vdash \textbf{let } y = \textbf{succ}(w) \textbf{ in } u\colon \exists\vec\imath.\vec\sigma}$$

Indeed, by induction hypothesis:

  - $\Gamma \vdash w\colon \textbf{nat}(n)$ implies $\Gamma^{\diamond}; \Omega \vdash w^{\diamond}\colon \textbf{nat}(n)$ for any $\Omega$, hence $\Gamma^{\diamond}; \vec r : \vec\sigma\,' \vdash w^{\diamond}\colon \textbf{nat}(n)$ ;
  - $\Gamma, y\colon \textbf{nat}(\textbf{p}(n)) \vdash u\colon \exists\vec\imath.\vec\sigma$ implies $\Gamma, y\colon \textbf{nat}(\textbf{p}(n)); \vec r : \vec\sigma\,' \vdash u \rhd \exists\vec\imath.\vec r : \vec\sigma^{\diamond}$, and, by Lemma C.3, $\Gamma^{\diamond}, y\colon \textbf{nat}(\textbf{p}(n)); \vec r : \vec\sigma\,', z\colon \textbf{nat}(\textbf{p}(n)) \vdash (u)^{\diamond}_{\vec r} \rhd \exists\vec\imath.\vec r : \vec\sigma^{\diamond}$.

Then

$$\dfrac{\dfrac{\Gamma^{\diamond}, y\colon \textbf{nat}(\textbf{p}(n)); \vec r : \vec\sigma\,', z\colon \textbf{nat}(\textbf{p}(n)) \vdash (u)^{\diamond}_{\vec r} \rhd \exists\vec\imath.\vec r : \vec\sigma^{\diamond}}{\Gamma^{\diamond}; \vec r : \vec\sigma\,', z\colon \textbf{nat}(\textbf{p}(n)) \vdash \textbf{cst } y = z;\ (u)^{\diamond}_{\vec r} \rhd \exists\vec\imath.\vec r : \vec\sigma^{\diamond}}}{\Gamma^{\diamond}; \vec r : \vec\sigma\,', z\colon \textbf{nat}(n) \vdash \textbf{pred}(z);\ \textbf{cst } y = z;\ (u)^{\diamond}_{\vec r} \rhd \exists\vec\imath.\vec r : \vec\sigma^{\diamond}}$$

and

$$\dfrac{\Gamma^{\diamond}; \vec r : \vec\sigma\,' \vdash w^{\diamond}\colon \textbf{nat}(n) \quad \Gamma^{\diamond}; \vec r : \vec\sigma\,', z\colon \textbf{nat}(n) \vdash \textbf{pred}(z);\ \textbf{cst } y = z;\ (u)^{\diamond}_{\vec r} \rhd \exists\vec\imath.\vec r : \vec\sigma^{\diamond}}{\Gamma^{\diamond}; \vec r : \vec\sigma\,' \vdash \textbf{var } z := w^{\diamond};\ \textbf{pred}(z);\ \textbf{cst } y = z;\ (u)^{\diamond}_{\vec r} \rhd \exists\vec\imath.\vec r : \vec\sigma^{\diamond}}$$

- $(\textbf{let } \vec x = \textbf{rec}(w, \vec w, \lambda i.\lambda\vec y.t) \textbf{ in } u)^{\diamond}_{\vec r} \;=\; \textbf{var } \vec z := \vec w;\ \textbf{for } i := 0 \textbf{ until } w^{\diamond}\ \{\textbf{cst } \vec y = \vec z;\ t^{\diamond}_{\vec z}\}_{\vec z};\ \textbf{cst } \vec x = \vec z;\ (u)^{\diamond}_{\vec r}$

$$\dfrac{\dfrac{\Gamma \vdash w\colon \textbf{nat}(n) \quad \Gamma \vdash \vec w\colon \vec\tau[0/j] \quad \Gamma, i\colon \textbf{nat}(j), \vec y : \vec\tau \vdash t\colon \vec\tau[\textbf{s}(j)/j]}{\Gamma \vdash \textbf{rec}(w, \vec w, \lambda i.\lambda\vec y.t)\colon \vec\tau[n/j]} \qquad \Gamma, \vec x : \vec\tau[n/j] \vdash u\colon \exists\vec\imath.\vec\sigma}{\Gamma \vdash \textbf{let } \vec x = \textbf{rec}(w, \vec w, \lambda i.\lambda\vec y.t) \textbf{ in } u\colon \exists\vec\imath.\vec\sigma}$$

with $j \notin \mathcal{FV}(\Gamma)$. Indeed, by induction hypothesis:

  - $\Gamma \vdash w\colon \textbf{nat}(n)$ implies $\Gamma^{\diamond}; \Omega \vdash w^{\diamond}\colon \textbf{nat}(n)$ for any $\Omega$, hence $\Gamma^{\diamond}; \vec r : \vec\sigma\,', \vec z : \vec\tau^{\diamond}[0/j] \vdash w^{\diamond}\colon \textbf{nat}(n)$ ;
  - $\Gamma \vdash \vec w\colon \vec\tau[0/j]$ implies $\Gamma^{\diamond}; \Omega \vdash \vec w^{\diamond}\colon \vec\tau^{\diamond}[0/j]$ for any $\Omega$, hence $\Gamma^{\diamond}; \vec r : \vec\sigma\,' \vdash \vec w^{\diamond}\colon \vec\tau^{\diamond}[0/j]$ ;
  - $\Gamma, \vec x : \vec\tau[n/j] \vdash u\colon \vec\sigma$ implies $\Gamma^{\diamond}, \vec x : \vec\tau^{\diamond}[n/j]; \vec r : \vec\sigma\,' \vdash (u)^{\diamond}_{\vec r} \rhd \exists\vec\imath.\vec r : \vec\sigma^{\diamond}$, and, by Lemma C.3, $\Gamma^{\diamond}, \vec x : \vec\tau^{\diamond}[n/j]; \vec r : \vec\sigma\,', \vec z : \vec\tau^{\diamond}[n/j] \vdash (u)^{\diamond}_{\vec r} \rhd \exists\vec\imath.\vec r : \vec\sigma^{\diamond}$ ;
  - $\Gamma, i\colon \textbf{nat}(j), \vec y : \vec\tau \vdash t\colon \vec\tau[\textbf{s}(j)/j]$ implies $\Gamma^{\diamond}, i\colon \textbf{nat}(j), \vec y : \vec\tau^{\diamond}; \vec z : \vec\tau\,' \vdash t^{\diamond}_{\vec z} \rhd \vec z : \vec\tau^{\diamond}[\textbf{s}(j)/j]$ for any $\vec\tau\,'$, hence $\Gamma^{\diamond}, i\colon \textbf{nat}(j), \vec y : \vec\tau^{\diamond}; \vec z : \vec\tau^{\diamond} \vdash t^{\diamond}_{\vec z} \rhd \vec z : \vec\tau^{\diamond}[\textbf{s}(j)/j]$.

Then

$$\pi_1 = \frac{\Gamma^\diamond, i\colon \mathbf{nat}(j), \vec{y}: \vec{\tau}^\diamond; \vec{z}: \vec{\tau}^\diamond \vdash t^\diamond_{\vec{z}} \rhd \vec{z}: \vec{\tau}^\diamond[\mathbf{s}(j)/j]}{\Gamma^\diamond, i\colon \mathbf{nat}(j); \vec{z}: \vec{\tau}^\diamond \vdash \mathbf{cst}\ \vec{y} = \vec{z};\ t^\diamond_{\vec{z}} \rhd \vec{z}: \vec{\tau}^\diamond[\mathbf{s}(j)/j]}$$

and

$$\pi_2 = \frac{\Gamma^\diamond, \vec{x}: \vec{\tau}^\diamond[n/j]; \vec{r}: \vec{\sigma}', \vec{z}: \vec{\tau}^\diamond[n/j] \vdash (u)^\diamond_{\vec{r}} \rhd \exists \vec{\imath}.\vec{r}: \vec{\sigma}^\diamond}{\Gamma^\diamond; \vec{r}: \vec{\sigma}', \vec{z}: \vec{\tau}^\diamond[n/j] \vdash \mathbf{cst}\ \vec{x} = \vec{z};\ (u)^\diamond_{\vec{r}} \rhd \exists \vec{\imath}.\vec{r}: \vec{\sigma}^\diamond}$$

and

$$\pi = \frac{\Gamma^\diamond; \vec{r}: \vec{\sigma}', \vec{z}: \vec{\tau}^\diamond[0/j] \vdash w^\diamond\colon \mathbf{nat}(n) \qquad \pi_1 \qquad \pi_2}{\Gamma^\diamond; \vec{r}: \vec{\sigma}', \vec{z}: \vec{\tau}^\diamond[0/j] \vdash \mathbf{for}\ i := 0\ \mathbf{until}\ w^\diamond\ \{\mathbf{cst}\ \vec{y} = \vec{z};\ t^\diamond_{\vec{z}}\}_{\vec{z}};\ \mathbf{cst}\ \vec{x} = \vec{z};\ (u)^\diamond_{\vec{r}} \rhd \vec{r}: \vec{\sigma}^\diamond}$$

since $j \notin \mathcal{FV}(\Gamma^\diamond)$, and finally

$$\frac{\Gamma^\diamond; \vec{r}: \vec{\sigma}' \vdash \vec{w}^\diamond\colon \vec{\tau}^\diamond[0/j] \qquad \pi}{\Gamma^\diamond; \vec{r}: \vec{\sigma}' \vdash \mathbf{var}\ \vec{z} := \vec{w};\ \mathbf{for}\ i := 0\ \mathbf{until}\ w^\diamond\ \{\mathbf{cst}\ \vec{y} = \vec{z};\ t^\diamond_{\vec{z}}\}_{\vec{z}};\ \mathbf{cst}\ \vec{x} = \vec{z};\ (u)^\diamond_{\vec{r}} \rhd \exists \vec{\imath}.\vec{r}: \vec{\sigma}^\diamond}$$

- $(\mathbf{let}\ \vec{x} = w\ \vec{w}\ \mathbf{in}\ u)^\diamond_{\vec{r}} \ = \ \mathbf{var}\ \vec{z};\ w^\diamond(\vec{w}^\diamond; \vec{z});\ \mathbf{cst}\ \vec{x} = \vec{z};\ (u)^\diamond_{\vec{r}}$

$$\frac{\dfrac{\Gamma \vdash w\colon \forall \vec{\imath}\,(\vec{\tau} \Rightarrow \exists \vec{\jmath}\,(\vec{\tau}')) \quad \Gamma \vdash \vec{w}\colon \vec{\tau}[\vec{n}/\vec{\imath}]}{\Gamma \vdash w\ \vec{w}\colon \exists \vec{\jmath}\,(\vec{\tau}'[\vec{n}/\vec{\imath}])} \qquad \Gamma, \vec{x}: \vec{\tau}'[\vec{n}/\vec{\imath}] \vdash u\colon \exists \vec{\kappa}.\vec{\sigma}}{\Gamma \vdash \mathbf{let}\ \vec{x} = w\ \vec{w}\ \mathbf{in}\ u\colon \exists \vec{\kappa}.\vec{\sigma}}$$

with $\vec{\jmath} \notin \mathcal{FV}(\Gamma, \vec{\sigma})$. Indeed, by induction hypothesis:

  - $\Gamma \vdash w\colon \forall \vec{\imath}\,(\vec{\tau} \Rightarrow \exists \vec{\jmath}\,(\vec{\tau}'))$ implies $\Gamma^\diamond; \Omega \vdash w^\diamond\colon \mathbf{proc}\ \forall \vec{\imath}\,(\mathbf{in}\ \vec{\tau}^\diamond; \exists \vec{\jmath}\ \mathbf{out}\ \vec{\tau}'^\diamond)$ for any $\Omega$, hence $\Gamma^\diamond; \vec{r}: \vec{\sigma}'$, $\vec{z}: \vec{\top} \vdash w^\diamond\colon \mathbf{proc}\ \forall \vec{\imath}\,(\mathbf{in}\ \vec{\tau}^\diamond; \exists \vec{\jmath}\ \mathbf{out}\ \vec{\tau}'^\diamond)$ ;
  - $\Gamma \vdash \vec{w}\colon \vec{\tau}[\vec{n}/\vec{\imath}]$ implies $\Gamma^\diamond; \Omega \vdash \vec{w}^\diamond\colon \vec{\tau}^\diamond[\vec{n}/\vec{\imath}]$ for any $\Omega$, hence $\Gamma^\diamond; \vec{r}: \vec{\sigma}', \vec{z}: \vec{\top} \vdash \vec{w}^\diamond\colon \vec{\tau}^\diamond[\vec{n}/\vec{\imath}]$ ;
  - $\Gamma, \vec{x}: \vec{\tau}'[\vec{n}/\vec{\imath}] \vdash u\colon \exists \vec{\kappa}.\vec{\sigma}$ implies $\Gamma^\diamond, \vec{x}: \vec{\tau}'^\diamond[\vec{n}/\vec{\imath}]; \vec{r}: \vec{\sigma}' \vdash (u)^\diamond_{\vec{r}} \rhd \exists \vec{\kappa}.\vec{r}: \vec{\sigma}^\diamond$, and, by Lemma C.3, $\Gamma^\diamond, \vec{x}: \vec{\tau}'^\diamond[\vec{n}/\vec{\imath}]; \vec{r}: \vec{\sigma}', \vec{z}: \vec{\tau}'^\diamond[\vec{n}/\vec{\imath}] \vdash (u)^\diamond_{\vec{r}} \rhd \exists \vec{\kappa}.\vec{r}: \vec{\sigma}^\diamond$.

Then

$$\pi = \frac{\Gamma^\diamond, \vec{x}: \vec{\tau}'^\diamond[\vec{n}/\vec{\imath}]; \vec{r}: \vec{\sigma}', \vec{z}: \vec{\tau}'^\diamond[\vec{n}/\vec{\imath}] \vdash (u)^\diamond_{\vec{r}} \rhd \exists \vec{\kappa}.\vec{r}: \vec{\sigma}^\diamond}{\Gamma^\diamond; \vec{r}: \vec{\sigma}', \vec{z}: \vec{\tau}'^\diamond[\vec{n}/\vec{\imath}] \vdash \mathbf{cst}\ \vec{x} = \vec{z};\ (u)^\diamond_{\vec{r}} \rhd \exists \vec{\kappa}.\vec{r}: \vec{\sigma}^\diamond}$$

and

$$\frac{\dfrac{\Gamma^\diamond; \vec{r}: \vec{\sigma}', \vec{z}: \vec{\top} \vdash w^\diamond\colon \mathbf{proc}\ \forall \vec{\imath}\,(\mathbf{in}\ \vec{\tau}^\diamond; \exists \vec{\jmath}\ \mathbf{out}\ \vec{\tau}'^\diamond) \quad \Gamma^\diamond; \vec{r}: \vec{\sigma}', \vec{z}: \vec{\top} \vdash \vec{w}^\diamond\colon \vec{\tau}^\diamond[\vec{n}/\vec{\imath}] \quad \pi}{\Gamma^\diamond; \vec{r}: \vec{\sigma}', \vec{z}: \vec{\top} \vdash w^\diamond(\vec{w}^\diamond; \vec{z});\ \mathbf{cst}\ \vec{x} = \vec{z};\ (u)^\diamond_{\vec{r}} \rhd \exists \vec{\kappa}.\vec{r}: \vec{\sigma}^\diamond}}{\Gamma^\diamond; \vec{r}: \vec{\sigma}' \vdash \mathbf{var}\ \vec{z};\ w^\diamond(\vec{w}^\diamond; \vec{z});\ \mathbf{cst}\ \vec{x} = \vec{z};\ (u)^\diamond_{\vec{r}} \rhd \exists \vec{\kappa}.\vec{r}: \vec{\sigma}^\diamond}$$

since $\vec{\jmath} \notin \mathcal{FV}(\Gamma^\diamond, \vec{\sigma}', \vec{\sigma})$.

- $(\mathbf{let}\ \vec{x} = t\ \mathbf{in}\ u)^\diamond_{\vec{r}} \ = \ \mathbf{var}\ \vec{z};\ \{(t)^\diamond_{\vec{z}}\}_{\vec{z}};\ \mathbf{cst}\ \vec{x} = \vec{z};\ (u)^\diamond_{\vec{r}}$

$$\frac{\Gamma \vdash t\colon \exists \vec{\kappa}.\vec{\tau} \quad \Gamma, \vec{x}: \vec{\tau} \vdash u\colon \exists \vec{\imath}.\vec{\sigma}}{\Gamma \vdash \mathbf{let}\ \vec{x} = t\ \mathbf{in}\ u\colon \exists \vec{\imath}.\vec{\sigma}}$$

Indeed, by induction hypothesis:

  - $\Gamma \vdash t\colon \exists \vec{\kappa}.\vec{\tau}$ implies $\Gamma^\diamond;\ \vec{z}: \vec{\tau}' \vdash (t)^\diamond_{\vec{z}} \rhd \exists \vec{\kappa}.\vec{z}: \vec{\tau}^\diamond$ for any $\vec{\tau}'$, hence, by Lemma C.3 $\Gamma^\diamond;\ \vec{r}: \vec{\sigma}',\ \vec{z}: \vec{\top} \vdash (t)^\diamond_{\vec{z}} \rhd \exists \vec{\kappa}.\vec{z}: \vec{\tau}^\diamond$ ;
  - $\Gamma, \vec{x}: \vec{\tau} \vdash u\colon \exists \vec{\imath}.\vec{\sigma}$ implies $\Gamma^\diamond, \vec{x}: \vec{\tau}^\diamond; \vec{r}: \vec{\sigma}' \vdash (u)^\diamond_{\vec{r}} \rhd \exists \vec{\imath}.\vec{r}: \vec{\sigma}^\diamond$, and, by Lemma C.3, $\Gamma^\diamond, \vec{x}: \vec{\tau}^\diamond; \vec{r}: \vec{\sigma}', \vec{z}: \vec{\tau}^\diamond \vdash (u)^\diamond_{\vec{r}} \rhd \exists \vec{\imath}.\vec{r}: \vec{\sigma}^\diamond$.

Then

$$\frac{\dfrac{\Gamma^\diamond; \vec{r}: \vec{\sigma}', \vec{z}: \vec{\top} \vdash (t)^\diamond_{\vec{z}} \rhd \exists \vec{\kappa}.\vec{z}: \vec{\tau}^\diamond \qquad \dfrac{\Gamma^\diamond, \vec{x}: \vec{\tau}^\diamond; \vec{r}: \vec{\sigma}', \vec{z}: \vec{\tau}^\diamond \vdash (u)^\diamond_{\vec{r}} \rhd \exists \vec{\imath}.\vec{r}: \vec{\sigma}^\diamond}{\Gamma^\diamond; \vec{r}: \vec{\sigma}', \vec{z}: \vec{\tau}^\diamond \vdash \mathbf{cst}\ \vec{x} = \vec{z};\ (u)^\diamond_{\vec{r}} \rhd \exists \vec{\imath}.\vec{r}: \vec{\sigma}^\diamond}}{\Gamma^\diamond; \vec{r}: \vec{\sigma}', \vec{z}: \vec{\top} \vdash \{(t)^\diamond_{\vec{z}}\}_{\vec{z}};\ \mathbf{cst}\ \vec{x} = \vec{z};\ (u)^\diamond_{\vec{r}} \rhd \exists \vec{\imath}.\vec{r}: \vec{\sigma}^\diamond}}{\Gamma^\diamond; \vec{r}: \vec{\sigma}' \vdash \mathbf{var}\ \vec{z};\ \{(t)^\diamond_{\vec{z}}\}_{\vec{z}};\ \mathbf{cst}\ \vec{x} = \vec{z};\ (u)^\diamond_{\vec{r}} \rhd \exists \vec{\imath}.\vec{r}: \vec{\sigma}^\diamond}$$

□

## C.5 Expressiveness

**Definition C.7.** *The translation of a type* $\tau \in \Sigma_{\mathbf{FD}^c}$ *into a type* $\tau^\natural \in \Sigma_{\mathbf{FD}^c}$ *is defined by the following rules:*

$$\begin{array}{rcl} \mathbf{nat}(n)^\natural & = & \mathbf{nat}(n) \\ (n=m)^\natural & = & (n=m) \\ (\forall \vec{\imath}\,(\sigma \Rightarrow \tau))^\natural & = & \forall \vec{\imath}\,(\sigma^\natural \Rightarrow \tau^\natural) \\ (\exists \vec{\imath}\,(\tau_1 \wedge \ldots \wedge \tau_n))^\natural & = & \exists \vec{\imath}\,(\tau_1^\natural \wedge \ldots \wedge \tau_n^\natural) \\ \\ \bot^\natural & = & \bot \end{array}$$

**Proposition C.8.** *For any functional term* $t$*, if* $\Gamma \vdash t\colon \tau$ *in* $\mathbf{FD}^c$ *then* $\Gamma^\natural \vdash t^\natural\colon \tau^\natural$ *in* $\mathbf{FD}^c$.

**Proof.** Straightforward induction on $t$. □

## C.6 CPS translation

**Lemma C.9.** *The following typing rules are derivable in* **FD**:

$$\frac{\Gamma \vdash u\colon \varphi}{\Gamma \vdash \mathbf{val}\ u\colon \nabla\varphi} \qquad \frac{\Gamma \vdash u\colon \nabla\varphi \qquad \Gamma, x\colon \varphi \vdash t\colon \nabla\psi}{\Gamma \vdash \mathbf{let\ val}\ x = u\ \mathbf{in}\ t\colon \nabla\psi}$$

**Proof.** Indeed,

$$\dfrac{\dfrac{\dfrac{}{\Gamma, z\colon \varphi \Rightarrow o \vdash z\colon \varphi \Rightarrow o} \qquad \Gamma \vdash u\colon \varphi}{\Gamma, z\colon \varphi \Rightarrow o \vdash z\ u\colon o}}{\Gamma \vdash \lambda z.(z\ u)\colon (\varphi \Rightarrow o) \Rightarrow o}$$

and,

$$\dfrac{\dfrac{\dfrac{\Gamma \vdash u\colon (\varphi \Rightarrow o) \Rightarrow o}{\Gamma, z\colon \psi \Rightarrow o \vdash u\colon (\varphi \Rightarrow o) \Rightarrow o} \qquad \dfrac{\dfrac{\dfrac{\Gamma, x\colon \varphi \vdash t\colon (\psi \Rightarrow o) \Rightarrow o}{\Gamma, z\colon \psi \Rightarrow o, x\colon \varphi \vdash t\colon (\psi \Rightarrow o) \Rightarrow o} \qquad \dfrac{}{\Gamma, z\colon \psi \Rightarrow o, x\colon \varphi \vdash z\colon \psi \Rightarrow o}}{\Gamma, z\colon \psi \Rightarrow o, x\colon \varphi \vdash (t\ z)\colon o}}{\Gamma, z\colon \psi \Rightarrow o \vdash \lambda x.(t\ z)\colon \varphi \Rightarrow o}}{\Gamma, z\colon \psi \Rightarrow o \vdash (u\ \lambda x.(t\ z))\colon o}}{\Gamma \vdash \lambda z.(u\ \lambda x.(t\ z))\colon (\psi \Rightarrow o) \Rightarrow o}$$

□

**Lemma C.10.** *Abbreviations callcc and throw are typable in* **FD** *as follows:*

$$\begin{array}{rcl} \mathit{callcc} & : & ((\varphi^\circ \Rightarrow o) \Rightarrow \nabla\varphi^\circ) \Rightarrow \nabla\varphi^\circ \\ \mathit{throw} & : & ((\varphi^\circ \Rightarrow o) \wedge \varphi^\circ) \Rightarrow \nabla\psi^\circ \end{array}$$

**Proof.** Indeed (with $\Gamma' = \Gamma, h\colon (\varphi^\circ \Rightarrow o) \Rightarrow ((\varphi^\circ \Rightarrow o) \Rightarrow o), k\colon \varphi^\circ \Rightarrow o$),

$$\dfrac{\dfrac{\dfrac{\dfrac{\dfrac{}{\Gamma' \vdash h\colon (\varphi^\circ \Rightarrow o) \Rightarrow ((\varphi^\circ \Rightarrow o) \Rightarrow o)} \qquad \dfrac{}{\Gamma' \vdash k\colon \varphi^\circ \Rightarrow o}}{\Gamma' \vdash (h\ k)\colon (\varphi^\circ \Rightarrow o) \Rightarrow o} \qquad \dfrac{}{\Gamma' \vdash k\colon \varphi^\circ \Rightarrow o}}{\Gamma' \vdash (h\ k\ k)\colon o}}{\Gamma, h\colon (\varphi^\circ \Rightarrow o) \Rightarrow ((\varphi^\circ \Rightarrow o) \Rightarrow o) \vdash \lambda k.(h\ k\ k)\colon (\varphi^\circ \Rightarrow o) \Rightarrow o}}{\Gamma \vdash \lambda h.\lambda k.(h\ k\ k)\colon ((\varphi^\circ \Rightarrow o) \Rightarrow ((\varphi^\circ \Rightarrow o) \Rightarrow o)) \Rightarrow (\varphi^\circ \Rightarrow o) \Rightarrow o}$$

and,

$$
\dfrac{\dfrac{\dfrac{\dfrac{}{\Gamma, k\colon \varphi^\circ \Rightarrow o, a\colon \varphi^\circ, k'\colon \psi^\circ \Rightarrow o \vdash k\colon \varphi^\circ \Rightarrow o} \qquad \dfrac{}{\Gamma, k\colon \varphi^\circ \Rightarrow o, a\colon \varphi^\circ, k'\colon \psi^\circ \Rightarrow o \vdash a\colon \varphi^\circ}}{\Gamma, k\colon \varphi^\circ \Rightarrow o, a\colon \varphi^\circ, k'\colon \psi^\circ \Rightarrow o \vdash k\ a\colon o}}{\Gamma, k\colon \varphi^\circ \Rightarrow o, a\colon \varphi^\circ \vdash \lambda k'.(k\ a)\colon (\psi^\circ \Rightarrow o) \Rightarrow o}}{\Gamma \vdash \lambda(k, a).\lambda k'.(k\ a)\colon ((\varphi^\circ \Rightarrow o) \wedge \varphi^\circ) \Rightarrow (\psi^\circ \Rightarrow o) \Rightarrow o}
$$

□

## C.7 Labels and jumps

**Proposition.** *The following typing rules are derivable.*

$$
\dfrac{\Gamma, k\colon \neg\vec{\sigma}; \vec{z} : \vec{\tau} \vdash s \rhd \vec{z} : \vec{\sigma} \qquad \Gamma; \Omega, \vec{z} : \vec{\sigma} \vdash s' \rhd \Omega'}{\Gamma; \Omega, \vec{z} : \vec{\tau} \vdash k\colon \{s\}_{\vec{z}};\ \ s' \rhd \Omega'}
$$

$$
\dfrac{\Gamma; \Omega, \vec{z} : \vec{\tau} \vdash k\colon \neg\vec{\sigma} \qquad \Gamma; \Omega, \vec{z} : \vec{\tau} \vdash \vec{e} : \vec{\sigma}}{\Gamma; \Omega, \vec{z} : \vec{\tau} \vdash \mathbf{jump}(k, \vec{e}\,)_{\vec{z}} \rhd \Omega', \vec{z} : \vec{\omega}\,'}
$$

**Proof.** Indeed, given the type of **callcc** and **throw**, we have

$$
\dfrac{\dfrac{\dfrac{\dfrac{\dfrac{\Gamma, k\colon \neg\vec{\sigma}; \vec{z} : \vec{\tau} \vdash s \rhd \vec{z} : \vec{\sigma}}{\Gamma, \vec{z}\,'\colon \vec{\tau}, k\colon \neg\vec{\sigma}; \vec{z} : \vec{\tau} \vdash s \rhd \vec{z} : \vec{\sigma}}}{\Gamma, \vec{z}\,'\colon \vec{\tau}, k\colon \neg\vec{\sigma}; \vec{z} : \vec{\top} \vdash \vec{z} := \vec{z}\,';\ \ s \rhd \vec{z} : \vec{\sigma}}}{\Gamma, \vec{z}\,'\colon \vec{\tau}; \Omega, \vec{z} : \vec{\tau} \vdash \mathbf{proc}\ (\mathbf{in}\ k; \mathbf{out}\ \vec{z}\,)\ \{\vec{z} := \vec{z}\,';\ \ s\}_{\vec{z}}\colon \mathbf{proc}\ (\mathbf{in}\ \neg\vec{\sigma}; \mathbf{out}\ \vec{\sigma}\,)} \qquad \Gamma; \Omega, \vec{z} : \vec{\sigma} \vdash s' \rhd \Omega'}{\Gamma, \vec{z}\,'\colon \vec{\tau}; \Omega, \vec{z} : \vec{\tau} \vdash \mathbf{callcc}(\mathbf{proc}\ (\mathbf{in}\ k; \mathbf{out}\ \vec{z}\,)\ \{\vec{z} := \vec{z}\,';\ \ s\}_{\vec{z}}; \vec{z}\,);\ \ s' \rhd \Omega'}}{\Gamma; \Omega, \vec{z} : \vec{\tau} \vdash \mathbf{cst}\ \vec{z}\,' = \vec{z};\ \ \mathbf{callcc}(\mathbf{proc}\ (\mathbf{in}\ k; \mathbf{out}\ \vec{z}\,)\ \{\vec{z} := \vec{z}\,';\ \ s\}_{\vec{z}}; \vec{z}\,);\ \ s' \rhd \Omega'}
$$

$$
\dfrac{\Gamma; \Omega, \vec{z} : \vec{\tau} \vdash k\colon \neg\vec{\sigma} \qquad \Gamma; \Omega, \vec{z} : \vec{\tau} \vdash \vec{e} : \vec{\sigma}}{\Gamma; \Omega, \vec{z} : \vec{\tau} \vdash \mathbf{throw}(k, \vec{e}\,; \vec{z}\,) \rhd \vec{z} : \vec{\omega}\,'}
$$

□

# Appendix D Examples of imperative programs

In this appendix, we adopt Prawitz style natural deduction for proof trees. Moreover, we will use the substitution rule (in both functional and imperative typing derivations) without explicitly displaying the equations, but only its number.

To begin with, we recall usual axioms of Peano's arithmetic for $+, \times$:

$$\begin{array}{lll} (1) & x+0 & = x \\ (2) & x+\mathbf{s}(i) & = \mathbf{s}(x+i) \\ (3) & x\times 0 & = 0 \\ (4) & x\times\mathbf{s}(i) & = (x\times i)+x \end{array}$$

## D.1 Multiplication

### D.1.1 Multiplication in FD

Let $\mathcal{D}_s$ be the derivation:

$$\dfrac{\dfrac{\dfrac{add\colon \forall p(\mathbf{nat}(p)\Rightarrow\forall q(\mathbf{nat}(q)\Rightarrow \mathbf{nat}(p+q))) \quad z\colon \mathbf{nat}(n\times u)}{(add\ z)\colon \forall q(\mathbf{nat}(q)\Rightarrow\mathbf{nat}(n\times u+q))} \quad x\colon\mathbf{nat}(n)}{(add\ z\ x)\colon \mathbf{nat}((n\times u)+n)}}{(add\ z\ x)\colon\mathbf{nat}(n\times\mathbf{s}(u))}(4)$$

Then:

$$\dfrac{\dfrac{\dfrac{y\colon\mathbf{nat}(m) \quad \dfrac{0\colon\mathbf{nat}(0)}{0\colon\mathbf{nat}(n\times 0)}(3) \quad \mathcal{D}_s}{\mathbf{rec}(y,0,\lambda i.\lambda z.(add\ z\ x))\colon\mathbf{nat}(n\times m)}}{\lambda y.\mathbf{rec}(y,0,\lambda i.\lambda z.(add\ z\ x))\colon\forall m(\mathbf{nat}(m)\Rightarrow\mathbf{nat}(n\times m))}}{\lambda x.\lambda y.\mathbf{rec}(y,0,\lambda i.\lambda z.(add\ z\ x))\colon\forall n(\mathbf{nat}(n)\Rightarrow\forall m(\mathbf{nat}(m)\Rightarrow\mathbf{nat}(n\times m)))}$$

### D.1.2 Multiplication in ID

| | |
|---|---|
| **cst** $mult=$ **proc** (**in** $X,Y$; **out** $Z$) { | $-\ (X\colon\mathbf{nat}(x), Y\colon\mathbf{nat}(y))[Z\colon\top]$ |
| $Z := 0$; | $\mid\ \ [Z\colon\mathbf{nat}(x\times 0)]$ by (3) |
| | $\mid$ |
| **for** $I := 0$ **until** $Y$ { | $\mid\ \ -\ (I\colon\mathbf{nat}(i))[Z\colon\mathbf{nat}(x\times i)]$ |
| $add(Z,X;Z)$; | $\mid\ \ \mid\ \ [Z\colon\mathbf{nat}(x\times\mathbf{s}(i))]$ by (4) |
| $\}_Z$; | $\mid\ \ [Z\colon\mathbf{nat}(x\times y)]$ |
| | $\mid$ |
| $\}_Z$; | $(mult\colon\mathbf{proc}\ \forall x,y(\mathbf{in}\ \mathbf{nat}(x),\mathbf{nat}(y);\mathbf{out}\ \mathbf{nat}(x\times y)))$ |

## D.2 Factorial

Here follows the equations defining the factorial function:

$$\begin{array}{lll} (1) & 0! & = \mathbf{s}(0) \\ (2) & \mathbf{s}(n)! & = n!\times\mathbf{s}(n) \end{array}$$

### D.2.1 Factorial in FD

Let $\mathcal{D}_s$ be the derivation:

$$\dfrac{\dfrac{\dfrac{\mathit{mult}\colon \forall n(\mathbf{nat}(n) \Rightarrow \forall m(\mathbf{nat}(m) \Rightarrow \mathbf{nat}(n \times m))) \quad z\colon \mathbf{nat}(u\mathbf{!})}{(\mathit{mult}\ z)\colon \forall m(\mathbf{nat}(m) \Rightarrow \mathbf{nat}(u\mathbf{!} \times m))} \qquad \dfrac{i\colon \mathbf{nat}(u)}{\mathbf{s}(i)\colon \mathbf{nat}(\mathbf{s}(u))}}{(\mathit{mult}\ z\ \mathbf{s}(i))\colon \mathbf{nat}(u\mathbf{!} \times \mathbf{s}(u))}}{(\mathit{mult}\ z\ \mathbf{s}(i))\colon \mathbf{nat}(\mathbf{s}(u)\mathbf{!})}\,(2)$$

Then:

$$\dfrac{\dfrac{x\colon \mathbf{nat}(p) \qquad \dfrac{1\colon \mathbf{nat}(\mathbf{s}(0))}{1\colon \mathbf{nat}(0\mathbf{!})}\,(1) \qquad \mathcal{D}_s}{\mathbf{rec}(x, 1, \lambda i.\lambda z.(\mathit{mult}\ z\ \mathbf{s}(i)))\colon \mathbf{nat}(p\mathbf{!})}}{\lambda x.\mathbf{rec}(x, 1, \lambda i.\lambda z.(\mathit{mult}\ z\ \mathbf{s}(i)))\colon \forall n(\mathbf{nat}(p) \Rightarrow \mathbf{nat}(p\mathbf{!}))}$$

### D.2.2 Factorial in ID

| | |
|---|---|
| **cst** *fact* = **proc** (**in** $X$; **out** $Z$) { | – $(X\colon \mathbf{nat}(n))[Z\colon \top]$ |
| $Z := 1$; | \| $[Z\colon \mathbf{nat}(0\mathbf{!})]$ by (1) |
| | \| |
| **for** $I := 0$ **until** $X$ { | \| – $(I\colon \mathbf{nat}(i))[Z\colon \mathbf{nat}(i\mathbf{!})]$ |
| **var** $Y := I$; | \| \| $[Y\colon \mathbf{nat}(i)]$ |
| **inc**$(Y)$; | \| \| $[Y\colon \mathbf{nat}(\mathbf{s}(i))]$ |
| *mult*$(Z, Y; Z)$; | \| \| $[Z\colon \mathbf{nat}(\mathbf{s}(i)\mathbf{!})]$ by (4) |
| $\}_Z$; | \| $[Z\colon \mathbf{nat}(n\mathbf{!})]$ |
| | \| |
| $\}_Z$; | $(\mathit{fact}\colon \mathbf{proc}\ \forall n(\mathbf{in}\ \mathbf{nat}(n); \mathbf{out}\ \mathbf{nat}(n\mathbf{!})))$ |

## D.3 Ackermann function

Here follows the equations defining a version of the Ackermann function (from [49]):

$$\begin{array}{lll} (1)\ \mathbf{a}(0, n) & = & \mathbf{s}(n) \\ (2)\ \mathbf{a}(\mathbf{s}(z), 0) & = & \mathbf{s}(\mathbf{s}(0)) \\ (3)\ \mathbf{a}(z, \mathbf{a}(\mathbf{s}(z), u)) & = & \mathbf{a}(\mathbf{s}(z), \mathbf{s}(u)) \end{array}$$

### D.3.1 Ackermann function in FD

Here follows an annotated version of the proof given in [49]. Let $\mathcal{D}_s$ be the derivation:

$$\dfrac{\dfrac{y\colon \mathbf{nat}(n) \qquad \dfrac{\dfrac{\dfrac{0\colon \mathbf{nat}(0)}{S(0)\colon \mathbf{nat}(\mathbf{s}(0))}}{S(S(0))\colon \mathbf{nat}(\mathbf{s}(\mathbf{s}(0)))}}{S(S(0))\colon \mathbf{nat}(\mathbf{a}(\mathbf{s}(z), 0))}\,(2) \qquad \dfrac{\dfrac{f\colon \forall n(\mathbf{nat}(n) \Rightarrow \mathbf{nat}(\mathbf{a}(z, n))) \quad k\colon \mathbf{nat}(\mathbf{a}(\mathbf{s}(z), u))}{(f\ k)\colon \mathbf{nat}(\mathbf{a}(z, \mathbf{a}(\mathbf{s}(z), u)))}}{(f\ k)\colon \mathbf{nat}(\mathbf{a}(\mathbf{s}(z), \mathbf{s}(u)))}\,(3)}{\mathbf{rec}(y, S(S(0)), \lambda j.\lambda k.(f\ k))\colon \mathbf{nat}(\mathbf{a}(\mathbf{s}(z), n))}}{\lambda y.\mathbf{rec}(y, S(S(0)), \lambda j.\lambda k.(f\ k))\colon \forall n(\mathbf{nat}(n) \Rightarrow \mathbf{nat}(\mathbf{a}(\mathbf{s}(z), n)))}$$

Then:

$$\dfrac{\dfrac{x\colon \mathbf{nat}(m) \qquad \dfrac{\dfrac{\dfrac{y\colon \mathbf{nat}(\mathbf{s}(n))}{y\colon \mathbf{nat}(\mathbf{a}(0,n))}(1)}{S(y)\colon \mathbf{nat}(\mathbf{a}(0,n))}}{\lambda y.S(y)\colon \forall n(\mathbf{nat}(n)\Rightarrow \mathbf{nat}(\mathbf{a}(0,n)))} \qquad \mathcal{D}_s}{\mathbf{rec}(x,\lambda y.S(y),\lambda i.\lambda f.\lambda y.\mathbf{rec}(y,S(S(0)),\lambda j.\lambda k.(f\ k)))\colon \forall n(\mathbf{nat}(n)\Rightarrow \mathbf{nat}(\mathbf{a}(m,n)))}}{\lambda x.\mathbf{rec}(x,\lambda y.S(y),\lambda i.\lambda f.\lambda y.\mathbf{rec}(y,S(S(0)),\lambda j.\lambda k.(f\ k)))\colon \forall m(\mathbf{nat}(m)\Rightarrow \forall n(\mathbf{nat}(n)\Rightarrow \mathbf{nat}(\mathbf{a}(m,n))))}$$

## D.4 Typing derivations for shift and reset in ID$^c$

### D.4.1 Typing derivation for reset

```
proc(in p; out r)_mk{
    k: {
        cst m = mk;

        mk := proc(in r; out z){
            jump(k, r, m)_z;
        }_z;

        var y;
        p(; y)_mk;
        jump(mk, y)_{r,mk};
    }_{r,mk};
}_{r,mk};
```

```
– (p: proc(in ¬α; out β, ¬β), mk′: ¬γ)[r: ⊤, mk: ¬γ]
|   – (k: cont(α, ¬γ))[r: ⊤, mk: ¬γ]
|   |   (m: ¬γ)
|   |
|   |   – (r: α)[z: ⊤]
|   |   |   [z: ⊥]
|   |   [mk: ¬α]
|   |
|   |   [y: ⊤]
|   |   [y: β, mk: ¬β]
|   |   [r: α, mk: ¬γ]
|   [r: α, mk: ¬γ]
proc(in proc(in ¬α; out β, ¬β), ¬γ; out α, ¬γ)
```

### D.4.2 Typing derivation for shift

```
proc(in p; out r)_mk{

    k: {

        proc q(in v; out r)_mk{

            reset(proc(out z)_mk{
                jump(k, v, mk)_{z,mk};
            }_{z,mk};
            r)_mk;
        }_{r,mk};

        var y;
        p(q; y)_mk;
        jump(mk, y)_{r,mk};
    }_{r,mk};
}_{r,mk}
```

```
– (p: proc(in proc(in α, ¬β; out γ, ¬β), ¬δ; out ϵ, ¬ϵ))
– (mk′: ¬δ)[r: ⊤, mk: ¬δ]
|   – (k: ¬(α, ¬γ))[r: ⊤, mk: ¬δ]
|   |
|   |   – (v: α, mk′: ¬β)[r: ⊤, mk: ¬β]
|   |   |
|   |   |   – (mk′: ¬γ)[z: ⊤, mk: ¬γ]
|   |   |   |   [z: η, mk: ¬η]
|   |   |   proc(in ¬γ; out η, ¬η)
|   |   |   [r: γ, mk: ¬β]
|   |   (q: proc(in α, ¬β; out γ, ¬β))
|   |
|   |   [y: ⊤]
|   |   [y: ϵ, mk: ¬ϵ]
|   |   [r: α, mk: ¬γ]
|   [r: α, mk: ¬γ]
proc(in proc(in proc(in α, ¬β; out γ, ¬β), ¬δ;
                out ϵ, ¬ϵ), ¬δ;
        out α, ¬γ)
```

# Appendix E Shift and reset in state-passing style

```
signature CONT = sig

  type void
  type 'a K = 'a → void

  val callcc: ('a K → 'a) → 'a
  val throw: 'a K → 'a → 'b

end

functor ShiftReset(Cont: CONT) = struct

  open Cont

  val reset: ('a K → 'c * 'c K) * 'd K → 'a * 'd K =
   fn (p, mk') ⇒
      let
        val (r, mk) = ((), mk')
        val (r', mk') = (r, mk)
        val (r, mk) =
            callcc (fn k ⇒
                    let
                      val (r, mk) = (r', mk')
                      val m = mk
                      val mk = fn r ⇒
                               let val z = throw k (r, m)
                               in z end
                      val (y, mk) = p(mk)
                      val (r, mk) = throw mk y
                    in (r, mk) end)
      in (r, mk) end

  val shift: (('a * 'b K → 'c * 'b K) * 'd K → 'e * 'e K) * 'd K → 'a * 'c K =
   fn (p, mk') ⇒
      let
        val (r, mk) = ((), mk')
        val (r', mk') = (r, mk)
        val (r, mk) =
            callcc (fn k ⇒
                    let
                      val (r, mk) = (r', mk')
                      val q = fn (v, mk) ⇒
                              let val (r, mk) =
                                      reset (fn mk ⇒
                                             let val (z, mk) = throw k (v, mk)
                                             in (z, mk) end, mk)
                              in (r, mk) end
                      val (y, mk) = p(q, mk)
                      val (r, mk) = throw mk y
                    in (r, mk) end)
      in (r, mk) end
end
```

# Table of contents